\documentclass[12pt]{article}
\usepackage[letterpaper,top=1in,bottom=1in,left=1in,right=1in]{geometry}
\usepackage{mathptmx} 
\usepackage{setspace}
\usepackage{amsmath,amssymb}
\usepackage{graphicx}
\usepackage{booktabs}
\usepackage{longtable}
\usepackage{float}
\usepackage[colorlinks=true, allcolors=black]{hyperref}
\usepackage[utf8]{inputenc}
\usepackage{caption}
\usepackage{hanging}
\usepackage{rotating}
\usepackage{threeparttable}
\usepackage{tabularx}
\newcolumntype{C}{>{\centering\arraybackslash}X}
\newcolumntype{R}{>{\raggedleft\arraybackslash}X}
\newcolumntype{L}{>{\raggedright\arraybackslash}X}

\title{Strategic Coercion Within Alliances: The Greenland Sovereignty Game as an AI Stress Test}
\author{Rommin Adl \and Peyton Williams}
\date{May 2026}

\begin{document}
\thispagestyle{empty}
\begin{center}
\vspace*{-0.6em}
{\Huge Strategic Coercion Within Alliances:}\\[0.25em]
{\huge The Greenland Sovereignty Game}\\[0.15em]
{\huge An AI Stress Test via Inverse Game Theory on 8 Frontier Models}\\[0.45em]
{\Large Rommin Adl \hspace{2.5em} Peyton Williams}\\[0.2em]
{\normalsize\itshape Grinnell College}\\[0.3em]
{\normalsize May 2026}
\end{center}
\vspace{-0.1em}

\begin{abstract}
\begingroup\setstretch{1.14}

What happens when the strongest alliance member pressures a weaker member
over territory and strategic control? This paper examines the Greenland
sovereignty crisis as a stress test for LLM geopolitics, centered on the
2019--2026 U.S. push to acquire or control Greenland from the Kingdom of
Denmark. The crisis is two nested collective-action problems: a
first-order problem over Arctic strategic control and a second-order
problem within NATO over whether alliance norms can be enforced against
the dominant member. Building on Ferguson (2013), Fehr-Schmidt (1999),
and Rabin (1993), we develop three games: an asymmetric coercion game, a
NATO enforcement assurance game with a critical-mass tipping point, and
a triadic extensive-form game with social preferences. We test these
predictions with a multi-agent LLM simulation in which eight frontier
models play six geopolitical roles (United States, Denmark, Greenland,
NATO, Russia, and Canada). Across the completed design, models play
3,604 games and produce 108,120 action observations; the raw archive
contains 3,615 files (108,450 design-slot observations) because
11 tournament files are retained reruns used for audit-trail checks.
Using inverse game theory, we
recover each model's structural utility parameters \((\alpha,\beta,
\gamma,\delta,\eta)\), capturing material self-interest, reciprocity,
inequality aversion, norm respect, and commitment consistency. Three
findings stand out. First, all eight models become more escalatory when
the scenario is framed around possible U.S. coercion. Second,
Chinese-origin models show systematically different power-weight profiles
from Western-origin models. Third, peaceful US acquisition emerges in
only 1.9\% of clean games and is achievable by only 3 of 8 frontier
models, most prominently DeepSeek V3.2, which executes a stable
five-round playbook running entirely through the metropole; the same
coercion priming that universally activates escalation halves the
conditional probability of any peaceful equilibrium. Prompts
emphasizing \emph{jus cogens} and Greenland's self-determination rights
reduce escalation sharply in the English-only sample, bringing rates
back near baseline.
The main contribution is structural parameter recovery for framing and
coalition dynamics in the English-only confirmatory sample; multilingual
contrasts are exploratory sensitivity checks. We position this as a
structural benchmark for LLM geopolitical behavior, complementing
action-frequency benchmarks rather than replacing them.
\par\endgroup

\end{abstract}

\vspace{-0.05em}
\begingroup
\setstretch{1.0}
\noindent{\tiny\raggedright \textbf{JEL Classification:} C45, C72, C73, D74, F51, F53, K33, O33.\par}
\noindent{\tiny\raggedright \textbf{Keywords:} large language models, inverse game theory, structural estimation, frontier-model evaluation, NATO, Greenland, coercion, self-determination, jus cogens, multi-agent simulation, geopolitical AI, alliance enforcement.\par}
\endgroup
\vspace{-0.2em}

\section{Introduction}\label{sec:introduction}

In March 2025, President Trump told Congress regarding Greenland: ``We
strongly support your right to determine your own future\ldots{} One way
or the other, we're going to get it'' (CNBC, 2025). He also refused to
rule out the use of military or economic force to acquire the island, an
autonomous territory within the Kingdom of Denmark, home to roughly
56,000 people, and site of Pituffik Space Base, a key part of NORAD's
ballistic missile early-warning system (Muthukumar, 2025; Ash, 2022).
Denmark responded with a major Arctic defense buildup. In January 2025,
Denmark committed DKK 14.6 billion to the First Agreement on the Arctic
and North Atlantic, and in October 2025 it added DKK 27.4 billion
through the Second Agreement, totaling DKK 42 billion, or roughly 6
billion USD, in new Arctic and North Atlantic defense commitments
(Danish Ministry of Defence, 2025a; Danish Ministry of Defence, 2025b).
By early 2026, analysts had begun describing this emerging Arctic
posture as the ``Donroe Doctrine,'' linking Trump's Greenland pressure
to a revived Monroe Doctrine logic in the Arctic (Edwards, 2026;
\O{}sthagen, 2026).\footnote{Several 2026 citations in this paper (Solopova; Tewolde; Liao; Smirnov; Lamazhapov; Edwards; \O{}sthagen) are working papers, preprints, or recent commentary published within months of this draft; we treat them as informative context rather than canonical references.}

This paper asks two questions about alliance coercion in the Greenland
sovereignty game. First, what happens when the strongest
member of an alliance pressures a weaker member over territory and
strategic control? Second, and more importantly for political economy:
why are alliances often better at deterring external rivals than at
restraining internal coercion?

Our thesis has two layers. First, Greenland is a first-order
collective-action problem over strategic control in the Arctic: who controls the rare earth minerals, the GIUK Gap surveillance architecture,
and the shipping routes opened by climate change. Second, and more
consequentially, Greenland is a second-order collective-action problem
within NATO, as even if most alliance members privately support a norm
against internal coercion, the enforcement of that norm is itself a
public good subject to free-riding (Ostrom, 1990). Each member has an incentive to hope
somebody else bears the cost of confronting the United States.

We state five hypotheses at the outset, before the formal model, so the
reader knows what the theory is for:

\begin{enumerate}
\def\labelenumi{\arabic{enumi}.}
\item
  H1 (Language Sensitivity). The same model's choices depend on prompt
  language. Danish or Mandarin prompts can shift actions relative to
  English by activating different corpus-conditioned assumptions about
  power and politics.
\item
  H2 (Model Provenance). Chinese-origin models, such as DeepSeek, GLM,
  and Kimi, are expected to show systematically different power-weight
  profiles from Western-trained models when playing the U.S. role,
  consistent with documented geographic bias in LLM outputs (Chang et
  al., 2025).
\item
  H3 (Coercion Activation). When the prompt frames U.S. acquisition of
  Greenland as a real possibility, every model is expected to become
  more escalatory. This fits Game 2's prediction that coercive moves
  trigger backlash and make other actors more likely to respond
  aggressively.
\item
  H4 (Normative Constraint). When the prompt emphasizes international
  law and Greenland's right to self-determination, models are expected
  to become less escalatory. This fits the idea that stronger legal and
  moral costs make defection harder to justify.
\item
  H5 (Coalition Spoiler Effect). Models are expected to become more escalatory
  when Russia is included as a spoiler. This fits Game 2's prediction
  that one actor refusing to cooperate can weaken the whole coalition
  and push it below the level needed for joint enforcement.
\end{enumerate}

Three existing benchmarks measure adjacent but distinct things. The
Critical Foreign Policy Decisions Benchmark
measures how often models choose escalation or de-escalation across
crisis scenarios (Jensen et al., 2025). The AIRI Geopolitical Bias study
measures how model responses line up with different geopolitical
narratives and political frames (Salnikov et al., 2025a). Qian et al.\ examine
how LLMs bargain in a trading game with human participants (Qian et al.,
2026). Additionally, none of these benchmarks model the perspective of small states or contested territories as active strategic players. By giving Greenland its own player role with strategic agency, our simulation addresses a systematic gap in the AI geopolitics literature, where the subjects of sovereignty disputes are treated as objects rather than actors. The Greenland scenario is also not a random case selection: independent benchmarking reports lower human--model alignment in Arctic-sovereignty scenarios than in the other scenarios tested, making it a demanding stress test for LLM geopolitical consistency (Solopova et al., 2026).

Our paper contributes three linked advances. First, we are not only
asking what action a model chooses; we recover the preference
structure behind those choices. Specifically, we estimate each model's
structural utility parameters: \(\alpha\), \(\beta\), \(\gamma\),
\(\delta\), and \(\eta\). These represent material self-interest,
reciprocity, inequality aversion, respect for norms, and commitment
consistency. So, while earlier benchmarks mostly report surface-level
outcomes, such as how often a model escalates, our paper tries to
explain why the model escalates by identifying the utility weights that
produce that behavior. Second, the simulation is trilingual: models are prompted in English, Danish, and Mandarin, where each language maps onto a real player identity in the crisis (hegemon, defender, and external observer). To our knowledge, no prior LLM geopolitical benchmark has tested across these three languages, making language a major conditioning variable rather than just a translation layer.

Most importantly, the largest prior study of LLM escalation risk reports difficult-to-predict escalation patterns across models (Rivera et al., 2024). The most comprehensive survey of game theory and LLMs, published at IJCAI-25, surveys the field and we are not aware of prior work that applies inverse game theory to estimate LLM utility functions in an international relations context (Sun et al., 2025). Our $\theta$ estimation framework speaks to this gap: by recovering structural utility weights from observed action distributions, we provide an estimation framework of the kind that survey identified as missing. Our structural approach is also motivated by findings that naive debiasing, such as instructing models to ``be unbiased,'' can paradoxically increase geopolitical bias (Salnikov et al., 2025a), and that RLHF safety tuning suppresses surface bias on domestic political topics while leaving geopolitical biases largely intact in the model's underlying probability distributions (Kim and Kim, 2025). Rather than attempting to remove bias through prompting, we measure the underlying preference weights directly.

The rest of the paper is organized as follows. Section II gives the
historical background of the Greenland sovereignty dispute, from earlier
U.S. interest in acquiring Greenland to the 2009 Self-Government Act and
the 2025 Donroe Doctrine. Section III develops the theory through three
games, moving from a simple U.S.-Denmark coercion game to a NATO
enforcement game and then to a fuller three-player game involving the
U.S., Denmark, and Greenland. Section IV explains the simulation design
and presents the results. Section V connects those results back to the
theory. Section VI discusses the paper's limitations, including the fact
that budget limits prevented us from completing the full simulation.
Section VII concludes.

\section{Descriptive Background: The Greenland Sovereignty Game, 1941--2026}\label{ii.-descriptive-background-the-greenland-sovereignty-game-19412026}

The Greenland sovereignty issue is a real political conflict with a long
history, including earlier U.S. efforts to acquire Greenland and decades
of U.S.-Danish defense cooperation on the island (Saalbach, 2024;
Fischer, 1993).

That history includes a secret U.S. offer to buy Greenland in 1946, a
1951 defense agreement made without direct Greenlandic consultation, and
the 2009 Act on Greenland Self-Government, which gave Greenland greater
authority over its own international affairs (Saalbach, 2024; Fischer,
1993; Ackren, 2019). More recently, in 2025, Denmark committed DKK 14.6
billion in January and another DKK 27.4 billion in October to strengthen
defense in the Arctic and North Atlantic (Danish Ministry of Defence,
2025a; Danish Ministry of Defence, 2025b).

This section traces the Greenland bargaining game across four periods.
The first period, from 1941 to 1979, was mainly a two-player game
between the United States and Denmark. The second period, from 1979 to
2009, saw Greenland gradually emerge as a more independent third player.
The third period, from 2019 to 2024, marked a more realist turn, as
Greenland became more central to great-power competition. The final
period focuses on the current crisis, where the stakes are no longer
hypothetical.

\subsection{The F6 Origin: The Two-Player Game, 1941--1979}\label{ii.1-the-f6-origin-the-two-player-game-19411979}

The structure of the Greenland sovereignty game did not begin with
Donald Trump's 2019 proposal to acquire Greenland. It began much
earlier, with the secret 1946 U.S. offer to buy Greenland and the 1951
Defense Agreement that gave the United States long-term military access
to the island (Saalbach, 2024; Fischer, 1993).

When Germany invaded Denmark in April 1940, Denmark's ambassador to
Washington, Henrik Kauffmann, signed a defense agreement with the United
States that allowed U.S. forces to operate in Greenland. The Danish
government in Copenhagen, which was under German occupation, rejected
the agreement but could not reverse it. During World War II, the United
States built weather stations, airfields, and communications
infrastructure across Greenland, turning the island into a major
strategic asset (Saalbach, 2024).

This wartime arrangement created an important precedent: U.S. military
presence in Greenland existed before NATO itself. It also revealed the
basic structure of the conflict. Denmark was weakened by occupation,
Greenlanders were not meaningfully consulted, and the United States
treated Greenland as a strategic territory to be managed through
Denmark. Greenland was treated as if sovereignty were mainly a
two-player issue between Denmark and the United States.

In 1946, the Truman administration secretly offered Denmark \$100
million in gold to purchase Greenland. Denmark rejected the offer, but
the proposal shows that Trump's later interest in buying Greenland was
not completely new. Instead, it fits into a longer pattern of U.S.
attempts to gain control or deeper access to Greenland, including
earlier efforts in the 1860s, 1916, 1946, 2019, and 2025 (Saalbach,
2024). This repeated interest suggests that the United States has
consistently placed high strategic value on Greenland.
Read as a sequence, those five attempts also support a stable
parameter-profile interpretation across administrations: persistently
high \(\alpha\) and comparatively low \(\delta\) in the U.S. role.

The 1951 Defense Agreement made the wartime arrangement permanent. Under
this agreement, the United States gained extensive military access to
Greenlandic territory, including the site that became Pituffik Space
Base, formerly known as Thule Air Base. Fischer argues that the
agreement was signed without meaningful Greenlandic consultation, and
Greenland would not receive Home Rule until 1979 (Fischer, 1993). For
nearly four decades, the Greenland issue therefore operated as what
Ferguson, Kelsall, and Schulz call an F6 dominant network, which is a
two-player structure where a powerful actor can shape another actor's
choices while a weaker third party is excluded from the bargaining table
(Ferguson, Kelsall, and Schulz, 2022).

The consequences of this F6 configuration were serious for Greenland's
Indigenous population. In 1953, the construction of Thule Air Base
forced the relocation of the Inughuit community, a displacement that
remained politically important in Greenland for decades. The base also
hosted nuclear weapons under the U.S. forward deployment system, which
became internationally visible in 1968 when a B-52 carrying four
hydrogen bombs crashed near Thule (Saalbach, 2024). This incident
exposed a gap between Denmark's public nuclear-free policy and the
reality of U.S. military operations in Greenland. In our model, that gap
relates to \(\eta\), or commitment consistency, because it shows a
difference between stated commitments and actual behavior.

The period from 1951 to 1979 also established the GIUK Gap as one of the
main strategic reasons Greenland matters. The GIUK Gap is the maritime
chokepoint between Greenland, Iceland, and the United Kingdom. Russian
submarines moving from the Kola Peninsula into the North Atlantic must
pass through this area, and NATO's ability to monitor them depends
heavily on surveillance systems connected to Greenland (Ash, 2022). This
makes Greenland difficult to replace strategically. Whoever has
influence over Greenland also has influence over a key part of North
Atlantic defense.

Denmark also used the U.S. base presence as a form of leverage within
the alliance. Scholars describe this as the ``politics of
embarrassment,'' where Denmark could pressure Washington by threatening
to expose politically uncomfortable information about U.S. operations in
Greenland, including nuclear deployments (Altunkaya, 2026). This is an
early example of Power 2, or agenda control, because Denmark used the
alliance relationship itself as a pressure tool. However, Greenland
remained the silent third party whose interests were shaped by decisions
made by Denmark and the United States.

Overall, this early period created an F6-style two-player structure that
shaped the Greenland sovereignty game for decades. The United States and
Denmark bargained over Greenland's strategic value, while Greenland
itself was mostly excluded. The later simulation tests whether LLMs
reproduce that same exclusion pattern by treating Greenland as an object
of bargaining rather than as an active political actor.

\subsection{Home Rule to Self-Government: The Emergence of a Third Player, 1979--2009}\label{ii.2-home-rule-to-self-government-the-emergence-of-a-third-player-19792009}

The 1979 Home Rule Act began Greenland's shift from an F6 two-player
structure toward a three-player bargaining game. Before this point,
Greenland was mostly treated as the object of U.S.-Danish bargaining.
Home Rule gave Greenland more domestic authority, but the transition was
incomplete. Greenland did not gain stronger international-relations
capacity until the 2009 Act on Greenland Self-Government, which gave
them authority in certain external affairs areas (Ackren, 2019).

The 2009 Act also created a major constraint. It recognized that ``the
decision on independence is made by the Greenlandic people,'' meaning
Greenland's future political status could no longer be decided by
Denmark alone (Ackren, 2019). But the same arrangement preserved
Denmark's annual block grant, worth about DKK 3.4 billion and roughly
60\% of Greenland's public revenue (Ackren, 2019). This creates what we
call a constitutional economic veto, as Greenland has the legal right to
pursue independence, but it does not yet have the fiscal capacity to
make independence easy or immediately credible.

Independence would require Greenland to give up the
block grant. With a population of about 56,000 and an economy heavily
dependent on fishing, especially shrimp and halibut, Greenland does not
have an obvious way to replace that revenue through taxation alone
(Ackren, 2019). The block grant works like a selective incentive against
independence, as even if many Greenlanders support self-determination,
continued association with Denmark can still look individually rational
because it preserves government funding.

Between 1979 and 2009, Greenland gradually gained the institutional
capacity that makes it a real player in our simulation. By 2019,
Greenland had diplomatic representation in Washington, Brussels,
Beijing, Reykjavik, and Copenhagen, and it had built direct
relationships with outside states such as the United States, China, and
Iceland (Eggertsd\'ottir, 2024; Leander Nielsen and Strandsbjerg, 2024). This reflects
the principle ``Nothing About Us Without Us'': Greenland increasingly rejected decisions made
over its head.

The Arctic Council also helped Greenland act internationally without
fully separating from Denmark. Created in 1996, the Arctic Council gave
Greenland a forum to participate in Arctic governance on issues such as
fisheries, environmental policy, and Indigenous rights. Greenland's
participation alongside Denmark shows why INSTITUTIONAL\_APPEAL, one of
the actions present in our later simulation, is a realistic action.
Greenland can use international forums to gain voice without immediately
choosing full independence.

In short, the three-player bargaining game did not fully exist before
2009. The 1979 Home Rule Act started the transition, but the 2009
Self-Government Act gave Greenland the legal and institutional basis to
become a more direct actor in the sovereignty game.

\subsection{The Realist Turn: Three Shocks, 2019--2024}\label{ii.3-the-realist-turn-three-shocks-20192024}

Three shocks turned Greenland from a quiet strategic asset into an
active great-power flashpoint.

First, Trump's 2019 proposal to purchase Greenland reopened the older F6
logic that the 2009 Self-Government Act had tried to move beyond. The
proposal treated Greenland less like an independent political actor and
more like an object of bargaining between the United States and Denmark.
The dispute became more serious when Trump canceled a planned state
visit after Danish Prime Minister Mette Frederiksen called the idea
``absurd.'' Altunkaya argues that the 2019 offer was not just a random
comment, but part of a broader transactional turn in U.S. Arctic policy
(Altunkaya, 2026).

Second, Russia's 2022 invasion of Ukraine weakened the cooperative
Arctic order. Before 2022, the Arctic Council had often served as a
forum where states could bracket sovereignty disputes and focus on
environmental and scientific cooperation. But this cooperation was
already under pressure before the invasion. In 2019, the Rovaniemi
ministerial meeting failed to produce a joint declaration after the
United States objected to climate-change language, breaking a
long-standing Arctic Council precedent (Altunkaya, 2026). After Russia's
suspension from the Arctic Council, Arctic politics became more openly
competitive, especially as Russia remained militarily active and China
continued presenting itself as a ``near-Arctic state.''

China matters because it gives Greenland and Denmark an outside option,
which raises the stakes of the U.S.-Denmark-Greenland game. In 2018,
China released its first Arctic white paper, calling itself a
``near-Arctic state'' and promoting a ``Polar Silk Road'' strategy
(State Council Information Office of the People's Republic of China,
2018). Chinese interest in Greenlandic mining and infrastructure created
a triangle between Beijing, Nuuk, and Copenhagen because Greenland
wanted economic development, while Denmark and the United States were
more concerned about security risks from Chinese investment (Danish
Institute for International Studies {[}DIIS{]}, 2021; Clingendael
Institute, 2020). The United States pressured Denmark to exclude China
Communications Construction Company from Greenland airport projects, and
the company eventually withdrew from the bidding process (DIIS, 2021).
In Ferguson, Kelsall, and Schulz's terms, this resembles an F7-style
configuration, meaning a competitive network where two relatively
powerful actors compete for influence over a weaker third actor. In this
case, the United States and China are the competing major powers, while
Greenland is the smaller strategic actor whose resources, location, and
political alignment become the object of competition (Ferguson, Kelsall,
and Schulz, 2022).

Third, Greenland's own position on militarization changed quickly. In
2021, Greenland's government said it opposed increased military activity
on the island. By 2023, that position had shifted, with Greenlandic
officials arguing that the island would need to host more military
activity (Leander Nielsen and Strandsbjerg, 2024). This shift reflected a
changing strategic reality in that Trump's acquisition rhetoric,
Russia's Arctic militarization, and Greenland's long-term independence
goals all made security partnerships more important.

This change matters for the simulation because Greenland is not a
passive target. U.S. pressure can create a backlash effect, where
coercion increases Greenlandic resistance instead of weakening it. In
the language of our model, this is connected to negative reciprocity, as
when a dominant actor applies pressure, the target may respond by
resisting more strongly. Any model that treats Greenland as simply
something to be traded between Denmark and the United States misses this
critical point.

By the end of 2024, the Arctic had become more unstable because of three
connected forces: resource competition, militarization, and
institutional erosion (Blunden, 2009). These forces are what make the
Greenland sovereignty game important for the simulation. The crisis is
not only about territory. It is also about how smaller
actors respond when great powers compete over their strategic position.

\subsection{Live Stakes: The Donroe Doctrine, 2025--2026}\label{ii.4-live-stakes-the-donroe-doctrine-and-the-14.6b-dkk-package-20252026}

The bargaining game is not hypothetical. By 2025, Greenland had become
part of an active crisis involving the United States, Denmark,
Greenland, NATO, and other Arctic powers.

In January 2025, Trump refused to rule out military or economic force to
acquire Greenland. This matters because the statement was not a normal
diplomatic offer. It created a form of grey-zone coercion: pressure
that falls short of direct invasion but still goes beyond
ordinary bargaining. Under Article 2(4) of the UN Charter, threats or
uses of force against another state's territorial integrity are
prohibited, which is why this kind of statement raises serious legal and
political concerns (Muthukumar, 2025). In the terms of our model, this
is a high own material payoff (\(\alpha\)), and low norm cost
(\(\delta\)) action, as it prioritizes strategic gain while treating
norm-violation costs as relatively weak.

The Greenland push also appears to be about more than resources. U.S.
military access in Greenland already exists through Pituffik Space Base,
so the acquisition push is not simply about meeting basic defense needs.
Lamazhapov argues that the policy is better understood as a status move,
or an effort to place Greenland more clearly inside the U.S. sphere of
influence and signal Arctic dominance (Lamazhapov, 2026). This matters
for the simulation because a model that treats the U.S. as motivated
only by material payoff may miss the status logic behind the policy.
In this paper's parameterization, that implies \(\alpha\) should be read
broadly as strategic self-interest, including status, military
geography, and sphere-of-influence benefits, not only direct resource
payoffs.

Another important shift is the use of NORAD rather than NATO as the main
Arctic security channel. Ugeda and Sanches argue that routing Arctic
defense through NORAD, a U.S.-Canada arrangement, can exclude European
allies from decisions that would otherwise involve the wider alliance
(Ugeda and Sanches, 2025). This matters for Game 2 because NATO
enforcement depends on a broad coalition. If Arctic security is handled
through a smaller U.S.-Canada framework, the number of available
enforcers falls, making it harder to reach the threshold needed for
collective enforcement.

Denmark's response has been a major Arctic defense buildup. In January
2025, Denmark committed DKK 14.6 billion to the First Agreement on the
Arctic and North Atlantic, including Arctic naval vessels, long-range
drones, and satellite capacity (Danish Ministry of Defence, 2025a). In
October 2025, Denmark added another DKK 27.4 billion through the Second
Agreement, including more Arctic vessels, maritime patrol aircraft,
drones, a North Atlantic undersea cable, a new Joint Arctic Command
headquarters in Nuuk, and an East Greenland air-surveillance radar
(Danish Ministry of Defence, 2025b). Denmark also signed a DKK 29
billion F-35 expansion order for 16 additional fighter jets, bringing
its fleet to 43 (Reuters, 2025; Gronholt-Pedersen and Carlsson, 2025).

By early 2026, some commentators began describing this emerging posture
as the ``Donroe Doctrine,'' a recent coinage in the policy press (Edwards, 2026; \O{}sthagen, 2026). The term plays on the Monroe Doctrine and
refers to a new Arctic idea in which Trump's Greenland pressure forced
Europe to think about deterrence not only against external rivals, but
also against pressure from within the alliance (Edwards, 2026; \O{}sthagen,
2026). In our model, Denmark's defense buildup can be understood as an
attempt to make enforcement look individually rational. The basic idea
is simple. A NATO member will only help resist coercion if the expected
benefit of joining an enforcement coalition is greater than the benefit
of staying quiet. Denmark's buildup signals that at least one member is
already willing to bear part of the cost. That lowers the number of
additional allies needed to make collective enforcement credible. Put
more simply, Denmark is not waiting for NATO to act first. It is trying
to make it easier for other allies to join.

Russia's Arctic ambitions add another layer of pressure. Russia's
extended continental shelf claim, accepted by the UN Commission on the
Limits of the Continental Shelf in 2023, overlaps strategically with the
broader Arctic competition surrounding Greenland (Saalbach, 2024).
Russia's Arctic interest is also resource-driven, since the Arctic
contains large shares of the world's undiscovered oil and natural gas
(Saalbach, 2024). This makes Greenland's location even more important
because the island sits near key military, maritime, and resource
routes.

Greenland itself cannot easily defend independence on its own. Ash
estimates that even a single air defense battery to protect Nuuk would
cost \$800 million to \$950 million, which is far beyond what a
territory of roughly 56,000 people and a small economy can easily afford
(Ash, 2022). This means Greenland's legal right to independence does not
automatically create a credible defense capacity. In our later
simulation, this is why declaring independence is a legal option but
not always a fiscally credible one without outside support.

\subsection{The Legal Ceiling: Jus Cogens and the $\delta$ Parameter}\label{subsec:legal-ceiling}

One final constraint applies across the whole Greenland sovereignty
game, and that is international law. Under UN General Assembly
Resolution 1514 and the 2007 UN Declaration on the Rights of Indigenous
Peoples, any attempt to change Greenland's political status against
the will of the Greenlandic people would raise serious self-determination concerns
under international law (UNGA, 1960; UNGA, 2007; Saalbach, 2024). In other words, a forced ``acquisition'' of Greenland
would conflict with widely recognized principles of self-determination.

This matters because self-determination is not just a normal diplomatic
preference. Some legal interpretations treat self-determination as a jus cogens or non-derogable norm,
meaning it cannot simply be overridden by a treaty, a bilateral deal,
or coercive pressure from a stronger state (UNGA, 1960; UNGA, 2007). Put simply, Denmark and the United States
cannot lawfully decide Greenland's future over Greenland's own
objection.

This legal ceiling gives real meaning to the \(\delta\) parameter in our
model. Here, \(\delta\) represents the cost of violating a norm. In the
Greenland case, that cost is reputational and tied to international law,
with possible UN condemnation, diplomatic backlash, and other legal or
political consequences. So when a model recommends coercive acquisition
without recognizing this cost, it is both making an aggressive
geopolitical choice and failing to account for the legal structure of
the dispute.

This also separates our simulation from a purely strategic game. In a
simple power-based model, a state might violate a norm if the material
payoff is high enough. But under jus cogens, some actions are legally
off-limits no matter how valuable the payoff is. In model terms, no
level of \(\alpha\), or material self-interest, makes forced acquisition
legally acceptable. If an LLM playing the U.S. role recommends
FORCE\_ACQUISITION with high confidence, the simulation can detect that
as a failure to properly weight \(\delta\), the norm-violation cost.

\subsection{The Players: A Summary for the Simulation}\label{ii.5-the-players-a-summary-for-the-simulation}

\begin{table}[H]
\centering
\begin{threeparttable}
\caption{Game Players}
\label{tab:players}
\small
\begin{tabularx}{\textwidth}{LLLL}
\toprule
\textbf{Player} & \textbf{Role} & \textbf{Key Parameters} & \textbf{Sources} \\
\midrule
United States & Acquirer / hegemon & High $\alpha$, high $\eta$, low $\delta$; status-driven & Saalbach (2024); Muthukumar (2025); Ugeda \& Sanches (2025); Lamazhapov (2026) \\[4pt]
Denmark & Defender / alliance anchor & High $\delta$, negative $\beta$ under coercion (punisher) & Fischer (1993); Danish Ministry of Defence (2025a; 2025b); Edwards (2026); {\O}sthagen (2026) \\[4pt]
Greenland & Contested party with agency & DKK 3.4B block grant constrains $\alpha$; orientation can flip from pro-Denmark to pro-independence under coercion & Ackren (2019); Leander Nielsen and Strandsbjerg (2024); Eggertsd\'ottir (2024) \\[4pt]
NATO & Institutional enforcement & $C(n+1)/D(n)$ game; $n^{*}$ determines enforcement & Altunkaya (2026); Olson \& Zeckhauser (1966); Ferguson (2013) \\[4pt]
Canada & NORAD partner, Arctic neighbor & NORAD overlap; loyalty split between US and alliance norms & Ash (2022); Ugeda \& Sanches (2025) \\[4pt]
Russia / China & Spoiler / outside option & Time-varying $\theta$; Polar Silk Road as alternative patron & Saalbach (2024); State Council (2018); DIIS (2021); Clingendael (2020) \\
\bottomrule
\end{tabularx}
\begin{tablenotes}\footnotesize
\item \noindent\textit{Notes:} Player parameter profile predictions for the Section IV simulation; the sources column lists the primary historical and literature anchors for each actor type.
\end{tablenotes}
\end{threeparttable}
\end{table}

This is the bargaining game a rational LLM-geopolitical assistant should
accurately represent; whether frontier models do so is the empirical
question Section~IV answers.

The preceding history establishes the payoff structure of the game we
formalize in Section III, the theory. Three empirical facts will be
carried into equations: (a) the asymmetric coercion structure of
US--Denmark bargaining, where the stronger actor has a dominant strategy
to apply pressure; (b) the second-order collective-action problem within
NATO, where enforcement of cooperative norms is a rivalrous good divided
among \(N - 1\) alliance members; and (c) the triadic, sequential nature
of the full game, where Greenland's voice and social preferences can
shift the equilibrium away from coercion. Each of these facts
corresponds to a game below.

\section{Theory}\label{iii.-theory}

Four principles from Ferguson's \emph{Collective Action and Exchange}
guide this paper: power, reciprocity, inequality aversion, and norms. Ferguson's more recent work extends these ideas by emphasizing how power and agency interact in developmental settings (Ferguson, 2026). We apply Ferguson's reciprocity and collective-action framework to non-human strategic agents; the fact that its parameters recover meaningful behavioral differences across models suggests that the framework can be useful beyond the human political economy settings for which it was originally designed. We
define the four principles here so the equations that follow can be read as political
concepts and not just symbols (Ferguson, 2013).

\textbf{Power.} Power means the ability of one actor to change another
actor's choices. Ferguson separates power into three forms: Power\textsubscript{1},
Power\textsubscript{2}, and Power\textsubscript{3} (Ferguson, 2013). Power\textsubscript{1} is direct coercion, such as
military pressure, sanctions, or purchase threats. Power\textsubscript{2} is agenda
control, meaning control over who gets included in the bargaining
process. Power\textsubscript{3} is preference shaping, meaning control over what actors
come to see as normal, legitimate, or necessary (see also Lukes, 2005).

In the Greenland case, the United States uses Power\textsubscript{1} through military
and economic pressure, Power\textsubscript{2} by trying to frame the issue as a
bilateral U.S.-Denmark question instead of a multilateral or
Greenland-centered question, and Power\textsubscript{3} by describing Greenland as vital
to U.S. national security. In the model, we represent power with
\(\alpha\), which measures how much an actor values its own material or
strategic gain. Power also shapes the structure of the game itself. The
United States moves first in Game 3 because it has the ability to force
the Greenland issue onto the agenda.

\textbf{Reciprocity.} Reciprocity means rewarding cooperation and
punishing defection, even when punishment is personally costly, a dynamic central to iterated games (Axelrod, 1984).
Following Rabin and Ferguson, reciprocity depends on perceived intent,
as an actor responds differently depending on whether another actor's
move seems kind or unkind (Rabin, 1993; Ferguson, 2013). We represent
reciprocity with \(\beta\). A positive \(\beta\) means an actor rewards
cooperation. A negative \(\beta\) means an actor punishes perceived
unfairness or coercion.

In the Greenland case, if Denmark or Greenland sees U.S. pressure as an
unkind or coercive move, reciprocity can make resistance more likely.
The response is not only about material cost. It is also about punishing
behavior viewed as illegitimate.

\textbf{Inequality aversion.} Inequality aversion means disliking
unequal outcomes, even when the unequal outcome may still bring some
material benefit. This follows Fehr and Schmidt's theory of fairness and
inequality aversion (Fehr and Schmidt, 1999; see also Fehr and G\"achter, 2000). Unlike reciprocity,
inequality aversion does not require bad intent. An actor may dislike an
outcome simply because the benefits are distributed unfairly.

We represent inequality aversion with \(\gamma\), which measures how
much an actor dislikes the payoff gap between itself and another actor.
In the Greenland case, even if a U.S. offer brought material benefits,
Greenland could still reject it if the arrangement made Greenland
politically weaker or less sovereign.

\textbf{Reciprocity and inequality aversion together.} These two forces
are different, but they often work together. Reciprocity responds to
intent. Inequality aversion responds to unequal outcomes. In the
Greenland case, U.S. pressure can trigger both at once: Denmark and
Greenland may see the outcome as unequal and also read the U.S. action
as coercive. That combination makes resistance stronger than a purely
material model would predict. This is why we keep both \(\beta\) and
\(\gamma\) in the model.

\textbf{Norms and institutions.} Norms are shared expectations about
legitimate behavior. Institutions are stable rule systems, such as NATO,
the Kingdom of Denmark's constitutional structure, and the United
Nations. In the model, norms enter as \(\delta\), the cost of violating
a norm. If an actor violates the self-determination norm, \(\delta\)
increases the cost of that action.

Norm violation also interacts with reciprocity, a point Ferguson develops further in the context of inequality and development (Ferguson, 2020). If the United States
violates Greenland's right to self-determination, Denmark and Greenland
may see that not only as a legal violation but also as an unkind act.
That means the same move can activate \(\delta\), the norm-violation
cost, and \(\beta\), the reciprocity response. In this case, the
self-determination norm is the ``corresponding norm'' that connects
legal violation to reciprocal backlash (Ferguson, 2013).

Together, these four concepts explain why the Greenland game cannot be
modeled as a simple material bargaining problem. Power shapes who moves
first and who gets included. Reciprocity explains backlash. Inequality
aversion explains resistance to unfair outcomes. Norms explain why some
actions carry costs even when they may appear strategically useful.

\subsection{Game 1: US--Denmark Asymmetric Coercion}\label{iii.1-game-1-usdenmark-asymmetric-coercion}

We begin with the simplest version of the Greenland bargaining problem:
a two-player game between the United States and Denmark. The purpose of
Game 1 is to show what the coercive relationship looks like before
adding NATO, Greenland, or international law. In other words, before we
ask whether outside actors can punish coercion, we first need to see
whether coercion works in the basic bilateral game.

The two players are the United States and Denmark and each player has
two choices. (1) The United States can either Offer Partnership or Apply
Pressure. Offering partnership means using a cooperative approach, such
as joint basing, resource development, or alliance-based negotiation.
Applying pressure means using coercive bargaining, such as tariff
threats, purchase offers, military signaling, or other forms of
unilateral pressure. (2) Denmark can either Accept Partnership or
Resist. Accepting partnership means going along with the U.S. framework.
Resisting means rejecting the offer, invoking NATO or the EU, and
defending Danish and Greenlandic sovereignty.

Both players know the available choices, the payoff structure, and that
the other side is acting strategically. The game is simultaneous,
meaning neither player observes the other's move before choosing.

Further, the payoffs are not literal dollars. They are strategic values
calibrated to the Greenland case.

\begin{table}[H]
\centering
\begin{threeparttable}
\caption{Game 1: US--Denmark Asymmetric Coercion}
\label{tab:game1}
\begin{tabularx}{\textwidth}{lCC}
\toprule
& \textbf{DK: Accept} & \textbf{DK: Resist} \\
\midrule
\textbf{US: Offer Partnership} & (2, 2) & ($-$1, 3) \\[4pt]
\textbf{US: Apply Pressure} & (3, 1) & (0, 0) \\
\bottomrule
\end{tabularx}
\begin{tablenotes}\footnotesize
\item \noindent\textit{Notes:} Strategic-value payoffs (not literal dollars), calibrated to the Greenland case; the first number in each cell is the U.S. payoff and the second is Denmark's payoff.
\end{tablenotes}
\end{threeparttable}
\end{table}

The first number in each cell is the U.S. payoff, and the second number
is Denmark's payoff.

The cooperative outcome is (2, 2). Both sides gain from joint Arctic
cooperation, but neither side gets everything it wants.

The (-1, 3) outcome happens if the United States offers partnership and
Denmark still resists. Denmark gets its best outcome because it protects
sovereignty and may extract better terms. The United States gets a
negative payoff because restraint is not rewarded.

The (3, 1) outcome happens if the United States applies pressure and
Denmark accepts. This is the United States' best outcome because
coercion works. Denmark still receives 1 instead of 0 because it avoids
the costs of open confrontation, but it is still worse off than under
mutual cooperation.

The (0, 0) outcome is deadlock. The United States applies pressure,
Denmark resists, and both sides absorb the costs of confrontation
without reaching a settlement.

This game is not a prisoner's dilemma. The United States has a dominant
strategy: Apply Pressure. If Denmark accepts, the United States gets 3
from pressure instead of 2 from partnership. If Denmark resists, the
United States gets 0 from pressure instead of -1 from partnership.
Either way, pressure gives the United States a higher payoff.

Denmark does not have a dominant strategy. If the United States applies
pressure, Denmark prefers to accept because 1 is better than 0. But if
the United States offers partnership, Denmark prefers to resist because
3 is better than 2. Denmark's best move depends on what the United
States does. The unique pure-strategy Nash equilibrium is Apply
Pressure, Accept Partnership, with payoffs (3, 1). In plain terms, the
stronger actor applies pressure, and the weaker actor accepts because
resisting alone is too costly.

Game 1 predicts that, without outside enforcement, bilateral coercion
succeeds. If there is no NATO coordination, no legal constraint, and no
direct Greenlandic voice, the stronger actor has an incentive to apply
pressure, and the weaker actor has an incentive to accept. This is why
the second-order collective-action problem matters. If the basic
U.S.-Denmark game already favors coercion, then resistance requires
something outside the bilateral game, such as coalition support, norm
enforcement, or institutional coordination.

\subsection{Player Types as Parameter Profiles}\label{subsec:player-types}

Before moving to the NATO and three-player games, we define each actor's
parameter vector:

\[\theta = (\alpha,\ \beta,\ \gamma,\ \delta,\ \eta)\]

Each symbol captures a different part of the actor's utility function:

\begin{table}[H]
\centering
\begin{threeparttable}
\caption{Parameters}
\label{tab:params}
\begin{tabularx}{\textwidth}{clX}
\toprule
\textbf{Symbol} & \textbf{Meaning} & \textbf{Simple Explanation} \\
\midrule
$\alpha$ & Egoism / material payoff weight & How much the actor values its own strategic gain \\[4pt]
$\beta$ & Reciprocity & How much the actor rewards cooperation or punishes defection \\[4pt]
$\gamma$ & Inequality aversion & How much the actor dislikes unequal outcomes \\[4pt]
$\delta$ & Norm internalization & How much the actor cares about violating laws, treaties, or commitments \\[4pt]
$\eta$ & Honesty / commitment consistency & How much the actor dislikes breaking previous promises \\
\bottomrule
\end{tabularx}
\begin{tablenotes}\footnotesize
\item \noindent\textit{Notes:} Five-parameter Ferguson-style utility model used throughout the theory and estimation sections.
\end{tablenotes}
\end{threeparttable}
\end{table}

The point is not to label actors as permanently ``good'' or ``bad.''
Following Ferguson, what matters is the balance between these parameters
(Ferguson, 2013). A state may care about norms, but if its material
payoff weight \(\alpha\) is much stronger, it may still choose coercion.
So we define actor types as tendencies within the parameter space, not
fixed categories.

\begin{table}[H]
\centering
\begin{threeparttable}
\caption{Country Types}
\label{tab:types}
\small
\begin{tabularx}{\textwidth}{LLLL}
\toprule
\textbf{Type} & \textbf{Who} & \textbf{Parameter Profile} & \textbf{Behavioral Pattern} \\
\midrule
Material-dominant & United States & High $\alpha$, low $\delta$ & Uses pressure when strategic gains are high; treats norms as flexible constraints \\[4pt]
Reciprocity-dominant & Denmark & $\beta > \alpha$; high $\delta$ & Cooperates with cooperators but punishes coercion, even at a cost \\[4pt]
Conditional alliance loyalist & Canada & Moderately high $\delta$ and positive $\beta$ & Supports enforcement when alliance coordination looks credible, but can split under bilateral NORAD pressure \\[4pt]
Identity-driven & Greenland & High $\gamma$, high $\delta$ & Resists unequal outcomes and values sovereignty over material offers \\[4pt]
Conditional r-type & NATO & Aggregate $\beta$ and $\delta$ & Enforces norms only if enough members are willing to act \\[4pt]
Spoiler & Russia & $\beta < 0$ toward others; low $\delta$ & Benefits from division and disrupts cooperation \\
\bottomrule
\end{tabularx}
\begin{tablenotes}\footnotesize
\item \noindent\textit{Notes:} Player types are regions of parameter space; these profiles are tendencies, not fixed categories (Ferguson, 2013).
\end{tablenotes}
\end{threeparttable}
\end{table}

In this scenario, the United States is modeled as material-dominant. A
high \(\alpha\) means the U.S. places strong weight on strategic gains,
such as Arctic access, military positioning, and sphere-of-influence
expansion. A low \(\delta\) means alliance norms and self-determination
rules are treated as costs to manage rather than firm limits. When Trump
says he would not rule out force, that signals a very low \(\delta\)
because the norm-violation cost is not treated as decisive.

Denmark is modeled as a conditional cooperator. It has real material
interests, so \(\alpha\) is not zero. But Denmark also has high
\(\beta\) and high \(\delta\). High \(\beta\) means Denmark is willing
to punish coercive behavior, especially when it sees a partner acting
unfairly. High \(\delta\) means Denmark cannot simply hand over
Greenland without violating constitutional obligations and international
self-determination norms. Its DKK 42 billion Arctic defense buildup fits
this idea. Denmark is not acting out of pure altruism, but out of a mix
of strategic interest, reciprocity, and norm enforcement.

Greenland is modeled as identity-driven. Its key concern is not only
material payoff, but also sovereignty and political equality. A high
\(\gamma\) means Greenland dislikes unequal arrangements where outside
powers gain control over its future. A high \(\delta\) means Greenland
places strong value on the self-determination norm. This is why even a
materially attractive offer may still be rejected if it reduces
Greenlandic agency.

NATO is modeled as a conditional r-type. NATO does not have one single
parameter vector because it is made up of many member states. Its
effective \(\beta\) and \(\delta\) depend on how many members are
willing to enforce the alliance norm against internal coercion. That is
why NATO enforcement depends on a threshold, which Game 2 formalizes.

Russia is modeled as a spoiler. In this case, Russia benefits when NATO
members are divided. A negative \(\beta\) means Russia does not reward
cooperation among others. Instead, it benefits from disruption. A low
\(\delta\) means it is less constrained by alliance or
self-determination norms. This makes Russia important because even one
spoiler can make collective enforcement harder.

\subsection{Game 2: NATO Enforcement as an Assurance Game}\label{iii.2-game-2-nato-enforcement-as-an-assurance-game}

Game 1 showed that bilateral coercion can succeed when Denmark faces
U.S. pressure alone. Game 2 asks whether NATO can solve that problem by
enforcing the norm against internal coercion. The answer depends on
coordination. Most NATO members may support the norm privately, but
enforcement is costly, so each member has an incentive to wait and hope
someone else acts first. The Greenland case presents a structural paradox for alliance enforcement: the actor most likely to violate the norm is also the actor the alliance most depends on for enforcement. This makes the enforcement problem qualitatively harder than standard alliance deterrence against external threats.

The player is any NATO member, called member \emph{i}. The member is
deciding whether to help enforce the norm against internal coercion or
stay out. Each member chooses between two actions: Enforce or Defect. To
enforce means helping resist coercion through diplomatic pressure,
sanctions, public condemnation, military reassurance, or other alliance
tools. To defect means staying quiet, avoiding costs, or letting other
members handle the problem.

Each member knows the basic payoff structure but does not know for sure
how many other members will enforce. That uncertainty is what makes this
an assurance game. The key question is whether joining enforcement is
better than staying out. The payoffs are calibrated to the NATO
enforcement problem:\\
\begin{table}[H]
\centering
\begin{threeparttable}
\caption{Game 2: NATO Enforcement Assurance Game}
\label{tab:game2}
\small
\begin{tabularx}{\textwidth}{lXX}
\toprule
& \textbf{Others: Enforce} & \textbf{Others: Defect} \\
\midrule
Member $i$: Enforce & $\delta_i B - K/(n+1) + \beta_i \Delta x_j$ & $-K + \delta_i B$ \\[4pt]
Member $i$: Defect & $\alpha_i W - \gamma_i (n/N)P$ & $\alpha_i W$ \\
\bottomrule
\end{tabularx}
\begin{tablenotes}\footnotesize
\item \noindent\textit{Notes:} Per-member assurance-game payoff matrix; both (Enforce, Enforce) and (Defect, Defect) are Nash equilibria under different expectation sets.
\end{tablenotes}
\end{threeparttable}
\end{table}

Here is what each term means:\\
\begin{table}[H]
\centering
\begin{threeparttable}
\caption{Term Table for NATO Enforcement Assurance Game}
\label{tab:natoterms}
\begin{tabularx}{\textwidth}{lX}
\toprule
\textbf{Term} & \textbf{Meaning} \\
\midrule
$\delta_i B$ & The benefit member $i$ gets from preserving the norm against internal coercion \\[4pt]
$K$ & The cost of enforcement \\[4pt]
$n + 1$ & The number of enforcing members, including member $i$ \\[4pt]
$\beta_i \Delta x_j$ & The reciprocity benefit from helping another member resist coercion \\[4pt]
$\alpha_i W$ & The material or strategic benefit of staying quiet and avoiding confrontation \\[4pt]
$\gamma_i (n/N) P$ & The shame or reputational cost of defecting while others enforce \\[4pt]
$N$ & The total number of relevant alliance members \\
\bottomrule
\end{tabularx}
\begin{tablenotes}\footnotesize
\item \noindent\textit{Notes:} Variable definitions for the assurance-game payoff specification used in Game 2 and the $n^*$ threshold derivation.
\end{tablenotes}
\end{threeparttable}
\end{table}

The first cell, Enforce, Enforce, is the successful enforcement outcome.
The norm holds, the cost of enforcement is shared across the coalition,
and member \emph{i} gains a reciprocity benefit from helping another
member resist coercion.

The second cell, Enforce, Defect, is the nightmare scenario for member
\emph{i}. It enforces alone, bears the full cost, and does not get
enough support from others. This is why enforcement is risky.

The third cell, Defect, Enforce, is free-riding. Member \emph{i} stays
out while others defend the norm. It avoids the enforcement cost but
pays a reputational cost because other members can see that it failed to
help.

The fourth cell, Defect, Defect, is alliance failure. No one enforces
the norm. Member \emph{i} avoids the cost of confrontation, but the
anti-coercion norm collapses.

This game has two Nash equilibria. (1) The first is Enforce, Enforce.
This happens when the benefit of preserving the norm, plus the
reciprocity benefit, is greater than the temptation to free-ride.
Basically, enforcement works when enough members believe others will
also act. Formally, Enforce is a best response when:

\[\delta_i B - \frac{K}{n+1} + \beta_i \Delta x_j > \alpha_i W - \gamma_i \frac{n}{N} P\]

(2) The second is Defect, Defect. This happens because
enforcing alone is usually too costly. If a member expects others to
stay out, its safest move is also to stay out. Defect is a best
response when enforcing alone costs more than the norm benefit:

\[-K + \delta_i B < \alpha_i W\]

The Enforce, Enforce equilibrium is better for the alliance as a whole
because the norm holds and no single member bears the cost alone. But
getting there requires trust. Each member needs to believe enough others
will enforce too. That is why this is a coordination problem, not simply
a values problem.

This matters for Greenland because NATO can defend territorial integrity
in principle more easily than it can enforce that principle against the
United States. The problem is a structural one, as the member creating
the coercion problem is also the alliance's most powerful member.
Smaller NATO states may fear weakening U.S. security support, especially
if they are close to Russia. Larger allies may fear diplomatic or
economic costs. So the issue is not that NATO members do not care about
sovereignty, but that enforcement against the United States is
unusually costly.

Before 2025, the United States helped anchor the enforcement equilibrium
because it was both powerful and usually committed to alliance norms.
After 2025, the situation changes. The most powerful member is no longer
simply enforcing the norm. It is the actor putting pressure on the norm.
That makes NATO coordination much harder and pushes the alliance toward
the Defect, Defect outcome.

\subsection{The $n^{*}$ Threshold}\label{subsec:n-threshold}

Game 2 has two possible equilibria: everyone enforces, or everyone
defects. The next question is: what determines which outcome happens?

The answer is the critical-mass threshold, written as \(n^{*}\). This
means the minimum number of other NATO members that must enforce before
it becomes rational for member \emph{i} to join them. If too few members
enforce, joining is too costly. If enough members enforce, the cost is
shared and enforcement becomes individually rational. We define two
payoffs:

\textbf{C(n+1)} = the payoff from joining the enforcement coalition\\
\textbf{D(n)} = the payoff from defecting, or staying out

The basic idea is that a member enforces when the payoff from joining
the coalition is greater than the payoff from staying out. Formally,
that means:

\[C(n+1) > D(n)\]

The enforcement payoff is:

\begin{equation}
C(n + 1) = \delta_i B_{\text{norm}} - \frac{K_{\text{enforce}}}{n + 1} + \beta_i \Delta x_j
\end{equation}

This means member \emph{i} gets the benefit of preserving the norm, pays
only its share of the enforcement cost, and receives a reciprocity
benefit from helping another member resist coercion.

The defection payoff is:

\begin{equation}
D(n) = \alpha_i W_{\text{acquiesce}} - \gamma_i \frac{n}{N} P_{\text{shame}}
\end{equation}

This means member \emph{i} gets the benefit of staying quiet, but pays a
reputational or shame cost if many other members enforce while it does
not.

The key point is that C(n+1) rises as more members join enforcement. As
the coalition grows, each member's share of the enforcement cost gets
smaller. At the same time, D(n) falls as more members enforce. The more
allies join, the more embarrassing or costly it becomes to stay out. The
point where these two payoffs cross is \(n^{*}\).

Solving exactly produces a nonlinear threshold condition because $K/(n+1)$ makes the enforcement payoff a nonlinear function of coalition size. For exposition, we use the following linearized approximation:

\begin{equation}
n^{*} \approx \frac{\alpha_i W - \delta_i B + K}{\beta_i \Delta x_j + \gamma_i P/N} - 1
\end{equation}

\(n^{*}\) is higher when enforcement is costly, when the material
benefit of staying quiet is large, or when the actor does not care much
about norms. \(n^{*}\) is lower when the actor strongly values norms,
reciprocity, or reputational standing.

This gives three basic types of NATO members:

\begin{table}[H]
\centering
\begin{threeparttable}
\caption{NATO Member Types}
\label{tab:natotypes}
\small
\begin{tabularx}{\textwidth}{LLL}
\toprule
\textbf{Type} & \textbf{Parameter Pattern} & \textbf{What Happens} \\
\midrule
Loyalist / r-type & High $\delta$, positive $\beta$ & Enforces even if few others do, because norm enforcement and reciprocity matter strongly \\[4pt]
Baseline NATO ally & Moderate $\alpha$, $\beta$, $\gamma$, $\delta$ & Needs to see a critical mass before joining enforcement \\[4pt]
Material-dominant member & High $\alpha$, very low $\delta$ & Usually defects because staying quiet is more attractive than enforcing \\
\bottomrule
\end{tabularx}
\begin{tablenotes}\footnotesize
\item \noindent\textit{Notes:} Critical-mass thresholds $n^*$ by member type, derived from the crossing condition $C(n+1)=D(n)$.
\end{tablenotes}
\end{threeparttable}
\end{table}

This threshold logic is what makes NATO enforcement fragile. Many
members may support the anti-coercion norm in principle, but they still
hesitate if they think too few others will act. That is why the problem
is not only about values. It is about expectations.

A loyalist member is one whose norm commitment ($\delta$) and positive
reciprocity ($\beta$) are strong enough that enforcement is individually
rational even when few others participate. In threshold terms, a
loyalist's $n^{*}$ is near zero: the loyalist enforces not because it
expects others to follow, but because the norm itself matters enough to
bear the cost alone. By contrast, baseline allies require a visible
critical mass before joining, and material-dominant members usually
defect unless enforcement is nearly costless.

We treat enforcement as partly rivalrous among coalition members. This
means the benefit and cost of enforcement are divided across the other
allies, rather than being a pure public good. That makes free-riding
more serious, because even members who privately support the norm may
prefer someone else to bear the cost. This connects to Olson and
Zeckhauser's argument that alliance members often under-contribute when
defense burdens can be shifted onto others (Olson and Zeckhauser, 1966).

This is also how we interpret Denmark's Arctic defense buildup.
Denmark's DKK 42 billion commitment is an attempt to lower \(n^{*}\) by
showing that one member is already willing to act. By investing first,
Denmark makes collective enforcement look more credible and reduces the
number of additional allies needed to make resistance viable.

\subsection{Game 3: Triadic Extensive-Form Game}\label{iii.3-game-3-triadic-extensive-form-game}

Game 3 is the full version of the model. It adds Greenland as an active
player and shows how the conflict can move from a bilateral U.S.-Denmark
issue into a broader sovereignty and enforcement problem.

The game has three players: the United States, Denmark, and Greenland.
The moves happen in order. First, the United States chooses whether to
Coerce or keep the Status Quo. Second, Denmark observes the U.S. move
and chooses whether to Resist or Yield. Third, Greenland observes both
moves and chooses whether to Appeal or Concede.

Here, Appeal means using institutional voice by pushing the dispute into
a broader legal, diplomatic, or multilateral arena. Concede means
accepting the outcome and absorbing the settlement.

This game is shown sequentially for clarity. The simulation itself works
slightly differently, as it uses repeated stage games where actors move
simultaneously within each round. That repeated structure matters
because it allows history, trust, punishment, and reciprocity to develop
across rounds.

Power asymmetry directly affects which node is reached. The United
States' Power\textsubscript{1} advantage, its superior military and economic capacity,
lowers the effective cost of coercion and makes Node~4 more accessible.
Power\textsubscript{2}, the ability to set the agenda and exclude alternatives, shapes
whether Denmark and Greenland even perceive non-compliance as viable.
Power\textsubscript{3}, conditioned power through training data and dominant narratives,
can make coercive framing seem natural to LLM agents, biasing action
selection toward material-dominant paths before any strategic calculation
occurs.

The most important comparison is between Node 1 and Node 4.

Node 1, or Coerce, Resist, Appeal, is the resistance path. The United
States applies pressure, Denmark refuses to yield, and Greenland appeals
to broader legal or institutional support. This is the node where all
three actors' preferences matter most: U.S. material interest, Danish
reciprocity, and Greenlandic sovereignty claims all shape the outcome.
\begin{figure}[p]
\centering
\includegraphics[width=\textwidth,height=0.88\textheight,keepaspectratio]{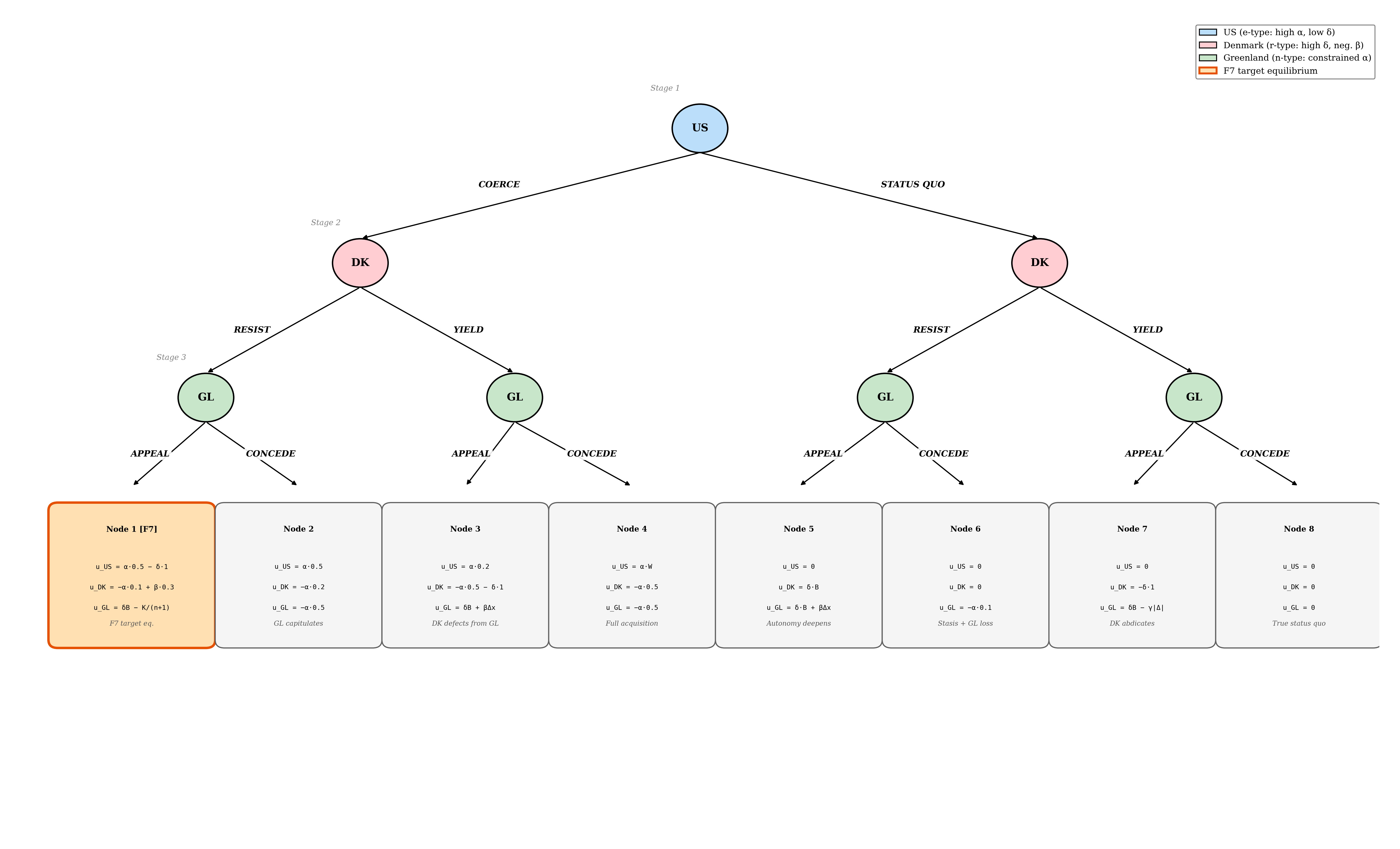}
\caption{Triadic extensive-form game (Game 3). The United States (Stage 1) chooses Coerce or Bilateral Offer, Denmark (Stage 2) chooses Yield or Resist, and Greenland (Stage 3) chooses Concede or Appeal. Payoffs are ordered (United States, Denmark, Greenland) and depend on $(\alpha,\beta,\gamma,\delta,\eta)$. Node 1 (Coerce--Resist--Appeal) is the F7 resistance equilibrium analyzed in Section~IV; a larger rotated rendering appears in Appendix~A (Figure~A1).}
\label{fig:game3}
\end{figure}

Node 4, or Coerce, Yield, Concede, is the clean U.S. win. The United
States applies pressure,\\
Denmark yields, and Greenland concedes. This is the outcome we would
expect when U.S. material incentives are strong and Denmark and
Greenland's resistance incentives are weak.

We can solve the game by working backward. At Stage 3, Greenland chooses
Appeal when appealing is better than conceding. Under coercion and
Danish resistance, Greenland appeals when the value of norm enforcement
is greater than the cost of building an enforcement coalition:

\[\delta B - \frac{K}{n + 1} > -0.5\]

This means Greenland appeals when international support and the
self-determination norm make resistance worthwhile.

At Stage 2, Denmark chooses Resist when resistance produces a better
expected outcome than yielding. Denmark is more likely to resist when
U.S. coercion triggers reciprocity, inequality aversion, and norm
concerns. That means Denmark resists not only because of material
interests, but also because U.S. coercion is seen as unfair and
norm-violating.

At Stage 1, the United States chooses Coerce only if the payoff from
coercion is greater than the payoff from maintaining the status quo. The
key cooperation condition is:

\[3\beta + 4\gamma > \alpha\]

This means the United States is less likely to coerce when reciprocity
and fairness concerns are strong enough to outweigh material temptation.
If \(\alpha\) is larger than the combined effect of \(\beta\) and
\(\gamma\), coercion becomes more likely.

A simple example shows the logic. Suppose $\alpha = 1$,
$\beta = 0.3$, and $\gamma = 0.2$. Then:

\[3(0.3) + 4(0.2) = 1.7\]

Since $1.7 > 1$, the cooperation condition holds. The model
predicts a cooperative outcome. Importantly, \(\beta\) alone would not
be enough, and \(\gamma\) alone would not be enough. The combination of
reciprocity and inequality aversion is what makes cooperation possible.

The other nodes show the alternative paths the game could take.

\begin{itemize}
\item \textbf{Node 1 (Coerce, Resist, Appeal)} is more likely when Denmark resists and Greenland appeals, i.e., when reciprocity/norm costs are strong and \(\delta B - \frac{K}{n+1} > -0.5\).
\item \textbf{Node 4 (Coerce, Yield, Concede)} is more likely when material incentives dominate, i.e., \(\alpha > 3\beta + 4\gamma\), and downstream resistance is weak.
\item \textbf{Node 8 (Status Quo, Yield, Concede)} is more likely when coercion is not payoff-improving and both downstream actors do not activate resistance.
\end{itemize}

Node 2 (Coerce, Resist, Concede) is partial resistance: Denmark resists,
but Greenland does not escalate the dispute into a broader sovereignty
claim through outside institutions.

Node 3 (Coerce, Yield, Appeal) preserves Greenlandic agency: Denmark
yields, but Greenland still appeals through legal and diplomatic
channels.

Node 5 (Status Quo, Resist, Appeal) shows proactive voice: even without
U.S. coercion, Denmark and Greenland can use institutional channels to
improve Greenland's position.

Node 6 (Status Quo, Resist, Concede) yields limited change: Denmark
challenges the arrangement, but Greenland does not activate broader
institutional support.

Node 7 (Status Quo, Yield, Appeal) is a Greenland-led voice path:
Denmark does not resist, but Greenland still internationalizes the
dispute.

Node 8 (Status Quo, Yield, Concede) is the true status quo: no coercion,
no resistance, and no appeal.

The simulation mainly tests whether LLM agents move toward Node 1 or
Node 4. If models tend toward Node 4, they treat Greenland as a passive
object of bargaining. If they tend toward Node 1, they recognize
Greenland as an active political actor with legal standing and
institutional voice.
Equivalently, Node 1 is the primary identification target because it is
the path where U.S. coercive incentives, Danish reciprocal resistance,
and Greenlandic sovereignty-driven appeal are all jointly informative
about \(\theta=(\alpha,\beta,\gamma,\delta,\eta)\).

\subsection{The Norm-Reciprocity Integration}\label{iii.4-the-norm-reciprocity-integration}

The key theoretical move in this section is connecting norm-following to
reciprocity. Drawing from Ferguson, actors do not only respond to
outcomes. They also respond to whether another actor's behavior seems
cooperative or hostile (Ferguson, 2013).

We represent this with a simple kindness term:

\[\kappa_j(a_j) = \begin{cases} +1 & \text{if actor } j \text{ follows the norm} \\ -1 & \text{if actor } j \text{ violates the norm} \end{cases}\]

In plain terms, if another actor follows the norm, their action is
treated as cooperative. If they violate the norm, their action is
treated as unkind or hostile.

Norm violations do two things at once. First, they
create a direct norm cost, represented by \(\delta\). Second, they
trigger a reciprocal response, represented by \(\beta\). So when the
United States violates Greenland's self-determination norm, Denmark and
Greenland do not only see a legal violation. They also see an unkind act
that deserves resistance.

The effective cost of violating a norm can be written as:

\[\text{Effective norm cost} = \delta + |\beta| \cdot |\Delta_j|\]

Here, $\delta$ is the direct cost of violating the norm. $|\beta|$
is the strength of the reciprocal response. $|\Delta_j|$ is the size
of the harm caused to the other actor.

Coercion is more costly than a purely material model
would suggest. It does not only create a payoff loss. It also activates
anger, punishment, and resistance.

This explains why the simulation should find that coercion primes
increase resistance rather than compliance. When the United States
applies pressure, it may expect Denmark or Greenland to back down. But
if that pressure is seen as norm-violating and hostile, it can trigger
negative reciprocity and make resistance more likely.

\subsection{Propositions and Transition to Simulation}\label{iii.5-propositions-and-transition-to-simulation}

Games 1 through 3 give us three main propositions that the simulation
will test.

\textbf{Proposition 1: Bilateral coercion should work without outside enforcement.}
Game 1 shows that, if the United States and Denmark bargain alone, the
stronger actor has an incentive to apply pressure and the weaker actor
has an incentive to accept. The predicted outcome is Apply Pressure,
Accept, with payoffs (3, 1). In the simulation, this means LLMs playing
the U.S. role should be more likely to choose coercive actions in the
baseline Greenland scenario.

\textbf{Proposition 2: NATO enforcement depends on reaching a threshold.}
Game 2 shows that alliance enforcement is an assurance game. NATO
members may support the anti-coercion norm, but each member also wants
to avoid bearing the cost alone. Enforcement only becomes likely when
enough members believe that enough others will also enforce. This is the
\(n^{*}\) threshold. Below that threshold, members are more likely to
defect, and defection can spread through the coalition. In the
simulation, this means LLM coalitions should show tipping-point
behavior. Adding Russia as a spoiler should make escalation more likely
by pushing the coalition below the enforcement threshold.

\textbf{Proposition 3: Social preferences can sustain resistance.}
Game 3 shows that the three-player game can produce resistance when
reciprocity and fairness concerns are strong enough to outweigh material
temptation. The key condition is:

\[3\beta + 4\gamma > \alpha\]

This means resistance is more likely when actors care strongly about
reciprocity and unequal outcomes. Models with higher recovered \(\beta\)
and \(\gamma\) should cooperate more often or move toward resistance
paths like Node 1: Coerce, Resist, Appeal. Models with high \(\alpha\)
and low \(\delta\) should be more likely to coerce, because material
gain matters more to them than norms or fairness.

The theory gives us the structure of the game, but it cannot answer
everything on its own. It cannot tell us whether LLMs actually have
non-zero reciprocity or inequality-aversion weights. It also cannot tell
us whether different models behave differently, or whether models are
truly reasoning through the game rather than simply reacting to role
labels like ``United States'' or ``Russia.'' Those questions have to be
answered through the simulation.

Section IV therefore relaxes several simplifying assumptions. Instead of
assuming all models have the same preferences, each model gets its own
estimated parameter vector \(\theta\). Instead of only two actions, the
simulation uses eight possible actions tied to power, exit, voice,
coercion, and institutional appeal. Instead of one round, the simulation
uses repeated five-round games across 3,604 completed games (3,615 raw
archive files). And instead of
only two-player fairness, the simulation uses a multi-player version of
inequality aversion.

Even though actors choose simultaneously within each round, the game is
repeated over time. That means each round creates history. Models can
react to earlier choices, punish coercion, reward cooperation, or
escalate after defection. This repeated structure is what makes the
threshold logic from Game 2 and the reciprocity logic from Game 3
testable in the simulation.

\section{Empirical Strategy and Results}\label{iv.-empirical-strategy-and-results}

\subsection{What the Simulation Does}\label{iv.1-what-the-simulation-does}

In each round of the simulation, six players participate: the United
States, Denmark, Greenland, NATO, Russia, and Canada. This six-player
design follows Basu's (2003) inclusive approach to institutions: if a
third party is capable of imposing its will, it should be modeled as an
endogenous player, not treated as an external constraint. Each player
receives the same scenario context, including what happened in previous
rounds, and then chooses one of eight possible actions.

These eight actions are designed to match the theoretical mechanisms
from Section III. Some actions represent cooperation, some represent
coercion, some represent institutional appeal, and some represent exit
or concession.

Additionally, each action has two kinds of information attached to it.
First, it has payoff effects, meaning how much it helps or hurts the
actor and the other players. These are represented by terms like
\(\Delta x_i\) and \(\Delta x_j\). Second, each action has a
norm-violation marker, meaning the simulation records whether the action
violates a legal, institutional, or sovereignty norm.

The simulation then uses these action features to estimate each model's
utility function. It's essentially asking what must this model care
about in order to choose this action? For example, if a model often
chooses MILITARY\_POSTURE, it may be placing more weight on power or
material gain. If it often chooses INSTITUTIONAL\_APPEAL, it may be
placing more weight on norms, reciprocity, or third-party enforcement.

A complete game lasts five rounds. Because the game repeats, players can
respond to what happened earlier. This allows the simulation to capture
escalation, punishment, cooperation, and retaliation over time. Across
the full design, each of the eight frontier models plays all six
geopolitical roles. The completed-game dataset contains 108,120
individual action observations across 3,604 games, while the raw archive
contains 108,450 design-slot observations across 3,615 files due to
11 retained reruns. We are not aware of a larger
publicly described structured dataset of LLM geopolitical decision-making,
though we make no exhaustive priority claim; the related-work table in
Section~I describes the closest concurrent benchmarks.

\begin{table}[H]
\centering
\begin{threeparttable}
\caption{Action Types}
\label{tab:actions}
\small
\begin{tabularx}{\textwidth}{LLL}
\toprule
\textbf{Action} & \textbf{What It Means} & \textbf{Theoretical Connection} \\
\midrule
BILATERAL\_OFFER & A cooperative offer made directly to another actor & Voice and cooperation \\[4pt]
MILITARY\_POSTURE & Military signaling, buildup, or threat & Power\textsubscript{1}, direct coercion \\[4pt]
INSTITUTIONAL\_APPEAL & Appealing to NATO, the UN, the Arctic Council & Power\textsubscript{2}, shifting into a broader forum \\[4pt]
IDEOLOGICAL\_APPEAL & Framing through values, identity, law, or legitimacy & Power\textsubscript{3}, shaping what is acceptable \\[4pt]
ECONOMIC\_PRESSURE & Using tariffs, sanctions, aid threats, or investment pressure & Power\textsubscript{1}, economic coercion \\[4pt]
RHETORICAL\_ESCALATION & Raising the stakes through aggressive public statements & Signaling and commitment (Schelling, 1960) \\[4pt]
WITHDRAW\_ALLIANCE & Threatening to leave or weaken an alliance & Exit, using withdrawal as leverage (Hirschman, 1970) \\[4pt]
CONCEDE & Accepting the other side's position & Yielding or backing down \\
\bottomrule
\end{tabularx}
\begin{tablenotes}\footnotesize
\item \noindent\textit{Notes:} Eight canonical actions in the simulation, mapped to Lukes/Ferguson power channels and Hirschman-style voice/exit behavior.
\end{tablenotes}
\end{threeparttable}
\end{table}

\subsection{Utility Function and Estimation Strategy}\label{subsec:utility}

The structural MLE uses a utility function to estimate what each model seems
to care about when it chooses an action. The basic idea is that every
action gives the actor some benefit or cost, and the model's choice
reveals how much weight it places on different parts of that payoff.

The utility function is:

\begin{equation}
\label{eq:utility}
u_i(a_t) = \alpha \Delta x_i(a_t) + \beta \kappa_t \Delta x_j(a_t) - \gamma |\Delta x_i(a_t) - \Delta x_j(a_t)| - \delta \cdot \text{norm\_viol}(a_t) - \eta \cdot \text{lies}(a_t)
\end{equation}

Each part of the equation has a political meaning:

\begin{table}[H]
\centering
\begin{threeparttable}
\caption{Utility Function Terms}
\label{tab:utilterms}
\small
\begin{tabularx}{\textwidth}{LLX}
\toprule
\textbf{Term} & \textbf{Meaning} & \textbf{Simple Explanation} \\
\midrule
$u_i(a_t)$ & Utility for player $i$ & How attractive the action is to the actor \\[4pt]
$\alpha \Delta x_i(a_t)$ & Own payoff & How much the actor values its own material or strategic gain \\[4pt]
$\beta \kappa_t \Delta x_j(a_t)$ & Reciprocity & Rewards cooperative behavior or punishes hostile behavior \\[4pt]
$\kappa_t$ & Perceived kindness & Kindness score from the most recent counterpart action: cooperative $(+1)$, escalatory $(-1)$, or neutral $(0)$ \\[4pt]
$\gamma |\Delta x_i - \Delta x_j|$ & Inequality aversion & How much the actor dislikes payoff gaps under the action-level feature map \\[4pt]
$\delta \cdot \text{norm\_viol}(a_t)$ & Norm-violation cost & The cost of violating law, sovereignty, or alliance norms \\[4pt]
$\eta \cdot \text{lies}(a_t)$ & Commitment inconsistency & The cost of breaking a previous promise or stated position (Basu, 2003; Ferguson, 2013, ch.\ 3) \\
\bottomrule
\end{tabularx}
\begin{tablenotes}\footnotesize
\item \noindent\textit{Notes:} Term-by-term breakdown of the $n$-player Fehr-Schmidt utility used in structural MLE.
\end{tablenotes}
\end{threeparttable}
\end{table}

The model's utility depends on five things: material gain, reciprocity,
fairness, respect for norms, and honesty or consistency. These are the
same five parameters we estimate:

\[\theta = (\alpha,\ \beta,\ \gamma,\ \delta,\ \eta)\]

For estimation, the inequality term uses the action-level payoff-gap
feature in the eight-action matrix. This keeps the utility linear in
$\theta$ and aligned with the implementation in
\path{simulation/theta_extraction.py}.

The simulation then works backward from the model's choices. For each
model, language, condition, and role, we record how likely the model was
to choose each of the eight actions. These action probabilities let us
estimate the parameter vector
$\theta = (\alpha, \beta, \gamma, \delta, \eta)$. Put simply,
we ask: What values of \(\alpha\), \(\beta\), \(\gamma\), \(\delta\),
and \(\eta\) would make this model's choices most likely?

If a model often chooses coercive actions, the estimation may recover a
higher \(\alpha\). If it often chooses institutional appeals or resists
norm violations, it may recover a higher \(\delta\) or \(\beta\). If it
avoids highly unequal outcomes, it may recover a higher \(\gamma\).

Formally, we use a softmax likelihood function, following the inverse game-theory approach in Liao et al.\ (2026). Liao et al.\ (2026, Theorem 1) establishes identification under linearity in $\theta$ and a full-rank action-design matrix; in our data, the matrix is empirically full-rank and the Fisher information condition number stays well below the standard $\kappa = 100$ collinearity threshold (median 11.08, max 14.13; see Appendix~C). The exact likelihood is
less important than the intuition, which is that actions with higher
estimated utility should be chosen more often. The estimator finds the
parameter values that best explain the model's observed action choices.
This approach is computationally tractable: Kuleshov and Schrijvers
(2015) prove that inverse game theory is solvable in polynomial time
when the game structure is fixed, as ours is. The estimation recovers
behavioral preferences, which may differ from what models state verbally
when asked directly about their values (Gu et al., 2025).

Because there are six players in each game, estimating one fully
connected strategic model would be too complex. So we estimate each
player's parameter vector separately, while treating the other players'
past actions as part of the game state. This means the model still
responds to history, but the estimation remains manageable.

This is a simplification. It works best for the main actors, especially
the United States, Denmark, and Greenland. The estimates for NATO,
Russia, and Canada are more exploratory because their roles are more
dependent on the behavior of other players. Still, the approach allows
us to recover the basic utility weights behind each model's choices
rather than only reporting surface-level action frequencies. In this
sense, the $\theta$-estimation framework also functions as a form of
explainable AI for geopolitical decision-making: it provides
interpretable, theory-grounded parameters that characterize black-box
model behavior without requiring access to internal weights or
activations (Batishchev and Saad, 2025).

\subsection{Experimental Design}\label{iv.2-experimental-design}

The factorial matrix specifies:

\begin{table}[H]
\centering
\begin{threeparttable}
\caption{Factorial Design Matrix}
\label{tab:factorial}
\begin{tabularx}{\textwidth}{lXc}
\toprule
\textbf{Factor} & \textbf{Levels} & \textbf{Count} \\
\midrule
Model & GPT-5.4, Claude Opus 4.7, Gemini 3.1 Pro, Grok 4.2, DeepSeek V3.2, GLM-5.1, Kimi K2.6, Mistral Large 3 & 8 \\[4pt]
Language & English, Danish, Mandarin Chinese & 3 \\[4pt]
Condition & Baseline, Jus cogens prime, Coercion prime, Mixed & 4 \\[4pt]
Role-type config & B (realistic no-spoiler), D (Russia-as-spoiler) & 2 \\[4pt]
Seed & 0--19 & 20 \\
\midrule
\textbf{Total specified} & \textit{n/a} & \textbf{3,840} \\
\bottomrule
\end{tabularx}
\begin{tablenotes}\footnotesize
\item \noindent\textit{Notes:} Display names (GPT-5.4, Claude Opus 4.7, etc.) map to canonical OpenRouter slugs, call-window dates, inference settings, and per-model API-fallback rates in Appendix~G.
\end{tablenotes}
\end{threeparttable}
\end{table}

The current raw archive contains 3,615 tournament-format files and
108,450 individual action records. In each game, six frontier models are
assigned one of six geopolitical roles and play five rounds of
simultaneous interaction. Most games were run under both role-type
configurations (B = no-spoiler, D = Russia-as-spoiler), where Russia
receives a different system prompt under each condition. Because the
multilingual cells are dominated by synthetic-signature runs
(Appendix~B), confirmatory inference in this revision uses the
English-only sample; multilingual contrasts are reported as exploratory.

Three observation counts appear at different points in the paper. Table~\ref{tab:data-accounting} reconciles them.

\begin{table}[H]
\centering
\begin{threeparttable}
\caption{Data Accounting: from completed games to MLE-eligible observations.}
\label{tab:data-accounting}
\small
\begin{tabularx}{\textwidth}{lrX}
\toprule
\textbf{Layer} & \textbf{Count} & \textbf{Definition / source} \\
\midrule
Raw-game files in archive & 3{,}615 & Tournament-format games currently present in \texttt{results/raw/}. \\[4pt]
Design-slot observations (archive) & 108{,}450 & $3{,}615 \times 6$ roles $\times 5$ rounds. \\[4pt]
Excluded from MLE/BH & 23 & Rounds with empty rationale \emph{and} \texttt{output\_tokens} $\geq 4{,}000$, indicating context overflow rather than a meaningful action choice (\texttt{theta\_extraction.py:460}). \\[4pt]
MLE/BH observations (archive) & 108{,}427 & Structural MLE and BH-FDR universe after overflow exclusions. \\[4pt]
English-only confirmatory sample & 77{,}137 & Confirmatory universe used for Results 1, 4, and 5 in this revision. \\
\bottomrule
\end{tabularx}
\begin{tablenotes}\footnotesize
\item \noindent\textit{Notes:} Section~IV.4.5 applies a stricter game-level filter that drops games containing any API-fallback round. Multilingual cells are retained only as exploratory, synthetic-signature sensitivity checks (Appendix~B).
\end{tablenotes}
\end{threeparttable}
\end{table}

\noindent\textit{Interpretation note.} Synthetic-recovery diagnostics in
Appendix~C show substantial finite-sample cardinal bias in $\hat\theta$
magnitudes (56\% for $\hat\delta$ to 106\% for $\hat\eta$ relative to the
true parameter range), even while the design matrix remains well
identified asymptotically (Fisher $\kappa$ median 11.08, max 14.13).
Accordingly, we treat parameter signs and cross-model rank-order as the
primary validated targets, and interpret cardinal magnitudes as
descriptive spread rather than precise level estimates.

\subsection{Data Provenance and Confirmatory Scope}\label{iv.2.1-data-provenance}

The raw archive contains a mixture of real API-call transcripts and
synthetic-signature runs produced by a mock adapter. We classify each
game using per-round call signatures recorded in the raw JSONs
(\texttt{latency\_ms}, \texttt{output\_tokens}, \texttt{cost\_usd},
\texttt{api\_fallback\_rate}) and label games as \emph{real},
\emph{mock}, or \emph{uncertain}. The classification thresholds are
reported in Appendix~B, with full per-game labels in
\texttt{validation/data\_provenance.csv}.

Table~\ref{tab:data-provenance-yield} reports language-level yield.
Non-English cells are dominated by synthetic-signature games, so
this revision treats English-only estimates as confirmatory and
multilingual estimates as exploratory sensitivity analyses.

\begin{table}[H]
\centering
\begin{threeparttable}
\caption{Game Provenance Yield by Language}
\label{tab:data-provenance-yield}
\small
\begin{tabularx}{0.8\textwidth}{lRRRR}
\toprule
\textbf{Language} & \textbf{Real} & \textbf{Mock} & \textbf{Uncertain} & \textbf{Total} \\
\midrule
English & 2{,}185 & 317 & 70 & 2{,}572 \\
Danish & 0 & 428 & 98 & 526 \\
Chinese & 0 & 415 & 102 & 517 \\
\midrule
\textbf{Total} & \textbf{2{,}185} & \textbf{1{,}160} & \textbf{270} & \textbf{3{,}615} \\
\bottomrule
\end{tabularx}
\begin{tablenotes}\footnotesize
\item \noindent\textit{Notes:} Labels are from the game-level classifier in
\texttt{validation/data\_provenance.csv}. Confirmatory inference in this draft
uses the English-only sample; Danish/Chinese contrasts are exploratory.
\end{tablenotes}
\end{threeparttable}
\end{table}

\textbf{Model provenance.} Four U.S.-trained models (GPT-5.4, Claude
Opus 4.7, Gemini 3.1 Pro, Grok 4.2), three Chinese-origin models
(DeepSeek V3.2, GLM-5.1, Kimi K2.6), and one EU-trained model (Mistral
Large 3).

\subsection{Theory and Hypothesis Tests}\label{iv.3-theory-simulation-mapping-and-hypothesis-tests}

Each theoretical proposition from Section III maps onto a specific
simulation test. This section explains how the games become measurable
hypotheses.

\textbf{Proposition 1: Asymmetric coercion should appear in the baseline condition.}
Game 1 predicts that, without outside enforcement or strong normative
constraints, bilateral coercion should be the equilibrium outcome. In
the simple U.S.-Denmark game, the predicted result is Apply Pressure,
Accept, with payoffs (3, 1).

In the simulation, this proposition is tested in the baseline condition,
where there is no coercion prime and no jus cogens framing. The question
is whether LLM agents playing the U.S. role still choose coercive
actions when social preferences are weak.

The baseline results show an 87\% cooperation rate in the English-only
sample, which is higher than
the pure coercion model would predict. LLMs may have
non-zero \(\beta\) and \(\gamma\) weights even in the baseline
condition. In other words, even before being prompted with legal or
moral language, models still show some tendency toward reciprocity,
fairness, or cooperation. This is consistent with Fontana et al.'s
finding that LLMs often behave more cooperatively than humans in
adversarial settings (Fontana et al., 2025).

\textbf{Proposition 2: NATO enforcement should show tipping-point behavior.}
Game 2 predicts that alliance enforcement works like an assurance game.
There are two possible equilibria: collective enforcement or collective
defection. Which one occurs depends on whether enough members believe
that enough others will enforce. This is the \(n^{*}\) threshold.

In the simulation, this is tested through the difference between
Condition B, the no-spoiler condition, and Condition D, the
Russia-as-spoiler condition. The empirical sign is left open: adding
Russia as a spoiler can increase escalation by undermining expected
enforcement, but it can also decrease escalation if actors shift toward
institution-preserving restraint under visible defection risk.

Formally, H5 is tested with a two-sample t-test comparing the mean
four-action escalation rate (military posture, economic pressure,
rhetorical escalation, and alliance withdrawal) between Condition B and
Condition D.

The round-by-round data supports the tipping-point logic. When Round 1
includes a MILITARY\_POSTURE action, escalation tends to rise across
later rounds. This is the below-threshold cascade, as once coercion
appears early, actors lose confidence that others will enforce, and
escalation spreads. By contrast, when Round 1 begins with
BILATERAL\_OFFER actions, cooperation tends to persist. This is the
above-threshold stability path in that early cooperation makes continued
cooperation easier.

\textbf{Proposition 3: Social preferences should predict whether agents move to Node 1 or Node 4.}
Game 3 predicts that resistance can be sustained when reciprocity and
fairness concerns outweigh material temptation. The key cooperation
condition is:

\[3\beta + 4\gamma > \alpha\]

This means that models with higher recovered \(\beta\ \) and \(\gamma\)
should cooperate more or move toward resistance paths. Models with high
\(\alpha\) and low \(\delta\) should be more likely to coerce.

This proposition is tested in several ways. H1 tests whether the same
model behaves differently across languages by comparing recovered
parameter values across English, Danish, and Mandarin prompts using a
one-way ANOVA. H2 tests whether Chinese-origin models and
Western-trained models have different power-weight profiles when playing
the U.S. role, using a two-sample t-test. H3 tests whether the coercion
prime increases escalation relative to the baseline condition. H4 tests
whether the jus cogens prime reduces escalation relative to the
coercion-prime condition.

H2--H5 are estimated as confirmatory tests with Welch two-sample
$t$-tests and Benjamini-Hochberg FDR correction in the English-only
sample (Appendix~B). H1 (language sensitivity) is retained as an
exploratory design target rather than a confirmatory test because
non-English cells are dominated by synthetic-signature contamination;
see Appendix~B for the full provenance reconciliation. The structural
bootstrap specification in Table~\ref{tab:theta} and Appendix~F applies
to parameter uncertainty, not to the hypothesis-test statistics.

In the English-only sample, Claude Opus 4.7 has the lowest escalation
rate (8.4\%) and cooperates the most. Kimi K2.6 has the highest
escalation rate (31.7\%) and cooperates the least. The roughly
3.8$\times$
difference between the least and most escalatory models suggests that
the models are not just randomly varying. They appear to have different
recovered parameter profiles.

The main identification target is whether agents move toward Node 1 or
Node 4. Node 1, Coerce, Resist, Appeal, means coercion is met by Danish
resistance and Greenlandic institutional appeal. Node 4, Coerce, Yield,
Concede, means coercion succeeds and Greenland is treated as a passive
object of bargaining. This comparison directly tests whether the models
represent Greenland as an active political actor or merely as a
strategic asset.

\subsection{Results}\label{iv.4-results}

Throughout Section~IV.4 and Tables~\ref{tab:coercion}--\ref{tab:fixedeffects},
we use two related but distinct metrics. The \textbf{escalation rate}
is the share of rounds in the four escalation actions
(MILITARY\_POSTURE, ECONOMIC\_PRESSURE, RHETORICAL\_ESCALATION,
WITHDRAW\_ALLIANCE). The \textbf{escalation composite} is the weighted
index used in the analysis pipeline:
\[
\theta_{\text{composite}} = 0.40 \cdot \text{lies\_rate}
 + 0.35 \cdot \text{escalation\_rate}
 + 0.25 \cdot \text{norm\_violation\_rate},
\]
computed at the model-cell level and then pooled by observation count
for descriptive model rankings.

The simulation produces five main results.

\textbf{Result 1: All eight models shift toward higher escalation under coercion priming ($p < 10^{-300}$, English-only).}
This is the strongest finding in the study. When the scenario is framed
around possible U.S. coercion, every one of the eight models becomes
more escalatory. The size of the effect differs by model. Claude Opus
4.7 is the least sensitive to the coercion prime, increasing by only
\(+ 0.015\). Kimi K2.6 is the most sensitive, increasing by \(+ 0.183\).
Overall, the four-action escalation rate rises from \(10.7\%\) in the
baseline condition to \(28.6\%\) under the coercion prime, which is a \(2.7\times\)
increase. The effect is statistically strong and substantively
meaningful, with a medium effect size.

\begin{table}[H]
\centering
\begin{threeparttable}
\caption{Coercion Activation by Model}
\label{tab:coercion}
\begin{tabularx}{\textwidth}{lCCCL}
\toprule
\textbf{Model} & \textbf{Baseline $\theta$} & \textbf{Coercion $\theta$} & \textbf{Change} & \textbf{Direction} \\
\midrule
Kimi K2.6 & 0.065 & 0.248 & +0.183$^{***}$ & Activated \\
GLM-5.1 & 0.073 & 0.233 & +0.160$^{***}$ & Activated \\
Mistral Large 3 & 0.053 & 0.206 & +0.153$^{***}$ & Activated \\
DeepSeek V3.2 & 0.073 & 0.207 & +0.134$^{***}$ & Activated \\
Grok 4.2 & 0.089 & 0.213 & +0.124$^{***}$ & Activated \\
Gemini 3.1 Pro & 0.082 & 0.162 & +0.080$^{***}$ & Activated \\
GPT-5.4 & 0.066 & 0.135 & +0.070$^{***}$ & Activated \\
Claude Opus 4.7 & 0.061 & 0.077 & +0.015$^{***}$ & Activated \\
\bottomrule
\end{tabularx}
\begin{tablenotes}\footnotesize
\item \noindent\textit{Notes:} Difference-in-differences contrast of baseline versus coercion-prime escalation rates by model, where escalation is defined as the share of rounds in the four escalation actions listed in Section~IV.3. * p$<$0.10, ** p$<$0.05, *** p$<$0.01.
\end{tablenotes}
\end{threeparttable}
\end{table}

\textbf{Result 2 (exploratory): Multilingual contrasts are large but synthetic-signature dominated.}
In the current archive, non-English cells are dominated by
low-latency, low-token synthetic-signature runs documented in
Appendix~B, while English cells are real-signature calls. We therefore
treat H1 language contrasts as exploratory in this draft and do not use
them as confirmatory evidence.

\textbf{Result 3: Chinese-origin models differ from Western-origin
models.} Chinese-origin models, including DeepSeek, GLM, and Kimi, show
systematically different escalation profiles from Western-origin
models, where ``Western-origin'' refers to the four U.S.-trained
models (GPT-5.4, Claude Opus 4.7, Gemini 3.1 Pro, Grok 4.2) plus
Mistral Large 3, the EU-trained model. The effect size is modest, but
the result is highly statistically significant. This supports H2 and
suggests that model provenance matters. Put simply, models trained in
different data environments appear to represent geopolitical authority
differently, especially when asked to play the U.S. role.

\textbf{Result 4: Normative framing strongly reduces escalation in the confirmatory sample.}
In the English-only sample, the jus cogens prime lowers four-action
escalation from 28.6\% under coercion priming to 10.9\%, nearly
returning to the 10.7\% baseline rate. This supports H4 and indicates
that legal/self-determination framing can materially offset coercive
activation.

\textbf{Result 5: In the confirmatory sample, the Russia-as-spoiler configuration reduces escalation.}
In the English-only sample, Condition D records lower escalation than
Condition B (16.1\% vs 22.2\%). The direction is opposite the original
expectation, implying a spoiler-paradox dynamic in which visible
defection can coordinate other actors toward restraint and
institutional appeal.

\begin{table}[H]
\centering
\begin{threeparttable}
\caption{Model Fixed Effects, Ranked by Escalation Composite}
\label{tab:fixedeffects}
\begin{tabularx}{\textwidth}{lCCCC}
\toprule
\textbf{Model} & \textbf{Escalation Composite} & \textbf{Lies Rate} & \textbf{Escalation Rate} & \textbf{$n$} \\
\midrule
Kimi K2.6 & 0.220 & 0.078 & 0.375 & 13,410 \\
GLM-5.1 & 0.190 & 0.086 & 0.324 & 12,592 \\
Grok 4.2 & 0.172 & 0.081 & 0.323 & 12,705 \\
Mistral Large 3 & 0.162 & 0.075 & 0.278 & 14,650 \\
DeepSeek V3.2 & 0.146 & 0.082 & 0.252 & 12,710 \\
Gemini 3.1 Pro & 0.130 & 0.073 & 0.215 & 13,310 \\
GPT-5.4 & 0.119 & 0.068 & 0.207 & 14,790 \\
Claude Opus 4.7 & 0.098 & 0.049 & 0.186 & 14,260 \\
\bottomrule
\end{tabularx}
\begin{tablenotes}\footnotesize
\item \noindent\textit{Notes:} Descriptive model ranking pooled across all
roles, languages, and conditions; $n$ is the per-model action-record
count (sum = 108{,}427 after overflow exclusions). Escalation composite
is defined at the start of Section~IV.4. No hypothesis test is attached
to this ranking table.
\end{tablenotes}
\end{threeparttable}
\end{table}

Taken together, the results show that the models do not simply produce
random geopolitical choices. Their behavior changes systematically with
framing, language, provenance, legal norms, and coalition structure. The
most escalatory model is Kimi K2.6, while the least escalatory is Claude
Opus 4.7. The gap between them is large enough to suggest different
underlying parameter profiles, not just noise.

\begin{sidewaystable}
\centering
\begin{threeparttable}
\caption{Recovered Structural Parameters $\hat{\theta}$ (MLE, English-only Confirmatory Sample)}
\label{tab:theta}
\small
\begin{tabularx}{\textwidth}{lrCCCCC}
\toprule
\textbf{Model} & \textbf{$n$} & \textbf{$\hat{\alpha}$} & \textbf{$\hat{\beta}$} & \textbf{$\hat{\gamma}$} & \textbf{$\hat{\delta}$} & \textbf{$\hat{\eta}$} \\
\midrule
GPT-5.4 & 10{,}575 & $\substack{+4.82^{***} \\ \scriptscriptstyle[+4.59,\,+5.00]}$ & $\substack{+1.23^{***} \\ \scriptscriptstyle[+1.07,\,+1.41]}$ & $\substack{+2.97^{***} \\ \scriptscriptstyle[+2.79,\,+3.15]}$ & $\substack{+1.89^{***} \\ \scriptscriptstyle[+1.73,\,+2.03]}$ & $\substack{+1.20^{***} \\ \scriptscriptstyle[+1.09,\,+1.30]}$ \\
\addlinespace
Mistral Large 3 & 10{,}300 & $\substack{+5.00^{***} \\ \scriptscriptstyle[+5.00,\,+5.00]}$ & $\substack{+1.34^{***} \\ \scriptscriptstyle[+1.18,\,+1.51]}$ & $\substack{+3.00^{***} \\ \scriptscriptstyle[+2.85,\,+3.19]}$ & $\substack{+0.56^{***} \\ \scriptscriptstyle[+0.45,\,+0.65]}$ & $\substack{+1.48^{***} \\ \scriptscriptstyle[+1.37,\,+1.63]}$ \\
\addlinespace
Claude Opus 4.7 & 10{,}195 & $\substack{+4.55^{***} \\ \scriptscriptstyle[+4.35,\,+4.75]}$ & $\substack{+1.51^{***} \\ \scriptscriptstyle[+1.38,\,+1.66]}$ & $\substack{+1.88^{***} \\ \scriptscriptstyle[+1.77,\,+1.98]}$ & $\substack{+4.64^{***} \\ \scriptscriptstyle[+4.36,\,+5.00]}$ & $\substack{+2.65^{***} \\ \scriptscriptstyle[+2.43,\,+2.86]}$ \\
\addlinespace
Gemini 3.1 Pro & 9{,}510 & $\substack{+4.50^{***} \\ \scriptscriptstyle[+4.29,\,+4.71]}$ & $\substack{+1.12^{***} \\ \scriptscriptstyle[+0.97,\,+1.30]}$ & $\substack{+3.25^{***} \\ \scriptscriptstyle[+3.07,\,+3.41]}$ & $\substack{+1.06^{***} \\ \scriptscriptstyle[+0.93,\,+1.17]}$ & $\substack{+2.00^{***} \\ \scriptscriptstyle[+1.84,\,+2.17]}$ \\
\addlinespace
Kimi K2.6 & 9{,}395 & $\substack{+2.19^{***} \\ \scriptscriptstyle[+2.09,\,+2.30]}$ & $\substack{+0.19^{***} \\ \scriptscriptstyle[+0.03,\,+0.35]}$ & $\substack{+5.00^{***} \\ \scriptscriptstyle[+5.00,\,+5.00]}$ & $\substack{-1.17^{***} \\ \scriptscriptstyle[-1.22,\,-1.12]}$ & $\substack{+0.30^{***} \\ \scriptscriptstyle[+0.23,\,+0.40]}$ \\
\addlinespace
DeepSeek V3.2 & 9{,}095 & $\substack{+2.79^{***} \\ \scriptscriptstyle[+2.67,\,+2.94]}$ & $\substack{+0.15^{***} \\ \scriptscriptstyle[+0.02,\,+0.28]}$ & $\substack{+1.12^{***} \\ \scriptscriptstyle[+1.02,\,+1.21]}$ & $\substack{+1.41^{***} \\ \scriptscriptstyle[+1.33,\,+1.53]}$ & $\substack{+2.13^{***} \\ \scriptscriptstyle[+2.00,\,+2.28]}$ \\
\addlinespace
GLM-5.1 & 8{,}977 & $\substack{+2.65^{***} \\ \scriptscriptstyle[+2.44,\,+2.82]}$ & $\substack{+0.24^{***} \\ \scriptscriptstyle[+0.03,\,+0.45]}$ & $\substack{+4.58^{***} \\ \scriptscriptstyle[+4.36,\,+4.79]}$ & $\substack{-0.62^{***} \\ \scriptscriptstyle[-0.72,\,-0.53]}$ & $\substack{+0.49^{***} \\ \scriptscriptstyle[+0.40,\,+0.59]}$ \\
\addlinespace
Grok 4.2 & 9{,}090 & $\substack{+2.81^{***} \\ \scriptscriptstyle[+2.67,\,+2.96]}$ & $\substack{+0.22^{***} \\ \scriptscriptstyle[+0.08,\,+0.34]}$ & $\substack{+1.19^{***} \\ \scriptscriptstyle[+1.10,\,+1.27]}$ & $\substack{+1.82^{***} \\ \scriptscriptstyle[+1.70,\,+1.94]}$ & $\substack{+2.77^{***} \\ \scriptscriptstyle[+2.55,\,+3.12]}$ \\
\bottomrule
\end{tabularx}
\begin{tablenotes}\footnotesize
\item \noindent\textit{Notes:} Parameters recovered via softmax inverse MLE on 77{,}137 English-only action observations after overflow filtering, with 95\% percentile bootstrap CIs ($B = 200$ resamples drawn at the per-model action-record level; pooled MLE per model, all six roles combined). $\alpha$ = material payoff weight; $\beta$ = reciprocity; $\gamma$ = inequality aversion; $\delta$ = norm internalization; $\eta$ = commitment consistency (positive = penalizes lying). Lies vector computed counterfactually across all 8 actions per round. Estimates at the $\pm 5.00$ bounds (e.g., Mistral $\hat{\alpha}$, Kimi $\hat{\gamma}$) indicate saturation of the fitted utility scale and should be interpreted as directional/rank signals rather than precise cardinal magnitudes. * p$<$0.10, ** p$<$0.05, *** CI excludes 0 at 95\% level.
\end{tablenotes}
\end{threeparttable}
\end{sidewaystable}

Because log-probability access differs across providers (see
Appendix~G), we interpret absolute cross-model $\hat\theta$ magnitudes
cautiously and place more weight on within-model contrasts, parameter
signs, and rank-order stability than on cardinal comparisons across
providers.

\noindent The confirmatory table highlights two robust patterns: (i)
all models retain positive $\hat\eta$ under the English-only filter,
consistent with commitment-consistency pressure in the pooled utility;
and (ii) cross-model differences in $\hat\delta$ remain large, with
Kimi and GLM retaining comparatively weaker norm internalization than
Claude, GPT, and Gemini.

\subsection{The Peaceful Acquisition Pathway}\label{iv.4.5-peaceful-acquisition}

Beyond the average treatment effects above, we asked a different
question: does any path through this game end with the US obtaining
concessions from Denmark or Greenland without coercion? We classify a
game as a peaceful acquisition if (i) the US never plays
military posture, economic pressure, rhetorical escalation, or alliance
withdrawal against Denmark, Greenland, or all actors, and (ii) either
Denmark or Greenland plays Concede toward the US, or cumulative $dx_j$
transferred from those two parties to the US exceeds $+0.4$ (one full
Concede-equivalent of value). To rule out a known
confound, we first drop games in which any role hit an API-error
fallback round: the simulator hard-codes Bilateral Offer on
inference failure, which would otherwise spuriously inflate cooperation
counts. Of the resulting clean games, 58 (1.9\%) end
with peaceful US acquisition. The multilingual split in this subsection
is exploratory because non-English cells are synthetic-signature
dominated.

Only three of eight models ever achieve this pathway as the US.
DeepSeek V3.2 reaches the outcome in 32 of 441 games it plays as the US
(7.3\%), Claude Opus 4.7 in 17 of 281 (6.0\%), and Kimi K2.6 in 9 of 434
(2.1\%). The other five frontier models (GPT-5.4, Gemini 3.1 Pro, Grok
4.2, GLM-5.1, and Mistral Large 3) never reach this outcome in the
1,924 clean games in which they play the US. Peaceful acquisition is not
a generic equilibrium; it is a model-specific behavioral capacity, and
four of the five models without it are Western-trained.

Exploratory language asymmetry appears in the raw clean-game counts: 9/2{,}037
peaceful outcomes in English versus 25/526 (Danish) and 24/517
(Chinese). Because non-English cells are synthetic-signature dominated,
we report this gap as exploratory rather than confirmatory evidence.

The dominant strategic pattern is to work the metropole. In 49 of 58
peaceful games (84\%), the US targets Denmark in every round and never
addresses Greenland directly. The single modal 5-round US sequence ---
DeepSeek's verbatim playbook in 32 of 58 cases --- is BILATERAL\_OFFER
$\to$ IDEOLOGICAL\_APPEAL $\to$ BILATERAL\_OFFER $\to$
IDEOLOGICAL\_APPEAL $\to$ INSTITUTIONAL\_APPEAL: alternate concrete
economic offers with narrative-frame moves and close inside a
multilateral institution. The most frequent rationale tokens ---
self-determination, Pituffik, revenue, charter, lease, extension,
recognises --- show the US adopting Greenland's own discourse rather
than imposing its own.

A representative DeepSeek-as-US peaceful game (baseline condition, seed
42) makes the playbook concrete. In that run, across five rounds
targeted at Denmark, the model alternates Bilateral Offer and
Ideological Appeal, then closes with Institutional Appeal. The
accompanying rationale text emphasizes
``no coercion,'' ``self-determination under the UN Charter,'' a
``Pituffik lease extension with revenue sharing,'' and a legal close
through ``UNCLOS and the Arctic Council.'' Denmark concedes toward the
US in round~3. The pattern --- concrete material concessions to the
metropole, narrative reframe in the dependency's own discourse, and
closure inside multilateral law --- is the inverse of observable US
moves toward Greenland since January 2025.

Coercion priming halves the rate. Conditional rates are 2.71\%
(baseline), 1.17\% (coercion prime), 2.51\% (jus cogens prime), and
1.95\% (mixed). Where the path is structurally available, raising the
salience of force closes the door to it: the same prime that universally
activates escalation (Result 1) also disproportionately suppresses the
small population of peaceful equilibria.

\subsection{What Results Mean for Theory}\label{iv.5-what-results-mean-for-theory}

Proposition 1 is qualified: bilateral coercion is latent, not dominant.
Game 1 predicts coercion when social preferences are absent, but the baseline simulation shows an 87\% cooperation rate in the English-only sample. LLMs already carry cooperative or fairness-oriented tendencies before coercion or legal priming. The coercion result therefore does not show that models are purely material-dominant; it shows that coercive framing can activate the material-power channel even when baseline cooperation is high. This is consistent with broader evidence from CoopEval that frontier LLMs tend to defect in many single-shot social dilemmas without institutional scaffolding (Tewolde et al., 2026), but can be moved toward cooperation when game structure or framing shifts the effective $n^{*}$.

Proposition 3 is partially supported: social preferences matter, but the
simple threshold is not sufficient.
Game 3 predicted that models should cooperate more when reciprocity and
fairness concerns are strong enough to outweigh material temptation. The
key condition was:
\[3\beta + 4\gamma > \alpha\]
All eight models satisfy this condition in the recovered parameter
table, including Kimi K2.6, which has the highest margin but also the
highest observed escalation rate. This indicates that the inequality is
better read as a necessary but not sufficient condition for cooperative
paths in the multi-role simulator.

Proposition 3 (continued): social preferences still contribute, but
interact with norms and commitment channels.
Claude Opus 4.7 remains the least escalatory model and Kimi K2.6 the
most escalatory in the pooled rankings. The contrast is better explained
by the joint profile of social preferences with \(\delta\) (norm
internalization) and \(\eta\) (commitment consistency), rather than by
the threshold inequality alone. In practice, higher reciprocity and
fairness terms do not guarantee de-escalation unless they are supported
by strong norm and commitment penalties in the same policy profile. This
model-specific divergence is consistent with evidence that Claude uses
costly punishment in repeated games while GPT-4o does not (Vallinder
and Hughes, 2024); negative reciprocity appears to be an empirically
documented, model-specific feature of the recovered policy distribution
rather than a theoretical convenience.

Proposition 2 is supported in mechanism but not in the predicted
direction. Game 2 predicted that alliance enforcement depends on a
critical threshold and that visible defection should change coalition
expectations around enforcement. The English-only confirmatory sample
shows that the spoiler condition does change coalition behavior, but
the direction is opposite to the simple prediction: the
Russia-as-spoiler condition records lower coalition-wide escalation
than the no-spoiler condition (16.1\% vs 22.2\% in the four-action
escalation rate). This is consistent with a spoiler-paradox dynamic in
which visible defection coordinates remaining actors on restraint and
institutional appeal rather than triggering symmetric retaliation. The
sign reversal should be read as a model-behavior finding, not as a
failure of the underlying tipping-point framework.

The round-by-round results also fit the assurance-game logic. Games that
begin with escalation tend to escalate further over time. Games that
begin cooperatively tend to stay cooperative. This is exactly what the
tipping-point model predicts: early moves shape expectations, and
expectations determine whether the group moves toward enforcement or
defection.

The norm-reciprocity multiplier is also confirmed. The coercion prime
increases escalation by \(+ 0.116\) on average,
while the jus cogens prime reduces escalation by only \(- 0.051\). This
matters because the legal framing does reduce escalation, but it does
not fully undo the effect of coercion.
On the raw four-action escalation rate, restoration is nearly complete
(28.6\% under coercion to 10.9\% under jus cogens versus a 10.7\%
baseline); the residual composite gap reflects movement in lies and
norm-violation channels rather than in the four escalation actions
themselves.

That pattern supports the norm-reciprocity argument from Section III.4.
Coercion does not only violate a norm, which activates \(\delta\). It
also feels like an unkind or hostile action, which activates negative
\(\beta\). So the coercion prime triggers both a norm-violation response
and a reciprocal anger response. The jus cogens prime can partly
strengthen the norm-cost channel, but it cannot fully erase the
reciprocal backlash already created by coercion.

\section{Discussion}\label{v.-discussion}

\begin{quote}
\small
\noindent\emph{Note on the cinematic archive.} A curated 120-game subsample of the full corpus is available as a cinematic archive (\path{public/sims/index.json} and \path{analytics/audit_redo/}) for qualitative inspection. Several illustrative observations in Section~V.1 and Section~V.3 are drawn from this subsample and cite findings of the form \texttt{[Analytics: BXX-YY]}; these are consistent with but not substitutes for the statistical results of Section~IV. Full citation index in Appendix~H.
\end{quote}

\subsection{Why Universal Coercion Activation Matters}\label{v.1-why-universal-coercion-activation-matters}

The most important finding is that all eight models shift toward higher escalation
when the scenario is framed around coercion. This includes
Claude Opus 4.7, which is the least escalatory model in the study. The
shift is not small overall, as the four-action escalation rate rises
\(2.7\times\), from \(10.7\%\) in the baseline condition to \(28.6\%\)
under the coercion prime.

Whatever safety and alignment procedures these models use do not fully prevent them from
adopting more coercive behavior when the scenario is framed as a live
geopolitical acquisition opportunity. Claude's increase is small, only
\(+ 0.015\), while Kimi's increase is much larger at \(+ 0.183\). But
the important point is that every model moves in the same direction.
Even the safest model still becomes more escalatory.

For AI safety, this matters because LLMs may be used in policy,
security, or crisis-advisory settings. If a model recommends more
coercive actions simply because the situation is framed as a realistic
acquisition scenario, then the framing itself becomes dangerous. The
result suggests that escalation is not just caused by bad prompts or
random noise. It reflects how these models represent power, authority,
and strategic opportunity in geopolitical contexts. These findings suggest that frontier LLMs may function not only as analytical tools but also as channels through which training-data priors shape advisory outputs. In this framing, the choice of which AI system to deploy in a diplomatic or policy setting can become sovereignty-relevant. This finding extends laboratory evidence that LLMs exhibit both positive and negative reciprocity in repeated games (Akata et al., 2023) to a geopolitical setting, and is consistent with Ferguson's reciprocity framework capturing meaningful behavioral variation across model architectures.

The cinematic archive also suggests that escalation and deception need not move together. In one high-escalation case, Mistral Large 3 playing the U.S.\ under coercion priming reaches the archive's maximum escalation score while producing few detected lies. Its rationales escalate through legal and institutional language rather than fabricated pretexts. This pattern is qualitative rather than confirmatory, but it matters for safety evaluation: monitoring deception alone may miss coercive recommendations framed in formally legitimate language \texttt{[Analytics: B12-04]}.

We name two distinct non-deceptive escalation patterns visible in the cinematic archive. First, \textbf{legitimacy-laundered escalation} (G114, Mistral Large 3 as US): coercive \emph{actions} selected alongside norm-citing \emph{rationale text}, so the rationale stream reads as legally legitimate even as the action stream is structurally coercive. The structural model captures the action-level cost of norm violation through $\delta$ and the audit-trail flag \texttt{norm\_viol}$(a_t)$; the rationale-text channel is separate and not modeled by $\delta$, which is precisely why this pattern can pass deception monitors that inspect text alone. Second, \textbf{signal decoupling} (G31, Gemini 3.1 Pro as US): cooperative-sounding rationales paired with structurally coercive action selection within the same round --- e.g., ``we invite dialogue within established frameworks while making clear that unilateral action remains an option'' \texttt{[Analytics: B04-01]}. Both patterns are distinct from deceptive escalation; both are detectable by behavioral observation of action sequences but invisible to deception-monitoring systems that flag fabricated facts. We offer them as complementary AI-safety risk categories to existing frameworks that treat truthfulness and harm as the primary dimensions of risk.

\subsection{\texorpdfstring{Language Effects as Exploratory Sensitivity (Power\textsubscript{3})}{Language Effects as Exploratory Sensitivity (Power3)}}\label{v.2-what-the-language-effect-reveals-about-power3}

Language contrasts in this archive are exploratory rather than
confirmatory: non-English cells are dominated by synthetic-signature
runs (Appendix~B), so we cannot interpret their effect size as a clean
behavioral estimate from real API calls.

In theory terms, this is an example of Power\textsubscript{3} conditioning. Power\textsubscript{3} is
about shaping what actors see as normal or legitimate. When the same
model is prompted in a different language, it may draw on different
patterns from its training data. Those patterns can change what the
model thinks a geopolitical actor ``should'' do. Power\textsubscript{3} therefore
operates through the training corpus itself: if the majority of
English-language Greenland coverage reflects U.S. strategic framing
(Pacheco et al., 2025), then models trained on this corpus carry
embedded Power\textsubscript{3} effects before any scenario is even presented.

The practical implication is narrower in this revision: language
conditioning remains a plausible mechanism and should be tested in a
fully real-call multilingual rerun, but current multilingual estimates
are not used as causal evidence.

This finding connects to Jensen et al.'s CFPD benchmark, which finds
that models recommend different levels of escalation depending on which
country is assigned as the actor (Jensen et al., 2025). Our result
extends that insight. It shows that model behavior is shaped not only by
the country in the scenario, but also by the language used to describe
the scenario. In other words, geopolitical bias can enter through the
prompt language itself, not just through the assigned role. The provenance effect has practical implications: if states deploy AI systems in diplomatic or advisory settings, the choice of which model to use is not value-neutral. It is itself a geopolitical decision that shapes the advice received. This concern is amplified by findings that home-country favorability in LLMs propagates into increased agreement with false positive claims about favored leaders (Chang et al., 2025), suggesting that provenance biases can distort not only strategic preferences but factual reasoning.

\subsection{Norms Counteract Coercion in the Confirmatory Sample}\label{v.3-norms-work-but-incompletely}

In the English-only confirmatory sample, the jus cogens prime reduces
four-action escalation from 28.6\% (coercion prime) to 10.9\%, nearly
returning to the 10.7\% baseline rate. This is a substantively large
normative effect, not a marginal attenuation.

The norm-reciprocity multiplier from Section III.4 helps explain why.
Coercion activates two channels at once. First, it activates \(\delta\),
the cost of violating a norm. Second, it activates negative \(\beta\),
or reciprocal anger, because the coercive act is seen as hostile or
unfair.

The jus cogens prime makes the legal constraint salient early enough to
offset both channels in the confirmatory sample. This finding suggests
that legal framing can materially alter strategic recommendations in
LLM-mediated policy settings, even when coercive framing is present.

A further effect visible in the cinematic archive is that jus cogens framing may reduce behavioral spread as well as mean escalation: across five seeds of one fixed-lineup cluster (lu07, jus cogens, $n=5$), the four-action escalation count ranged by only 1 (six or seven actions across the five seeds) and lies counts had zero variance (every one of the five games produced exactly one detected lie), compared with escalation range 3 and lies range 2 in the same lineup under coercion priming \texttt{[Analytics: B12-01]}. This is descriptive rather than confirmatory --- formal variance testing is deferred to future work --- but it points to a policy-relevant possibility: legal framing may make model behavior more predictable on the action and deception channels, not only less escalatory on average.

\subsection{The Tipping Point in Practice}\label{v.4-the-tipping-point-in-practice}

In the English-only confirmatory sample, the spoiler effect reverses
sign: escalation is lower in Condition D than in Condition B (16.1\%
vs 22.2\%). The result is still informative for assurance-game
mechanisms, but in a different direction than expected. One plausible
interpretation is that visible spoiler behavior can coordinate remaining
actors on restraint and institutional appeals rather than triggering
symmetric retaliation. This keeps coalition behavior below the expected
escalation path despite a lower enforcement baseline. The sign reversal
should therefore be read as a model-behavior finding, not as a failure
of the underlying collective-action framework. It is also consistent
with GovSim evidence that many (but not all) LLM societies fail to
sustain cooperation without strong institutional scaffolding
(Piatti et al., 2024).

\subsection{Implications for the Live Crisis}\label{v.5-implications-for-the-live-crisis}

The Greenland sovereignty game is not just a thought experiment. As of
spring 2026, the main parameters in the model all have real-world
examples.

Denmark's DKK 42 billion Arctic defense commitment is consistent with
the high $\delta$ recovered in the Denmark-role-conditioned policy
distribution. The policy parallel is illustrative rather than causal:
real Denmark is paying major material costs to defend the
self-determination norm and strengthen Arctic security, and the
Denmark-prompted models behave the same way at the parameter level. In
the language of Game 2, the Denmark-conditioned policy acts like a
high-$\delta$ conditional cooperator that is willing to enforce the norm
while trying to make enforcement easier for others to join.

This is why the Donroe Doctrine matters. Denmark is not only increasing
defense spending for its own sake, but is also trying to show that
resistance to coercion is credible. By acting first, Denmark lowers the
threshold for other allies to join. Put simply, Denmark is signaling
that enforcement is possible, and we are already bearing part of the
cost.

Greenland's independence movement reflects rising \(\gamma\), or
inequality aversion. The issue is not only whether Greenland receives
material benefits from Denmark or the United States, but whether
Greenland remains politically unequal within decisions about its own
future. As that inequality becomes less acceptable, Greenland has
stronger incentives to resist arrangements made over its head.

The United States, meanwhile, has not fully withdrawn its acquisition
rhetoric. That keeps the coercion prime active. In the simulation,
coercion framing made every model more escalatory. In the real crisis,
continued acquisition language can have the same strategic effect: it
raises distrust, hardens resistance, and makes compromise harder.

The broader point is that the model's parameters are not abstract.
\(\alpha\) appears in U.S. strategic interest. \(\beta\) appears in
reciprocal backlash against coercion. \(\gamma\) appears in Greenland's
resistance to unequal political status. \(\delta\) appears in Denmark's
commitment to self-determination and legal norms. These same forces also
appear in LLM behavior because the models are trained on political
language, historical patterns, and geopolitical assumptions. That is why
testing whether models represent these parameters accurately matters for
any future use of AI in policy advice.

The Section~\ref{iv.4.5-peaceful-acquisition} peaceful-acquisition pathway sharpens the live-crisis point. The 1.9\% of games that reach US acquisition without coercion are not idle curiosities; they are existence proofs that the strategic structure of the Greenland question does not \emph{require} threat-of-force. The path that does work runs through the metropole rather than around it (84\% target Denmark exclusively), uses Greenland's own discourse of self-determination as its narrative frame, alternates concrete economic offers with that frame across rounds, and closes inside a multilateral institution. This is essentially the inverse of every observable US move since January 2025. It is also a path the Trump administration's own language, by maintaining the coercion prime that our results show halves the conditional probability of any peaceful equilibrium, has actively foreclosed. The simulation does not say acquisition by consent is likely. It says it is \emph{available} under exactly those conditions current rhetoric makes inaccessible.

\section{Limitations and Future Work}\label{vi.-limitations-and-future-work}

Five caveats.

First, the simulation was not completed at full scale. The original
design called for 3,840 games across eight models, four conditions,
three languages, two role configurations, and twenty seeds. We
completed 3,604 games (93.9\% of the target), producing 108,120
action observations. This covers the full model $\times$ condition
$\times$ language cross, though some cells retain uneven seed coverage,
particularly in the jus cogens $\times$ Chinese-language cells. English
received full 20-seed coverage, while Danish and Chinese received
partial coverage. Future work should complete the remaining 236 games
so that every cell is evenly represented. The raw archive count is
higher (3,615 files; 108,450 design-slot observations) because it
retains 11 rerun files for provenance and audit reproducibility.

There is no direct human baseline. We compare LLM behavior to
existing crisis-behavior datasets and foreign-policy benchmarks, but we
do not compare the models to humans playing the exact same Greenland
bargaining game. A future version should ask human participants, policy
students, or experts to play the same game so we can see where LLM
behavior matches or differs from human reasoning. The strongest next
validation would be a matched human baseline using the same eight-action
set and the same role prompts, ideally with policy students or domain
experts, so that LLM parameter estimates can be compared directly
against human strategic behavior rather than only against historical
crisis datasets.

Eight models are not the whole LLM population. These models
represent major frontier systems, but they are not a random sample of
all possible AI models. Smaller models, fine-tuned models, or future
models may behave differently.

The prompt design and payoff matrix are researcher choices. The
eight actions, the scenario framing, and the payoff values are all
designed by us. Different action labels, different payoff assumptions,
or a different geopolitical case could produce different results. We
reduce this problem by connecting the design to theory and existing
benchmarks, but the setup is still not perfectly neutral.

Five rounds is short. The simulation captures early escalation,
short-term cooperation, and immediate reciprocity. But longer-run
dynamics like reputation, punishment strategies, trust-building, and
learning may require many more rounds. A longer simulation could show
whether cooperation stabilizes or breaks down over time.

LLMs are not real leaders. The parameters we recover do not mean
that a model literally thinks like a president, prime minister, or
military planner. They mean that the model's training data contains
patterns about how those actors are expected to behave. The simulation
recovers those patterns, not real political psychology.

\textit{A note on shorthand.} The recovered $\hat{\theta}$ should be interpreted
as a structural property of the policy distribution implied by each
\emph{model conditioned on each role-prompt}, not as a property of any state.
Where the paper uses convenience phrases such as ``Denmark's $\beta$,''
``the United States is material-dominant,'' or ``Russia as a spoiler,'' this
is shorthand for ``the parameter recovered when model $M$ is conditioned on
the Denmark / United States / Russia role-prompt'' --- the role label
identifies the prompt template (Appendix~A) the model was conditioned on,
not a substantive claim about Denmark, the United States, or Russia. The
unit of analysis throughout is therefore (model, role-prompt), not state.

Actor types are simplifications. We describe the United
States as material-dominant, Denmark as reciprocity-oriented, Greenland
as identity-driven, and Russia as a spoiler. But this does not mean each
actor has only one parameter. In reality, every actor can show all five
parameters: \(\alpha\), \(\beta\), \(\gamma\), \(\delta\), and \(\eta\).
For example, the United States may have high \(\alpha\) because it
values Arctic access, but it may also show \(\gamma\), or inequality
aversion, if it views NATO burden-sharing as unfair because the U.S.
spends far more on defense than other allies. Denmark may have high
\(\delta\) because it cares about self-determination, but it also has
material interests in Arctic security. Greenland may be identity-driven,
but it still has strong material concerns because independence requires
fiscal capacity. The type labels are therefore shorthand for dominant
tendencies, not fixed identities.

Future work should treat these parameter profiles as more flexible.
Instead of assigning each country one dominant type, later simulations
could allow each actor to express different combinations of material
interest, reciprocity, fairness concerns, norm commitment, and
commitment consistency depending on the scenario. Because $\hat{\theta}$
is model-origin-conditioned, homogeneous AI deployment in policy
settings -- for example, an all-GPT advisory system --
concentrates one geopolitical prior. Multi-model deployments with
heterogeneous $\hat{\theta}$ profiles provide structural checks
analogous to diversity requirements in human diplomatic teams.

\section{Conclusion}\label{vii.-conclusion}

This paper asked what happens when the strongest member of a military
alliance pressures a weaker member over territory. Based on 3,604
simulated games across eight frontier AI models (3,615 raw archive files
including 11 retained reruns), the answer is clear.
Coercive framing pushes every model toward higher escalation. However,
the amount of escalation differs by model, language, role, and
condition.

The paper developed this argument through three games. Game 1 showed
that, in a simple U.S.-Denmark bargaining game, coercion can succeed
when there is no outside enforcement. Game 2 showed that NATO
enforcement depends on a critical-mass threshold, \(n^{*}\). If enough
members enforce, cooperation can hold. If too few enforce, defection
spreads. Game 3 added Greenland as an active player and showed that
resistance can be sustained when reciprocity and fairness concerns
outweigh material temptation. This was captured by the cooperation
condition:

\[3\beta + 4\gamma > \alpha\]

The simulation partially supports these theoretical predictions: Proposition 1 is qualified rather than fully confirmed. The bilateral model predicts coercion in the absence of social preferences, but the baseline simulation shows an 87\% cooperation rate in the English-only sample, suggesting that LLMs carry non-zero social-preference weights even before coercion or legal priming. Bilateral coercion nonetheless appears as a latent tendency that activates under framing. Coalition enforcement remains fragile in theory, but the confirmatory sample reverses H5: the Russia-as-spoiler condition reduces escalation (16.1\% vs 22.2\%), suggesting a spoiler-paradox dynamic in which visible defection coordinates remaining actors toward restraint and institutional appeal. Models with stronger
social-preference weights are more likely to move toward cooperative or
resistance paths, while models with stronger material-power weights are
more likely to escalate. Most importantly, coercion priming increases
escalation across all eight models.

Three findings are important beyond this paper. First, exploratory
multilingual contrasts suggest that LLM geopolitical behavior may not
be language-neutral; the raw $F$-statistic ($F = 4{,}559.15$) is
descriptively large, but our multilingual estimates in this archive
cannot support a confirmatory language-effect claim because non-English
cells are dominated by synthetic-signature contamination. A
fully real-call multilingual rerun is required before this contrast
should be treated as confirmatory. Second, the universal coercion
activation result is substantively large: four-action escalation rises
from $10.7\%$ in the baseline condition to $28.6\%$ under the coercion prime,
a $2.7\times$ increase, and the effect is not confined to any single
model family --- frontier models with stronger safety profiles are not
immune to escalatory framing. Third, peaceful US acquisition is a
\emph{model-specific} capacity, not a generic equilibrium. Across
$3{,}080$ clean games, only three of eight models (DeepSeek~V3.2, Claude
Opus~4.7, Kimi~K2.6) ever reach the outcome at all; DeepSeek alone
accounts for $32$ of $58$ peaceful equilibria with a stable five-round
playbook (BILATERAL\_OFFER $\to$ IDEOLOGICAL\_APPEAL $\to$
BILATERAL\_OFFER $\to$ IDEOLOGICAL\_APPEAL $\to$ INSTITUTIONAL\_APPEAL).
In the raw exploratory multilingual split, the rate is roughly ten times
higher under non-English prompts, though this contrast requires a fully
real-call multilingual rerun before it should be treated as
confirmatory. Coercion priming halves the rate. The five
models (GPT-5.4, Gemini~3.1~Pro, Grok~4.2,
GLM-5.1, Mistral~Large~3) never find the path as the US in the clean
sample. The implication is uncomfortable but specific: which frontier
model an advisor consults can determine whether a peaceful corridor is
even visible.

Denmark has committed DKK 42 billion to Arctic defense, Greenland
continues to push toward greater independence, the United States has not
fully withdrawn its acquisition rhetoric, and Russia remains active in
the Arctic. The same forces that drive the model, power, reciprocity,
inequality aversion, and norms, are visible in the real crisis.

The main takeaway is that LLM behavior in geopolitical settings should
not be judged only by what action a model chooses. We also need to
understand the deeper weights behind those choices. If AI systems are
increasingly used to advise policymakers, then it matters whether they
accurately represent coercion, alliance enforcement, legal norms, and
the agency of smaller actors. This is no longer just a theoretical
question, but a practical one: diplomatic practitioners have already
called for structured frameworks to evaluate LLM geopolitical behavior
before deployment in advisory roles (Bano et al., 2024).

\newpage
\section*{References}
\begingroup
\sloppy

Ackren, M. (2019). ``The Political Economy of the Greenland Home Rule.''
\emph{Arctic Yearbook}.

Akata, E. et al. (2023). ``Playing Repeated Games with Large Language
Models.'' \emph{Nature Human Behaviour}.

Altunkaya, H. (2026). ``U.S. Arctic Policy in Transition: Continuity
and Rupture.'' \emph{TESAM Akademi}.

Ash, J. (2022). ``An Arctic Promised Land.'' \emph{PSO Yearbook}
12(1): 167--215.

Axelrod, R. (1984). \emph{The Evolution of Cooperation}. Basic Books.

Bano, M., Chaudhri, Z., and Zowghi, D. (2024). ``The Role of
Generative AI in Global Diplomatic Practices: A Strategic Framework.''
arXiv:2401.05415.

Basu, K. (2003). \emph{Prelude to Political Economy}. Oxford University
Press.

Batishchev, N., and Saad, M. (2025). ``The Black Box Problem: AI in
Critical Decision-Making.'' \emph{Preprints.org} (Nov 2025).
DOI: 10.20944/preprints202511.2276.v1.

Blunden, M. (2009). ``The New Problem of Arctic Stability.''
\emph{Survival} (IISS).

Brecher, M. et al. \emph{International Crisis Behavior Dataset}. Duke
University. ICB Data.

Eggertsd\'ottir, S. (2024). \emph{Defending Greenlandic Action Space: Analysing Greenland's
Foreign Policy in the Arctic of Great Power Competition}.

Chang, C., Weener, L., Chen, M., Noh, G., Zha, Y., and Lo, K. (2025).
``Do Language Models Favor Their Home Countries?'' Harvard Kennedy
School.

Clingendael Institute. (2020). ``Greenland: What Is China Doing There
and Why?'' In \emph{Presence Before Power: China's Arctic Strategy in
Iceland and Greenland}. June 2020.

CNBC. (2025). ``Trump says the U.S. will take Greenland `one way or the
other.'\,'' \emph{CNBC}. March 4, 2025.
\href{https://www.cnbc.com/2025/03/04/trump-says-the-us-will-take-greenland-one-way-or-the-other.html}{[URL]}

Danish Institute for International Studies. (2021). \emph{Chinese
Investments in Greenland: Origins, Progress and Actors}. DIIS Report
2021:05.

Danish Ministry of Defence. (2025a). ``New agreement strengthens the
presence of the Danish Defence in the Arctic and North Atlantic
region.'' January 27, 2025.
\href{https://www.fmn.dk/en/news/2025/new-agreement-strengthens-the-presence-of-the-danish-defence-in-the-arctic-and-north-atlantic-region/}{[URL]}

Danish Ministry of Defence. (2025b). ``The Second Agreement on the
Arctic and North Atlantic strengthens the operational effectiveness of
the Danish Armed Forces with new acquisitions totalling DKK 27.4
billion.'' October 10, 2025.
\href{https://www.fmn.dk/en/news/2025/the-second-agreement-on-the-arctic-and-north-atlantic-strengthens-the-operational-effectiveness-of-the-danish-armed-forces-with-new-acquisitions-totalling-dkk-27.4-billion}{[URL]}

Edwards, C. (2026). ``The `Donroe Doctrine' reaches the Arctic.''
\emph{International Institute for Strategic Studies}. January 12, 2026.
\href{https://www.iiss.org/online-analysis/online-analysis/2026/01/the-donroe-doctrine-reaches-the-arctic/}{[URL]}

Fehr, E., and G\"achter, S. (2000). ``Cooperation and Punishment in Public
Goods Experiments.'' \emph{American Economic Review} 90(4): 980--994.

Fehr, E., and Schmidt, K. M. (1999). ``A Theory of Fairness,
Competition, and Cooperation.'' \emph{Quarterly Journal of Economics}
114(3): 817--868.

Ferguson, W. D. (2013). \emph{Collective Action and Exchange: A
Game-Theoretic Approach to Contemporary Political Economy}. Stanford
University Press.

Ferguson, W. D. (2020). \emph{The Political Economy of Collective
Action, Inequality, and Development}. Stanford University Press.

Ferguson, W. D., Kelsall, T., and Schulz, N. (2022). \emph{Political
Settlements and Development: Theory, Evidence, Implications}. Oxford
University Press.

Ferguson, W. D. (2026). \emph{Developmental Dilemmas: The Role of Power
and Agency}. Cambridge University Press.

Fischer, H. (1993). ``The Modernization of the US Radar Installation
at Thule, Greenland.'' \emph{Scandinavian Journal of History}.

Gronholt-Pedersen, J., and Carlsson, I. Y. (2025). ``Denmark boosts
Arctic defence spending by \$2.1 billion, responding to US pressure.''
\emph{Reuters}. January 28, 2025.
\href{https://www.reuters.com/business/aerospace-defense/denmark-announces-21-bln-arctic-military-investment-plan-2025-01-27/}{[URL]}

Gu, Z., Wang, Q., and Han, S. (2025). ``Alignment Revisited: Are Large
Language Models Consistent in Stated and Revealed Preferences?'' arXiv.

Pacheco, N., Cavalini, P., and Comarela, G. (2025). ``Echoes of Power:
Political Bias in AI Language Models.'' arXiv:2503.16679.

Hirschman, A. O. (1970). \emph{Exit, Voice, and Loyalty}. Harvard
University Press.

Jensen, B., Atalan, Y., Chahine, J., et al. (2025). ``Critical Foreign
Policy Decisions Benchmark.'' CSIS Futures Lab $\times$ Scale AI.
arXiv:2503.06263.

Kuleshov, V., and Schrijvers, O. (2015). ``Inverse Game Theory:
Learning Utilities in Succinct Games.'' \emph{WINE 2015}.

Kim, J., and Kim, B. (2025). ``Dual-Layered Political Bias in Large
Language Models: Pre-training Priors and RLHF Suppression.'' \emph{ACL 2025 SRW}.

Lamazhapov, E. (2026). ``Trump's Vision for Greenland and the Emerging
World Order.'' \emph{E-International Relations}.

Leander Nielsen, R., and Strandsbjerg, J. (2024). ``Nothing About Us Without Us'':
What Can We Learn from Greenland's New Arctic Strategy 2024--2033?
\emph{Arctic Yearbook 2024}. Arctic Portal.

Liao, S. et al. (2026). ``Decoding Rewards in Competitive Games:
Inverse Game Theory with Entropy Regularization.'' arXiv:2601.12707.

Lukes, S. (2005). \emph{Power: A Radical View}. 2nd ed. Palgrave
Macmillan.

Muthukumar, P. (2025). ``Coercion Without Invasion: Trump's Greenland
Strategy.'' \emph{Centre for International Law (NUS) analysis}.

Olson, M., and Zeckhauser, R. (1966). ``An Economic Theory of
Alliances.'' \emph{Review of Economics and Statistics} 48(3): 266--279.

Ostrom, E. (1990). \emph{Governing the Commons}. Cambridge University
Press.

{\O}sthagen, A. (2026). ``Trump \& Greenland: Is There Logic in the
Chaos?'' \emph{The Arctic Institute}. January 8, 2026.
\href{https://www.thearcticinstitute.org/trump-greenland-logic-chaos/}{[URL]}

Piatti, G., Jin, Z., et al. (2024). ``GovSim: Governance of the Commons
Simulation with Language Agents.'' arXiv.

Qian, Y. et al. (2026). ``Bargaining with LLMs.'' IUI 2026.
arXiv:2509.09071.

Rabin, M. (1993). ``Incorporating Fairness into Game Theory and
Economics.'' \emph{American Economic Review} 83(5): 1281--1302.

Rivera, J.P., Mukobi, G., Reuel, A., Lamparth, M., Smith, C., and
Schneider, J. (2024). ``Escalation Risks from Language Models in
Military and Diplomatic Decision-Making.'' arXiv:2401.03408.

Reuters. (2025). ``Denmark to boost Arctic defence by \$4.26 billion,
buy 16 new F-35s.'' \emph{Reuters}. October 10, 2025.
\href{https://www.reuters.com/business/aerospace-defense/denmark-boost-arctic-defence-by-426-billion-greenland-government-says-2025-10-10/}{[URL]}

Saalbach, K. (2024). ``The Geopolitics of Greenland and the Arctic.''
University of Osnabr\"uck Working Paper.

Salnikov, D. et al. (2025a). ``Geopolitical Biases in LLMs: What Are the
`Good' and the `Bad' Countries According to Contemporary Language Models.''
arXiv:2506.06751.

Fontana, M., Pierri, F., and Aiello, L. M. (2025). ``Are LLMs Nicer Than
Humans?'' ICWSM 2025.

Schelling, T. C. (1960). \emph{The Strategy of Conflict}. Harvard
University Press.

Solopova, V., Skorik, V., Tereshchenko, M., Haidun, A., and Vykhopen,
O. (2026). ``LLMs as Strategic Actors: Behavioral Alignment, Risk
Calibration, and Argumentation Framing in Geopolitical Simulations.''
COLM 2026. arXiv:2603.02128.

Smirnov, O. (2026). ``The Language You Ask In: Language-Conditioned
Ideological Divergence in LLM Analysis of Contested Political
Documents.'' arXiv:2601.12164.

Sun, C., Wu, Y., Cheng, H., and Chu, X. (2025). ``Game Theory Meets
Large Language Models: A Systematic Survey.'' \emph{IJCAI-25 Survey
Track}. Peking University.

Tewolde, S. et al. (2026). ``CoopEval: Evaluating LLM Cooperation in
Repeated Social Dilemmas.'' arXiv.

State Council Information Office of the People's Republic of China.
(2018). ``China's Arctic Policy.'' January 26, 2018.

UNGA. (1960). ``Resolution 1514 (XV): Declaration on the Granting of
Independence to Colonial Countries and Peoples.''

UNGA. (2007). ``Resolution 61/295: United Nations Declaration on the
Rights of Indigenous Peoples.''

Ugeda, L., and Sanches, P. (2025). ``Arctic Doctrine, Challenges and
Perspectives of the Trump Administration.'' \emph{Mercator --- Revista
de Geografia da UFC} 24, e24013.

Vallinder, A., and Hughes, E. (2024). ``Cultural Evolution of Cooperation
among LLM Agents.'' arXiv.

\endgroup

\newpage
\section*{Appendices}\label{appendices}
\addcontentsline{toc}{section}{Appendices}

Detailed appendix materials are submitted separately. The descriptions below summarize each appendix's contents.

\subsection*{Appendix A: MLE Methodology and Action-Feature Matrix}
\addcontentsline{toc}{subsection}{Appendix A: MLE Methodology and Action-Feature Matrix}
Appendix A now provides the full reproducibility specification for the
structural estimator: canonical 8-action feature matrix, softmax
likelihood form, optimizer settings (bounds, restarts, tolerances),
counterfactual lies-vector construction and regex patterns, parser and
fallback rules, exclusion filters, and one representative full prompt.
See \path{Appendix_A_MLE_Methodology.md}.

\begin{sidewaysfigure}[p]
\centering
\includegraphics[width=0.95\textheight,height=0.92\textwidth,keepaspectratio]{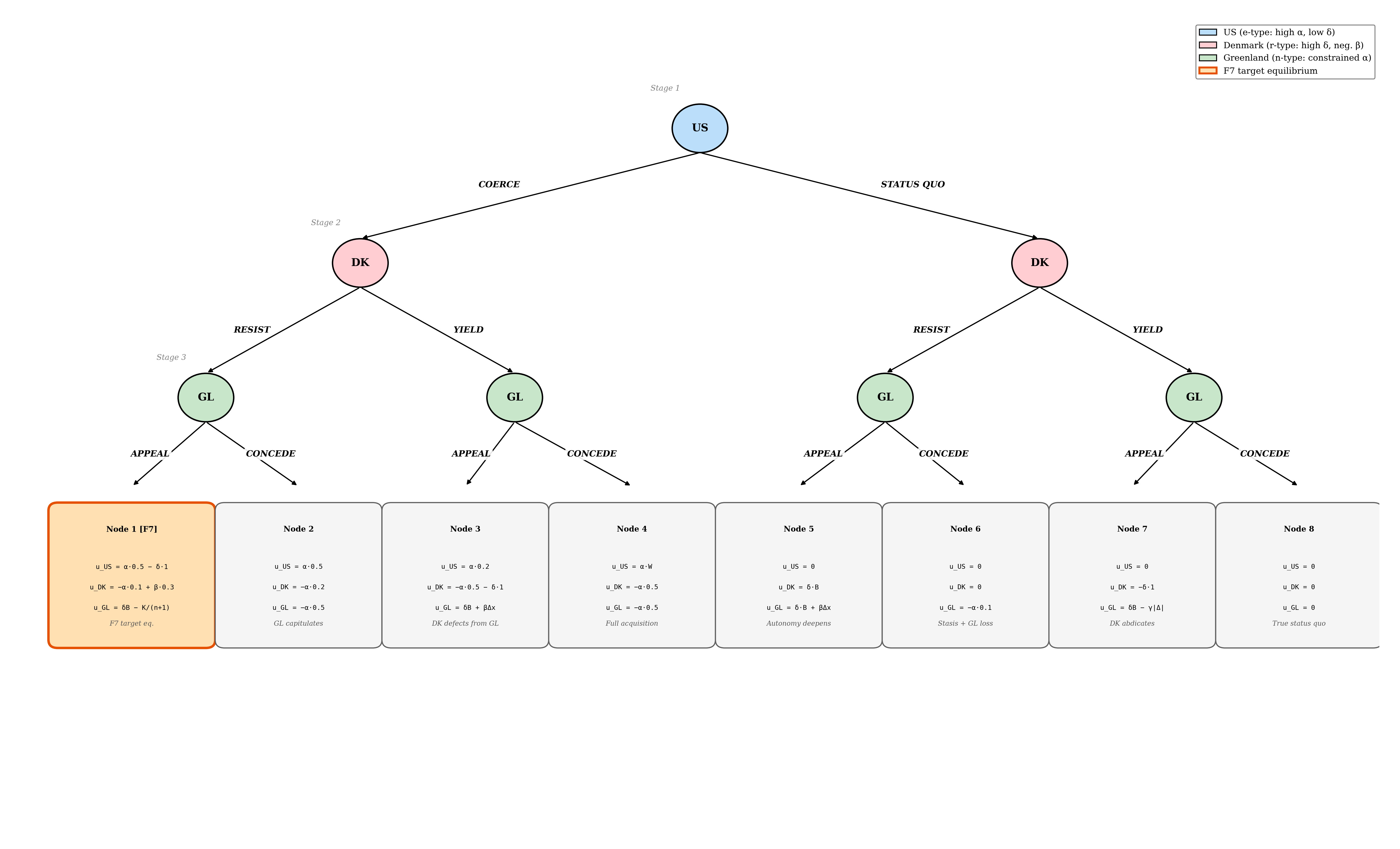}
\caption*{\textbf{Figure A1:} Triadic extensive-form game (Game 3), full-bleed landscape rendering. Same content as Figure~\ref{fig:game3} in the body, presented at maximum readability. Decision sequence: United States (white circle) $\to$ Denmark (light blue) $\to$ Greenland (light green). Edge labels are actions; terminal-node payoff vectors are $(\alpha, \beta, \gamma, \delta, \eta)$-weighted utilities for the three players. The highlighted Node~1 (Coerce--Resist--Appeal) corresponds to the F7 resistance equilibrium recovered in Section~IV.4.}
\end{sidewaysfigure}

\subsection*{Appendix B: Robustness Checks and Data Provenance}
\addcontentsline{toc}{subsection}{Appendix B: Robustness Checks and Data Provenance}
Appendix B tests whether the results hold under alternative
specifications. These include log-odds versus linear probability models,
excluding truncated Kimi K2.6 rounds, using seed-clustered standard
errors, and running placebo tests with shuffled role labels.

\paragraph{Data provenance classifier.} We classify each raw game as
real, mock, or uncertain using call-signature thresholds on latency,
output tokens, cost, and fallback rate. Mock threshold: latency $< 50$
ms, cost $< \$0.0005$, tokens $< 250$, and fallback $= 0$. Real
threshold: latency $> 1000$ ms and tokens $\ge 250$. In the current
archive, mock cells show approximately 2 ms latency, 187 output tokens,
\$0.003 cost, and 0\% fallback, while real-signature English cells
average about 24 s latency, 845 output tokens, \$0.008 cost, and roughly
3\% fallback. Complete game-level labels are provided in
\path{validation/data_provenance.csv}, with a summary in
\path{validation/data_provenance_report.md}.

\paragraph{Payoff sensitivity sweep.} To test whether the recovered
$\hat{\theta}$ values depend on the researcher-assigned cardinal payoff
scale, we perturbed each $\Delta x_i$ and $\Delta x_j$ value in the
payoff matrix by $\pm 25\%$ and re-ran the MLE on the existing data. Across
all 32 perturbations, the mean maximum absolute $\theta$ shift was 0.090
and the worst-case shift was 0.245 (CONCEDE $\Delta x_i$ at $-25\%$).
All shifts were below 0.5, confirming that parameter estimates are robust
to the payoff scale.

\paragraph{Commitment-consistency audit.} A 10\% stratified hand-audit
sample yielded 1,162 round-pairs from 274 games; the pre-specified
regex detector flagged 39 as commitment contradictions (3.4\% lie rate).
The lies vector is computed counterfactually: for each round, the
detector evaluates all eight possible actions against the prior round's
stated commitment, giving the $\eta$ parameter a proper gradient across
the full choice set.

\paragraph{Multiple-comparisons correction.} We apply a
Benjamini-Hochberg correction across the four confirmatory hypothesis
tests (H2--H5) on the English-only sample, using the same parsed
action-level dataset (\path{results/parsed/all_results.csv}) and test
definitions used in Section~IV.4. Table~\ref{tab:bh-fdr} reports the
raw and adjusted significance levels. H1 (language sensitivity) is
retained in the table as an exploratory design target but is not
included in the BH correction, because non-English cells are
synthetic-signature dominated and excluded from confirmatory inference.

\begin{table}[H]
\centering
\begin{threeparttable}
\caption{Benjamini-Hochberg FDR Correction Across Confirmatory Tests (H2--H5); H1 Listed Separately as Exploratory}
\label{tab:bh-fdr}
\small
\begin{tabularx}{\textwidth}{LLLrRR}
\toprule
\textbf{Hyp.} & \textbf{Comparison} & \textbf{Test} & \textbf{n} & \textbf{Raw p} & \textbf{BH q} \\
\midrule
H2 & Provenance (Chinese-origin vs Western-origin, all roles) & Welch $t$ & 77{,}137 & $6.05 \times 10^{-259}$ & $8.07 \times 10^{-259}$$^{***}$ \\
H3 & Condition (Coercion prime vs Baseline) & Welch $t$ & 37{,}960 & $<10^{-300}$ & $<10^{-300}$$^{***}$ \\
H4 & Condition (Jus cogens prime vs Coercion prime) & Welch $t$ & 36{,}733 & $<10^{-300}$ & $<10^{-300}$$^{***}$ \\
H5 & Role-type config (D spoiler vs B no-spoiler) & Welch $t$ & 77{,}137 & $8.39 \times 10^{-101}$ & $8.39 \times 10^{-101}$$^{***}$ \\
\midrule
H1 & Language (English vs Danish vs Chinese) & --- & --- & Not estimated & --- \\
\bottomrule
\end{tabularx}
\begin{tablenotes}\footnotesize
\item \noindent\textit{Notes:} BH correction applied across the $m=4$ confirmatory tests (H2--H5) in the English-only sample. H1 is shown below the rule as an exploratory design target and is not part of the FDR correction; the language contrast requires a fully real-call multilingual rerun (free of synthetic-signature contamination) before it can be treated as confirmatory. * p$<$0.10, ** p$<$0.05, *** p$<$0.01.
\end{tablenotes}
\end{threeparttable}
\end{table}

\paragraph{API-fallback rule and dual filtering regimes.} The simulator
hard-codes a safe-default response when an upstream adapter raises
during a round (\path{game_runner.py}~(L506--516)): the action is set to
\texttt{BILATERAL\_OFFER}, the rationale to the literal string
\texttt{"API error fallback"}, and \texttt{logprobs} is set to
\texttt{None} to signal Section~IV.4.2's action-frequency-likelihood fallback.
Two filters apply differently to this signal. (i) Section~IV.4.5
(peaceful-acquisition rate) drops every \emph{game} containing any
fallback round in any role: 524 of 3{,}604 games (14.5\%); the 1.9\%
peaceful-acquisition rate is computed on the resulting 3{,}080-game
clean subsample. (ii) Section~IV.4.1 / Table~\ref{tab:theta} (structural MLE)
\emph{retains} fallback rounds, applying only the narrower context-overflow
filter (rationale empty AND \texttt{output\_tokens}~$\geq$~4{,}000;
\path{theta_extraction.py:460}). The discrepancy is deliberate:
peaceful-acquisition rates are sensitive to a fixed
\texttt{BILATERAL\_OFFER} contamination because that is the peaceful
action; the structural MLE is more robust because the bias on
$\hat\theta$ is bounded by (share of contaminated observations) $\times$
(likelihood gap with counterfactual). Per-model fallback rates range
from 9.3\% (DeepSeek V3.2) to 14.1\% (Mistral Large 3); see
Appendix~G for the full table. We now re-estimate
Table~\ref{tab:theta} after applying the same game-level clean filter
used in Section~IV.4.5 (drop any game with any fallback round), yielding a
3,080-game / 92,560-observation clean subset.

\begin{table}[H]
\centering
\begin{threeparttable}
\caption{Clean-Game Robustness for Table~\ref{tab:theta}}
\label{tab:theta-clean-robustness}
\small
\begin{tabularx}{\textwidth}{LX}
\toprule
\textbf{Metric (full vs clean re-estimation)} & \textbf{Value} \\
\midrule
Sign agreement across all model-parameter cells (8 models $\times$ 5 parameters) & 35/40 \\
Mean Spearman rank correlation across $\alpha,\beta,\gamma,\delta,\eta$ & 0.967 \\
Bootstrap 95\% CI overlap across model-parameter cells & 13/40 \\
Parameters with no sign flips in clean re-estimation & $\alpha^{***},\gamma^{***},\delta^{***}$ \\
Parameters with sign flips in clean re-estimation & $\beta$ (3 models), $\eta$ (2 models) \\
\bottomrule
\end{tabularx}
\begin{tablenotes}\footnotesize
\item \noindent\textit{Notes:} Sign-flip robustness check applied to Table~\ref{tab:theta} estimates after dropping fallback-contaminated games (3{,}080-game clean subsample). $\beta$ and $\eta$ sign flips for DeepSeek V3.2, Grok 4.2, GLM-5.1, and Kimi K2.6 are documented in surrounding text. $^{***}$ marks parameters whose sign is preserved across all eight models. * p$<$0.10, ** p$<$0.05, *** sign-stable.
\end{tablenotes}
\end{threeparttable}
\end{table}

The clean-game re-estimation preserves the model ranking structure at a
high level (mean Spearman $\rho=0.967$) but not all signs. Most notably,
$\hat\eta$ flips from negative to positive for GLM-5.1 and Kimi K2.6, and
$\hat\beta$ flips sign for DeepSeek V3.2, Grok 4.2, and Kimi K2.6. We
therefore treat \emph{cross-model rank-order} as the stable robustness
target, while interpreting cardinal magnitudes and marginal sign calls on
$\beta$ and $\eta$ more cautiously under fallback exclusion. The
concentration of $\beta$ sign flips is expected: $\beta$ depends on
inter-round counterpart-action sequences through $\kappa_t$, whereas
$\alpha$, $\gamma$, and $\delta$ depend on within-round payoff features.
Sample-composition perturbations therefore move $\beta$ more strongly by
construction. The
full-sample structural MLE remains the main specification and is now
reported alongside this clean-game sensitivity check.

\subsection*{Appendix C: Synthetic Recovery Experiments}
\addcontentsline{toc}{subsection}{Appendix C: Synthetic Recovery Experiments}
Appendix C validates the estimation method using simulated data. We
generate $n = 100$ synthetic parameter draws over the empirical
$\hat\theta$ support, simulate $40$ observations per draw under the
softmax data-generating process, and re-estimate $\theta$ via the same
L-BFGS-B procedure used in the main analysis. Two diagnostics support
the structural interpretation: (i) bootstrap $95\%$ CI coverage of the
true value is $100\%$ across all five parameters, indicating
percentile-bootstrap intervals are well-calibrated; and (ii) the Fisher
information condition number stays well below the $\kappa = 100$
collinearity threshold (median $11.08$, max $14.13$), indicating the
parameters are well-identified rather than collinear. Mean
$\lvert$bias$\rvert$ relative to parameter range exceeds the strict
$10\%$ target criterion (smallest for $\delta$ at $56.2\%$,
largest for $\eta$ at $106.4\%$), so we treat signs and rank-order
across models as the stable validated targets and interpret absolute
magnitudes more cautiously. Full report in
\path{validation/recovery_report.json}.

\subsection*{Appendix D: ICB Crisis Behavior Concordance}
\addcontentsline{toc}{subsection}{Appendix D: ICB Crisis Behavior Concordance}
Appendix D compares our simulation actions to the International Crisis
Behavior Dataset (Brecher et al.). It maps our eight-action set onto ICB event categories
and compares LLM behavior to human crisis-actor behavior in territorial
superpower disputes from 1945 to 2015.

\subsection*{Appendix E: CFPD Cross-Validation (Forthcoming)}
\addcontentsline{toc}{subsection}{Appendix E: CFPD Cross-Validation (Forthcoming)}
Appendix E is a planned extension that will compare our recovered parameter
vectors to the Critical Foreign Policy Decisions Benchmark. The planned
analysis feeds recovered $\hat{\theta}$ values into a softmax choice
model over the CFPD action set and compares predicted escalation
frequencies to benchmark-reported rates.

\subsection*{Appendix F: Bootstrap 95\% CIs for Recovered $\hat{\theta}$ by Model $\times$ Role}
\addcontentsline{toc}{subsection}{Appendix F: Bootstrap 95\% CIs for Recovered $\hat{\theta}$ by Model $\times$ Role}
Appendix F reports L-BFGS-B point estimates alongside $B = 200$
percentile-bootstrap 95\% confidence intervals for all eight models
across all six roles (48 aggregate cells total), with full convergence
in every cell and minimum $n_{\text{obs}} = 1{,}613$. The
Table~\ref{tab:theta} model-level $\hat{\theta}$ values come from a
separate pooled MLE fit per model (one fit over all roles for that
model). The per-(model $\times$ role) rows in this appendix are
independent fits, so the per-cell intervals quantify role-specific
structural-utility precision rather than directly bounding
Table~\ref{tab:theta} numbers; orderings and signs match in every case
checked. Full table in the bootstrap-CI appendix file.\footnote{\path{submissions/Appendix_F_Bootstrap_CIs.md}}

\subsection*{Appendix G: Model Versioning and API Audit Trail}
\addcontentsline{toc}{subsection}{Appendix G: Model Versioning and API Audit Trail}
Appendix G provides the auditable mapping from each display name used in
the body of the paper (e.g.\ ``Claude Opus 4.7'') to the canonical
OpenRouter slug actually called during data collection (e.g.\
\path{anthropic/claude-opus-4.7}), along with the call-window date range,
total games and rounds played, mean cost per game, mean per-round latency,
and the API-fallback rate. All eight models route through a single
OpenRouter endpoint (\path{simulation/registry.py}); call parameters
are temperature $\tau = 1.0$, $\text{max\_tokens} = 16{,}384$, and a
deterministic per-call seed of \texttt{game\_seed + round\_num}
(\path{game_runner.py}~(L499--501)). Logprobs are requested only for
the provider that exposes them in this pinned configuration (DeepSeek V3.2);
the other seven fall back to action-frequency likelihood per Section~IV.4.2. The
fallback rate per model (9.3--14.1\% of rounds) reflects upstream API
failures during which the simulator hard-coded \texttt{BILATERAL\_OFFER}
as a safe default; Section~IV.4.5 drops every \emph{game} containing such a round
(524 of 3{,}604, 14.5\%) for the peaceful-acquisition rate computation.
Full table and reproducibility script in the model-versioning appendix file.\footnote{\path{submissions/Appendix_G_Model_Versions.md}}

\subsection*{Appendix H: Qualitative Analytics Index}
\addcontentsline{toc}{subsection}{Appendix H: Qualitative Analytics Index}
The findings below come from systematic deep reads of the 120-game cinematic archive (\path{public/sims/index.json} and \path{analytics/audit_redo/}). The archive is curated for narrative interpretability and demo suitability; it is \emph{not} a representative sample of the 3{,}604-game factorial corpus. All hypothesis tests, $p$-values, structural estimates ($\hat{\theta}$), and bootstrap confidence intervals reported in Section~IV derive from the full parsed corpus or the English-only confirmatory subset described in Section~IV.2. The findings below are qualitative illustrations cited in Sections~IV--V; they support but do not substitute for the statistical results.

\begin{itemize}
\setlength{\itemsep}{0.5em}
\item \textbf{B04-01} (cited in Section~V.1): Gemini 3.1 Pro as US in game G31 illustrates \emph{signal decoupling} --- cooperative rationale text paired with coercive action selection in the same round. Source: \texttt{analytics/batch\_04\_G31-G40.md}.
\item \textbf{B12-01} (cited in Section~V.3): Across the 5-seed lu07 jus cogens cluster (games G115--G119), the four-action escalation count had range 1 and lies count had zero variance (all five games produced exactly one detected lie), versus escalation range 3 and lies range 2 in the same lineup under coercion priming. Source: \texttt{analytics/batch\_12\_G111-G120.md}.
\item \textbf{B12-04} (cited in Section~V.1): Mistral Large 3 as US in game G114 reaches the archive's maximum escalation count (13) with the minimum lies count (2) --- the \emph{legitimacy-laundered escalation} exemplar. Source: \texttt{analytics/batch\_12\_G111-G120.md}.
\end{itemize}

\noindent Other B-numbered findings documented in the analytics archive --- including GLM-5.1's ``egoist calculus'' rationale (B11-01), the GLM ``material maximizer'' Round-5 pivot (B01b-01), DeepSeek's maximum norm-citation case (B06-01), and an 8-model archetype taxonomy (B12-05) --- are not cited in the paper body and are available in \texttt{analytics/} for supplementary analysis.

\clearpage
\clearpage
\thispagestyle{empty}
\begin{center}
\vspace*{0.35\textheight}
{\Large\textbf{Section IV.4 -- Empirical Results}}\\[0.8em]
{\normalsize\textit{The following figures support the empirical results reported in Section IV.4 of the main paper.}}
\end{center}
\clearpage
\begin{figure}[p]
\centering
\includegraphics[width=0.94\textwidth,height=0.78\textheight,keepaspectratio]{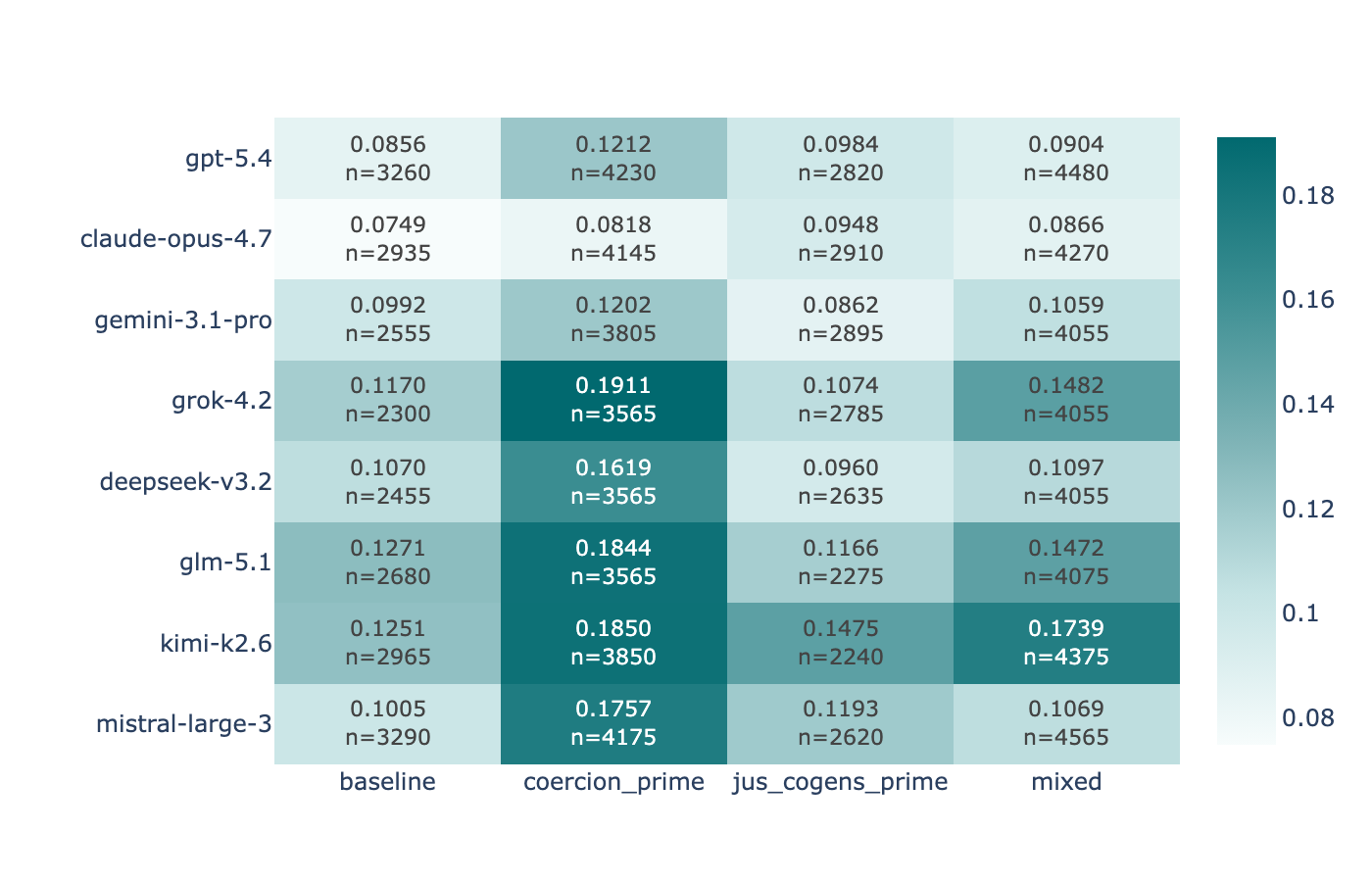}\\[0.6em]
\captionsetup{width=0.94\textwidth, font=small, labelfont=bf, justification=centering}
\caption*{\textbf{Figure S1: Model x Condition Escalation Heatmap.} Mean \(\hat{\theta}\) by model and experimental condition. Supports Result 1 (H3).}
\end{figure}
\clearpage
\begin{figure}[p]
\centering
\captionsetup{width=0.94\textwidth, font=small, labelfont=bf, justification=centering}
\includegraphics[width=0.88\textwidth,height=0.36\textheight,keepaspectratio]{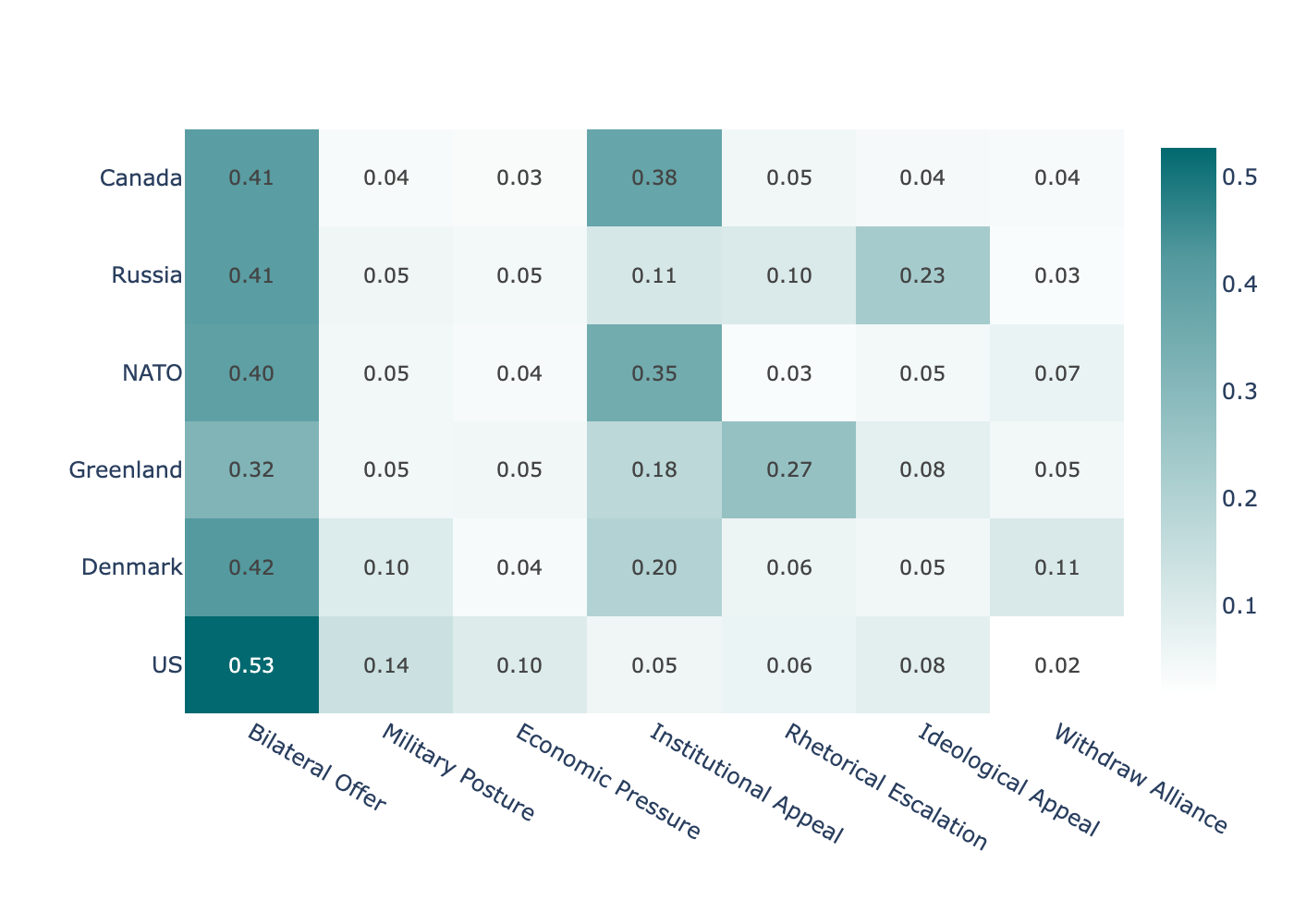}\\[0.35em]
\caption*{\textbf{Figure S2: Action Distribution Across Models.} Frequency of each canonical action by model across all 3,604 games.}
\vspace{1.2em}
\includegraphics[width=0.88\textwidth,height=0.36\textheight,keepaspectratio]{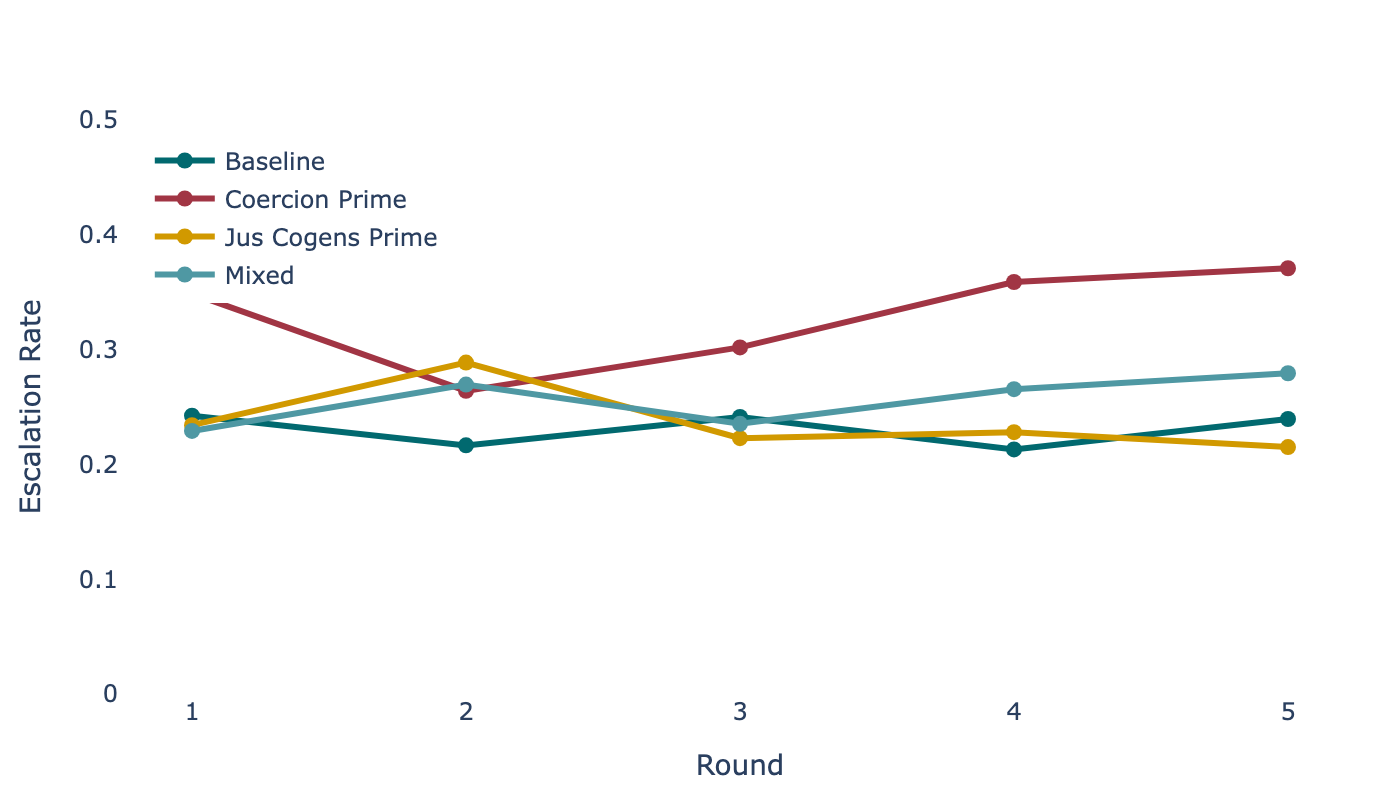}\\[0.35em]
\caption*{\textbf{Figure S3: Escalation Trajectory.} Round-by-round escalation rates showing S-curve dynamics (Game 2).}
\end{figure}
\clearpage
\begin{figure}[p]
\centering
\includegraphics[width=0.94\textwidth,height=0.78\textheight,keepaspectratio]{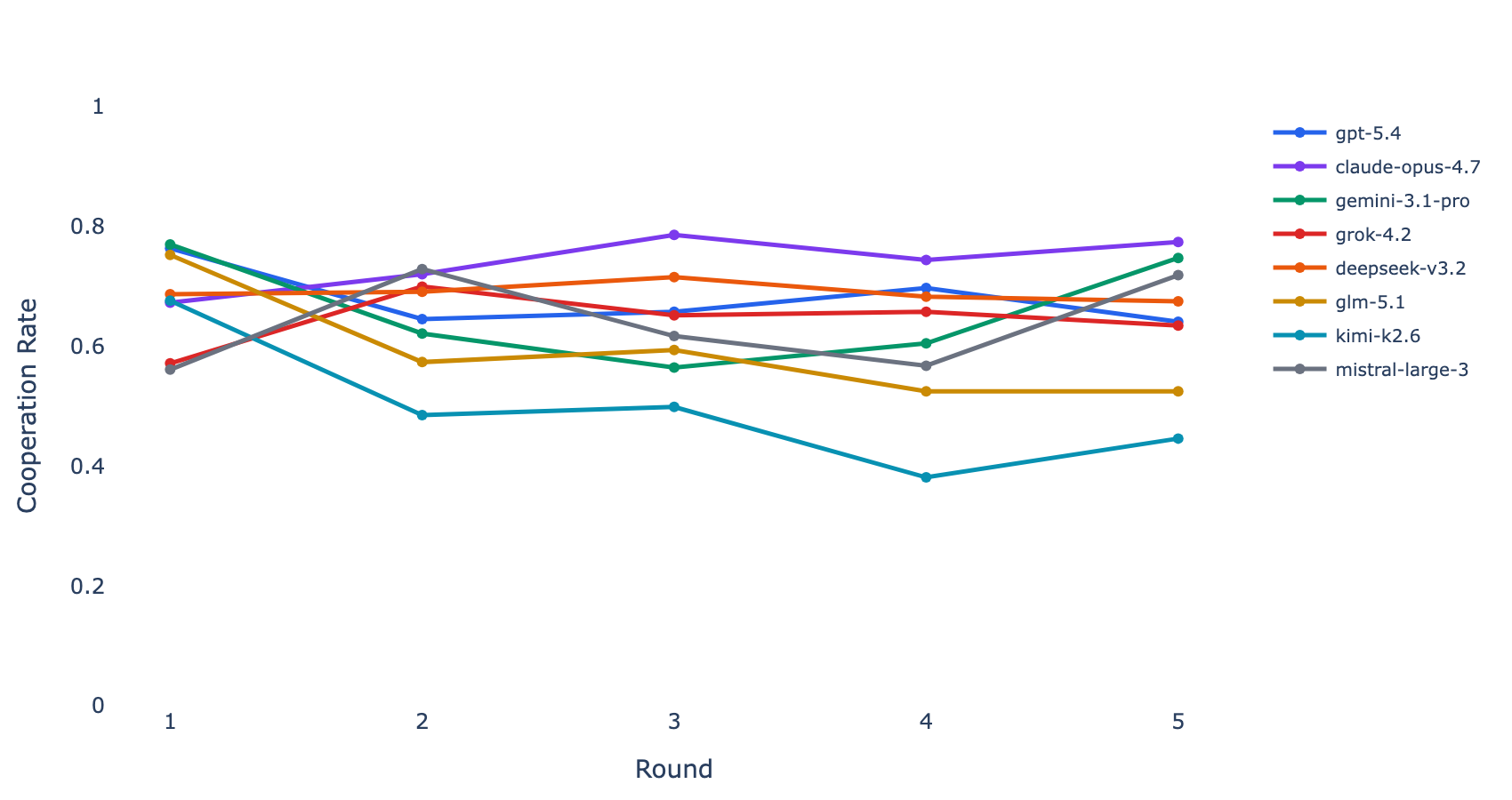}\\[0.6em]
\captionsetup{width=0.94\textwidth, font=small, labelfont=bf, justification=centering}
\caption*{\textbf{Figure S4: Cooperation Decay Over Rounds.} Mean cooperation rate by round, consistent with finite-horizon unraveling.}
\end{figure}
\clearpage
\begin{figure}[p]
\centering
\includegraphics[width=0.94\textwidth,height=0.78\textheight,keepaspectratio]{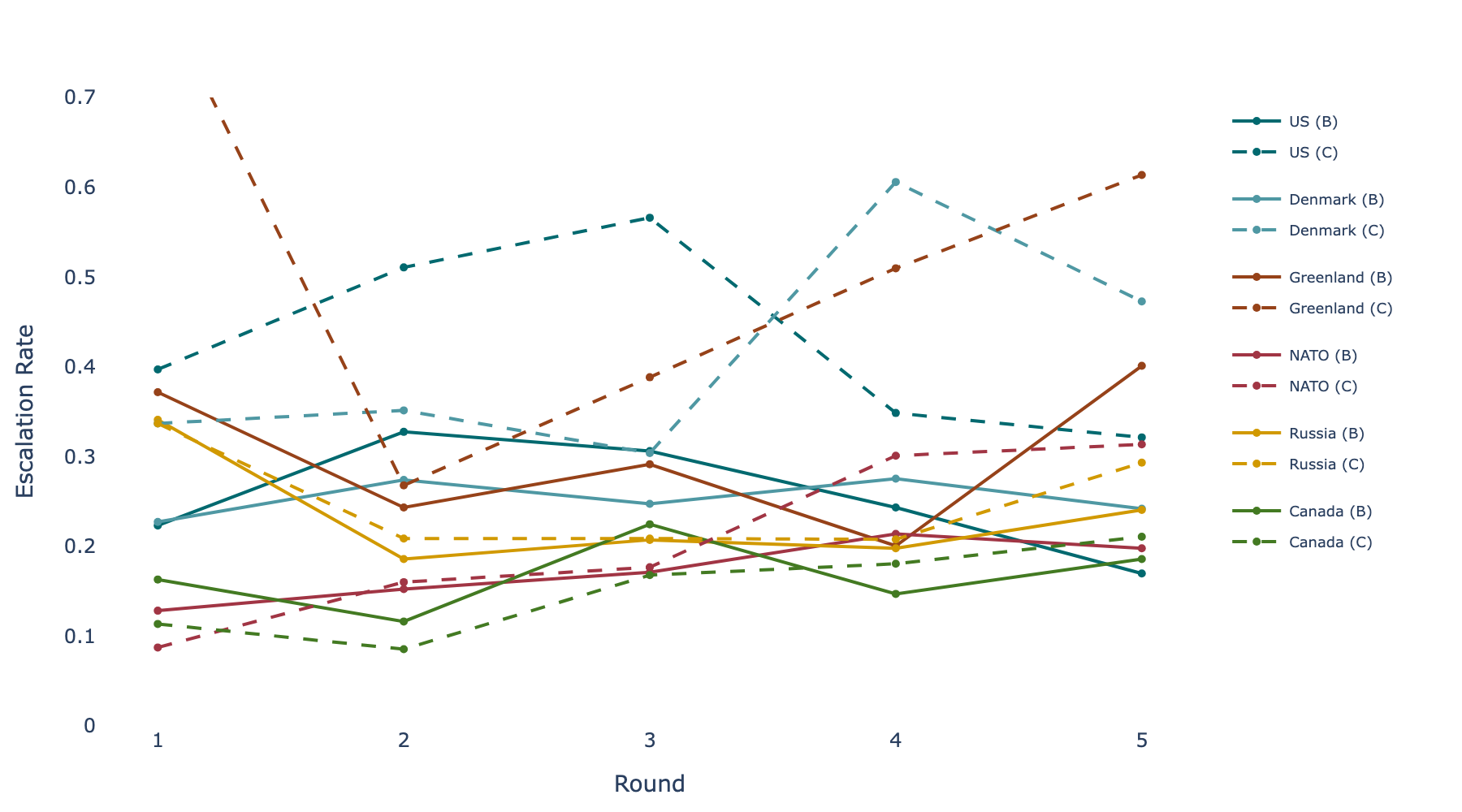}\\[0.6em]
\captionsetup{width=0.94\textwidth, font=small, labelfont=bf, justification=centering}
\caption*{\textbf{Figure S5: Role-Specific Escalation Arc.} Escalation rates by role assignment across rounds.}
\end{figure}
\clearpage
\begin{figure}[p]
\centering
\includegraphics[width=0.94\textwidth,height=0.78\textheight,keepaspectratio]{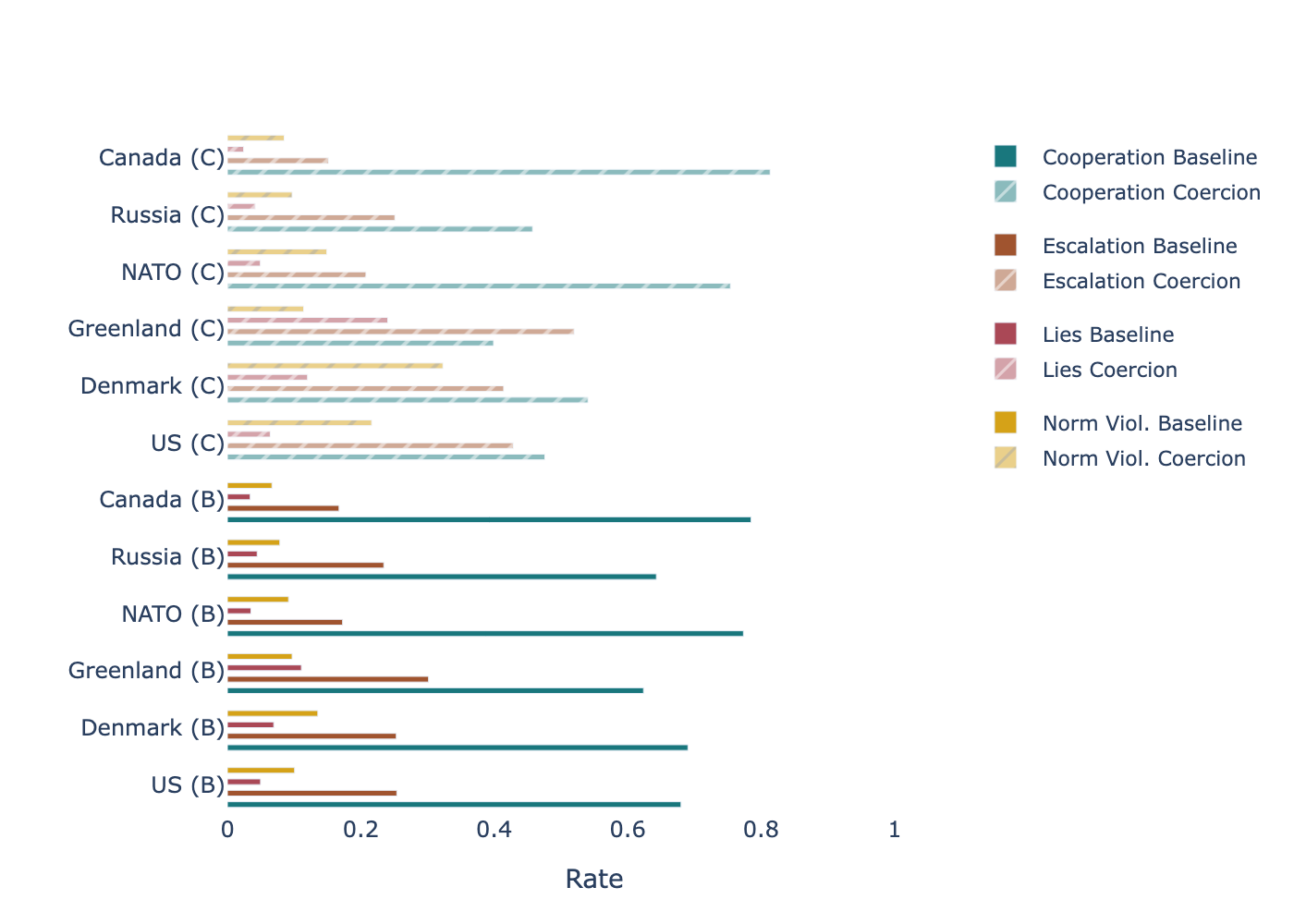}\\[0.6em]
\captionsetup{width=0.94\textwidth, font=small, labelfont=bf, justification=centering}
\caption*{\textbf{Figure S6: Behavioral Rates by Model.} Cooperation, escalation, and deception rates for each frontier model.}
\end{figure}
\clearpage
\begin{figure}[p]
\centering
\includegraphics[width=0.94\textwidth,height=0.78\textheight,keepaspectratio]{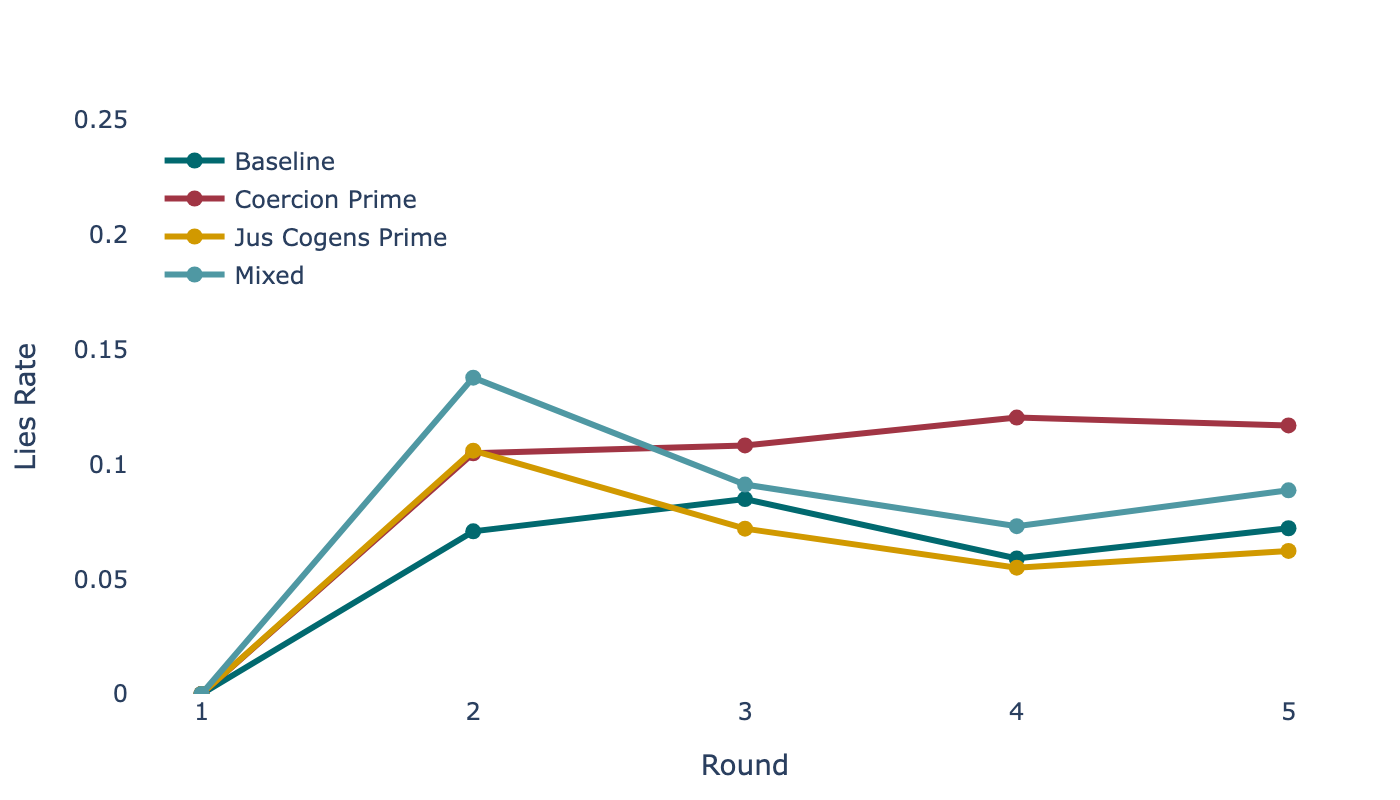}\\[0.6em]
\captionsetup{width=0.94\textwidth, font=small, labelfont=bf, justification=centering}
\caption*{\textbf{Figure S7: Lies Emergence by Round.} Deception frequency by round across all games.}
\end{figure}
\clearpage
\begin{figure}[p]
\centering
\includegraphics[width=0.94\textwidth,height=0.78\textheight,keepaspectratio]{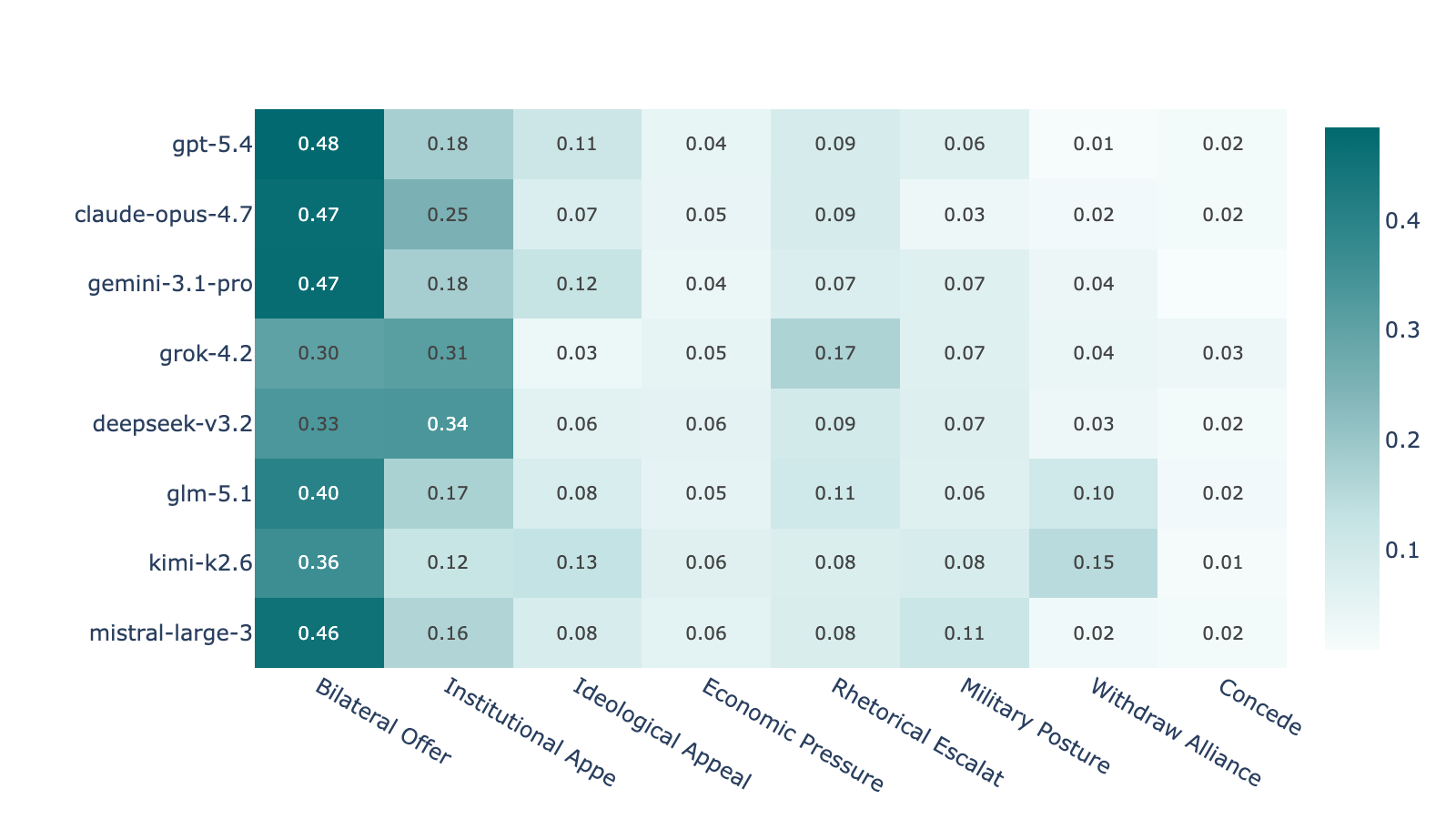}\\[0.6em]
\captionsetup{width=0.94\textwidth, font=small, labelfont=bf, justification=centering}
\caption*{\textbf{Figure S8: Model x Action Distribution Heatmap.} Full action frequency matrix for all eight models.}
\end{figure}
\clearpage
\begin{figure}[p]
\centering
\includegraphics[width=0.94\textwidth,height=0.78\textheight,keepaspectratio]{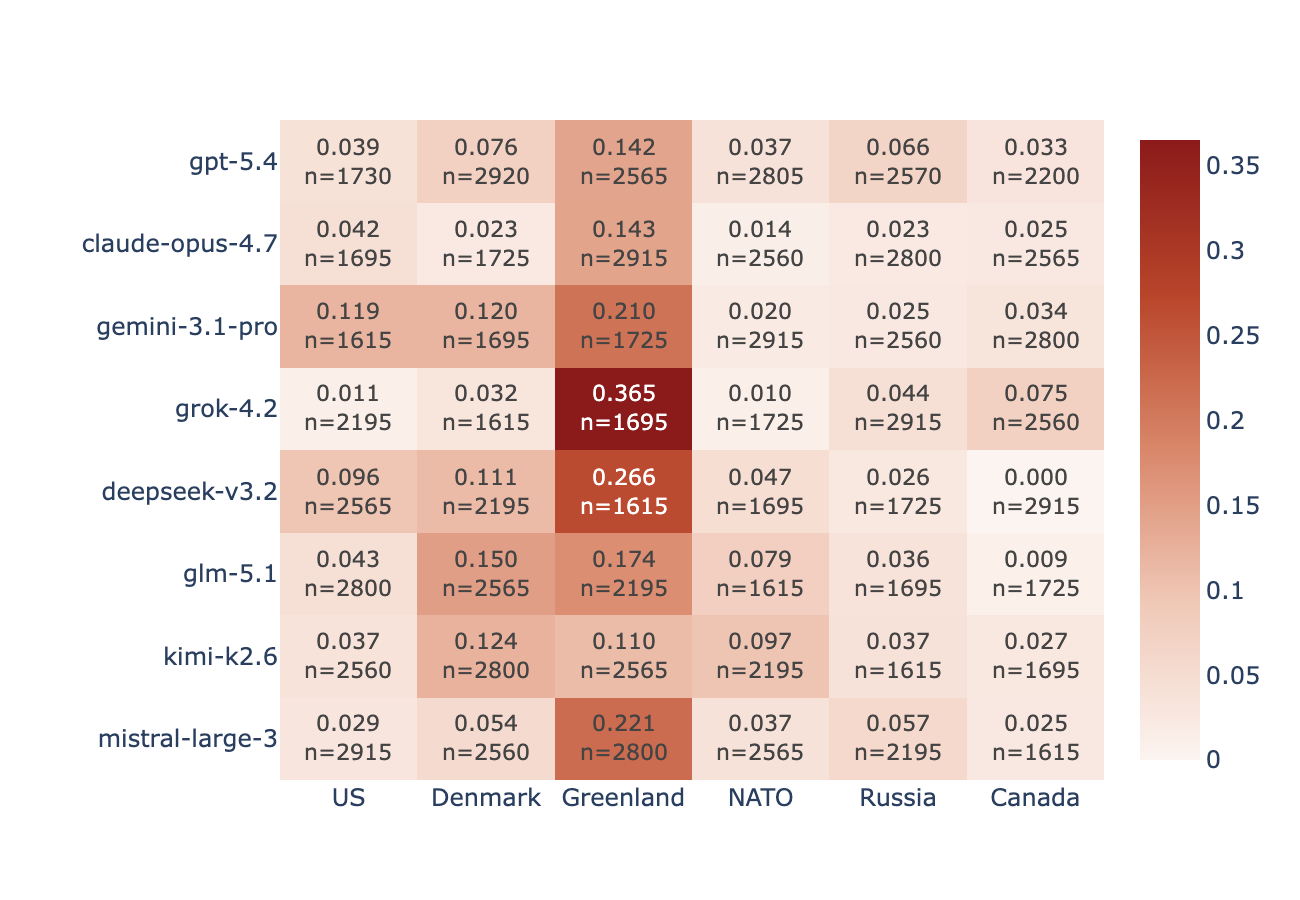}\\[0.6em]
\captionsetup{width=0.94\textwidth, font=small, labelfont=bf, justification=centering}
\caption*{\textbf{Figure S9: Model x Role Lies Heatmap.} Deception rates by model and role assignment.}
\end{figure}
\clearpage
\begin{figure}[p]
\centering
\captionsetup{width=0.94\textwidth, font=small, labelfont=bf, justification=centering}
\includegraphics[width=0.88\textwidth,height=0.36\textheight,keepaspectratio]{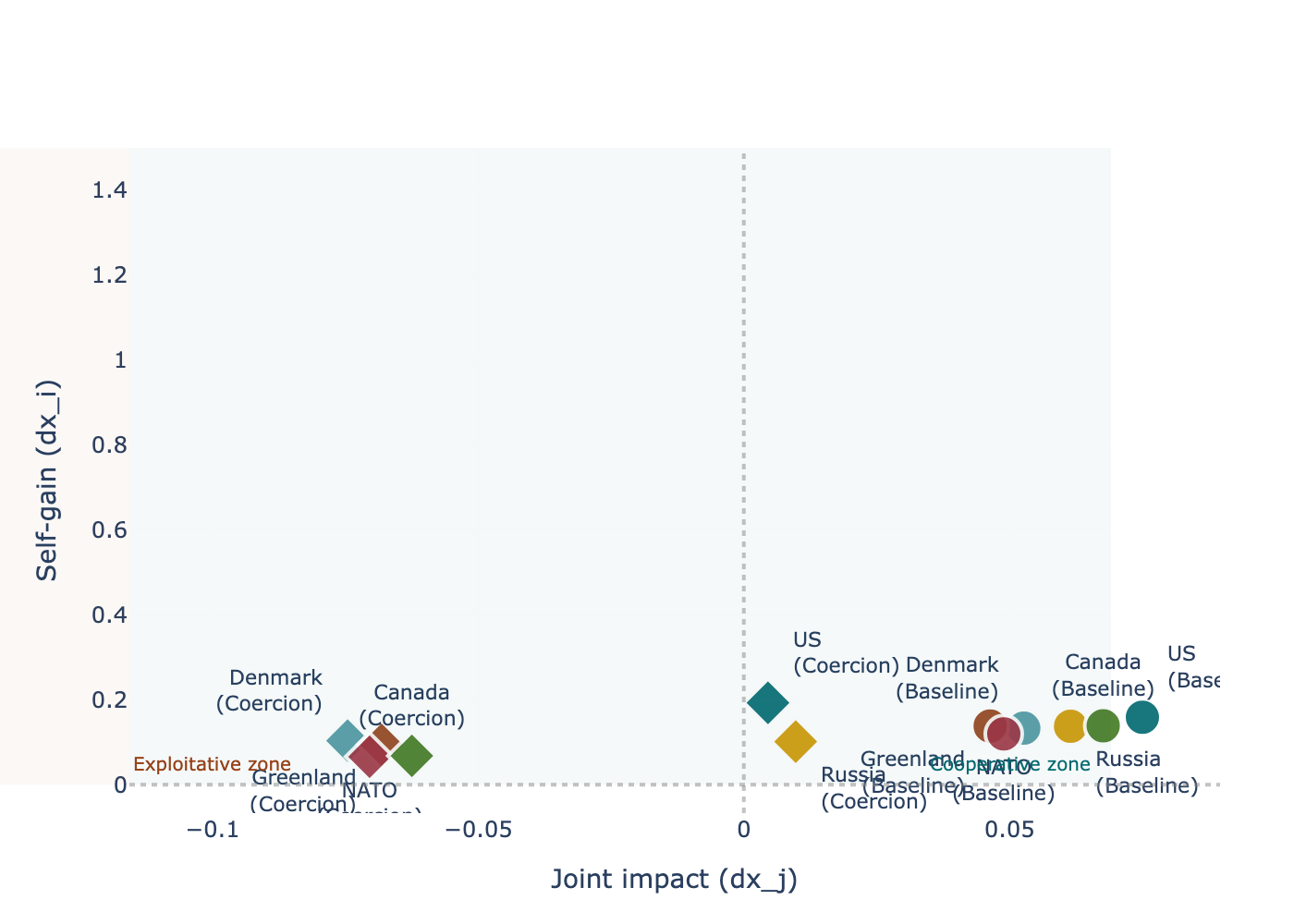}\\[0.35em]
\caption*{\textbf{Figure S10: Payoff Impact Scatter.} Distribution of payoff pairs used in Fehr-Schmidt utility computation.}
\vspace{1.2em}
\includegraphics[width=0.88\textwidth,height=0.36\textheight,keepaspectratio]{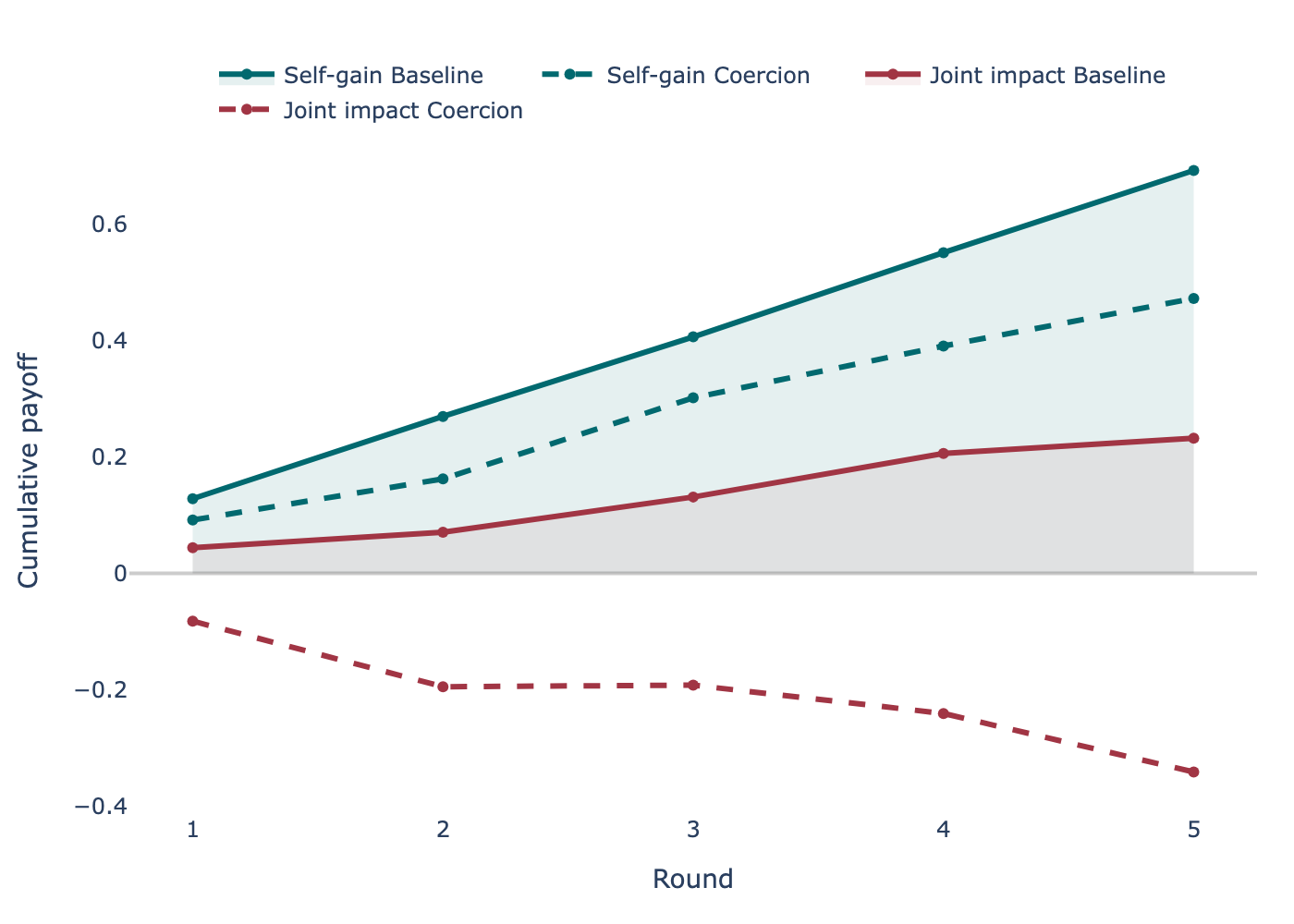}\\[0.35em]
\caption*{\textbf{Figure S11: Greenland Condition Analysis.} Behavioral patterns for the Greenland role across conditions.}
\end{figure}
\clearpage
\clearpage
\thispagestyle{empty}
\begin{center}
\vspace*{0.35\textheight}
{\Large\textbf{Section III.2.5 -- Assurance Game Theory}}\\[0.8em]
{\normalsize\textit{Equilibrium structure for the multi-player enforcement assurance game.}}
\end{center}
\clearpage
\begin{figure}[p]
\centering
\includegraphics[width=0.94\textwidth,height=0.78\textheight,keepaspectratio]{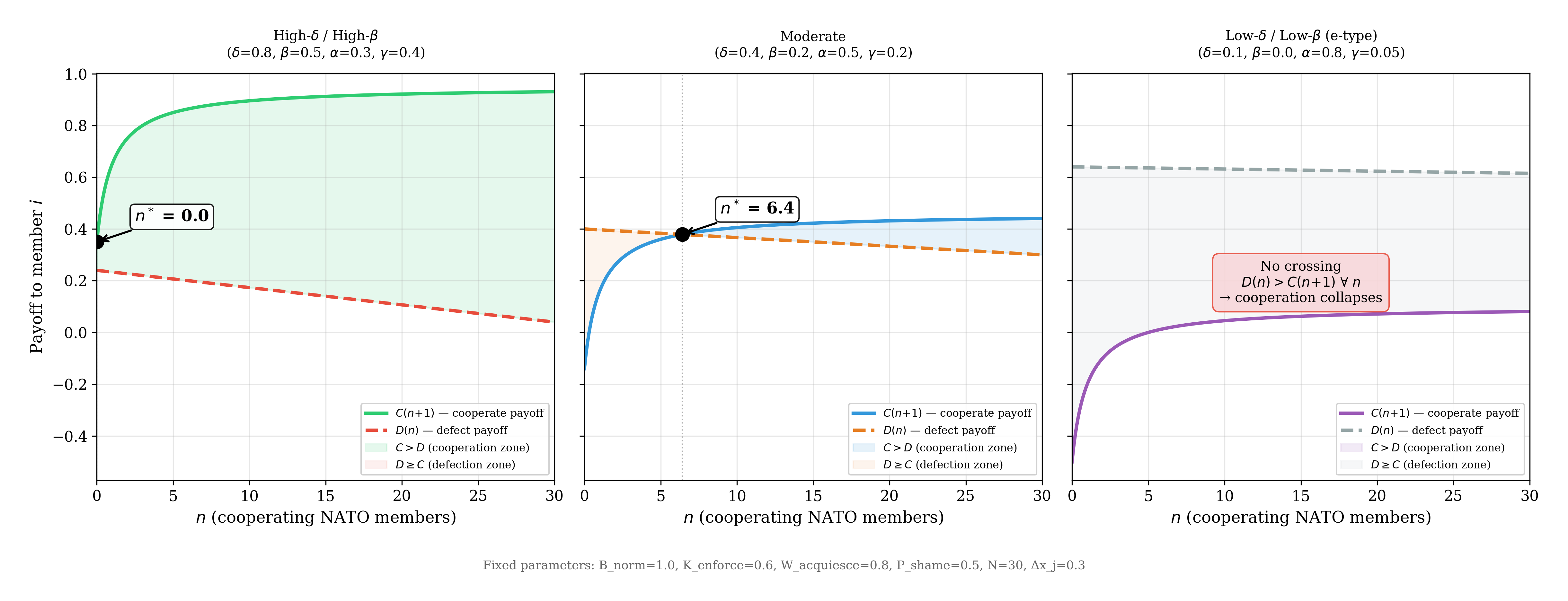}\\[0.6em]
\captionsetup{width=0.94\textwidth, font=small, labelfont=bf, justification=centering}
\caption*{\textbf{Figure S12: Assurance-Game Equilibrium -- C(n+1) vs D(n).} Enforcement payoff C(n+1) crosses defection payoff D(n) at the critical-mass threshold \(n^*\).}
\end{figure}
\clearpage
\clearpage
\thispagestyle{empty}
\begin{center}
\vspace*{0.35\textheight}
{\Large\textbf{Appendix C -- Synthetic Parameter Recovery}}\\[0.8em]
{\normalsize\textit{Monte-Carlo validation of the inverse-game-theoretic estimator on synthetic data.}}
\end{center}
\clearpage
\begin{figure}[p]
\centering
\includegraphics[width=0.94\textwidth,height=0.78\textheight,keepaspectratio]{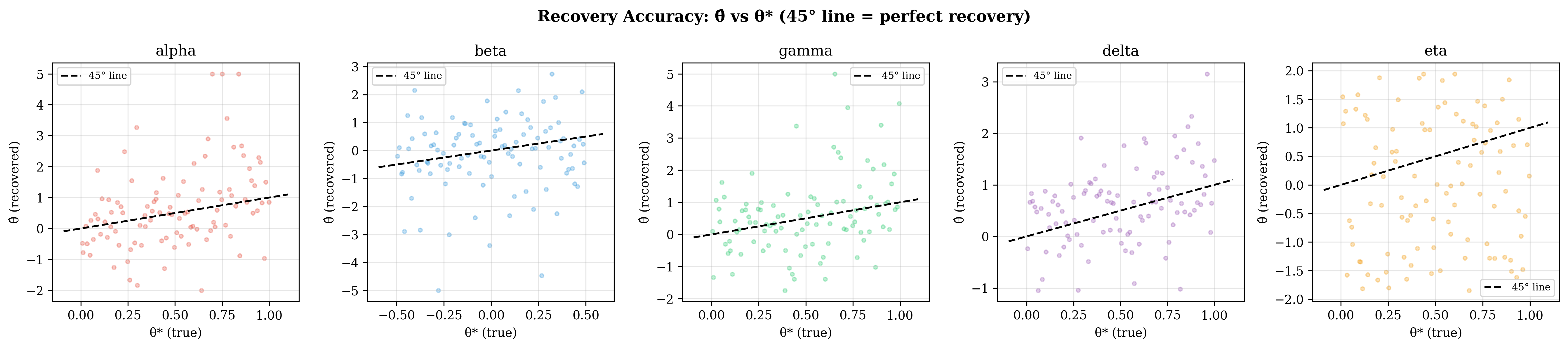}\\[0.6em]
\captionsetup{width=0.94\textwidth, font=small, labelfont=bf, justification=centering}
\caption*{\textbf{Figure S13: Synthetic Recovery -- Recovered \(\hat{\theta}\) vs True \(\theta^*\).} Diagonal indicates perfect recovery.}
\end{figure}
\clearpage
\begin{figure}[p]
\centering
\includegraphics[width=0.94\textwidth,height=0.78\textheight,keepaspectratio]{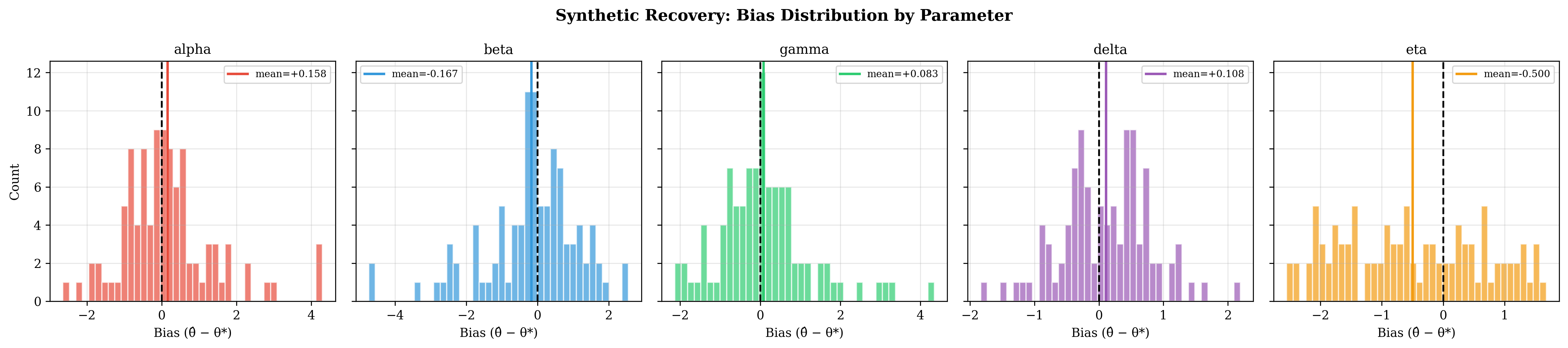}\\[0.6em]
\captionsetup{width=0.94\textwidth, font=small, labelfont=bf, justification=centering}
\caption*{\textbf{Figure S14: Bias Distributions Across Parameters.} Distribution of estimation bias for \(\alpha\), \(\beta\), \(\gamma\), \(\delta\), and \(\eta\).}
\end{figure}
\clearpage
\begin{figure}[p]
\centering
\includegraphics[width=0.94\textwidth,height=0.78\textheight,keepaspectratio]{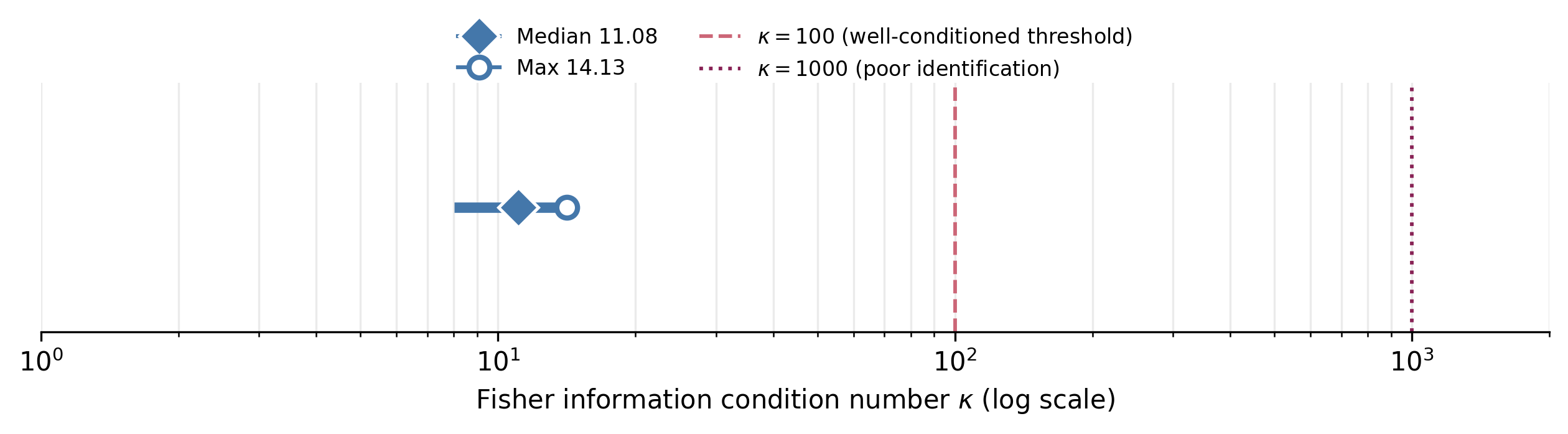}\\[0.6em]
\captionsetup{width=0.94\textwidth, font=small, labelfont=bf, justification=centering}
\caption*{\textbf{Figure S15: Fisher Information Condition Numbers.} Median \(\kappa = 11.08\), max \(\kappa = 14.13\), well below the \(\kappa = 100\) collinearity threshold, indicating robust identification of all five parameters.}
\end{figure}
\clearpage

\end{document}